\documentclass[twocolumn,twocolappendix]{aastex631} 
\usepackage{CJK}
\usepackage{multirow}
\usepackage{mathrsfs}

\shorttitle{X-ray spectrum and time lags}
\shortauthors{Hu et al.}
\graphicspath{{./}{figures/}}

\usepackage{mathrsfs}

\newcommand\XMM{XMM-Newton }

\begin{document}
\begin{CJK*}{UTF8}{gbsn}
\title{Integrated Study of X-ray Spectrum and Time Lags for HBL Mrk 421 within the Framework of the Multiple-Zone Leptonic Model}

\correspondingauthor{Wen Hu}
\email{huwen.3000@jgsu.edu.cn}

\correspondingauthor{Jun-Xian Wang}
\email{jxw@ustc.edu.cn}

\author{Wen Hu（胡文）} 
\affiliation{Department of Physics, Jinggangshan University, Ji'an, 343009, P. R. China\\}

\affiliation{CAS Key Laboratory for Research in Galaxies and Cosmology, Department of Astronomy, University of Science and Technology of China, Hefei, Anhui 230026, P. R. China\\}

\author{Jia-Lai Kang（康嘉来）} 
\affiliation{CAS Key Laboratory for Research in Galaxies and Cosmology, Department of Astronomy, University of Science and Technology of China, Hefei, Anhui 230026, P. R. China\\}
\affiliation{School of Astronomy and Space Science, University of Science and Technology of China, Hefei 230026, P. R. China\\}

\author{Zhen-Yi Cai（蔡振翼）}
\affiliation{CAS Key Laboratory for Research in Galaxies and Cosmology, Department of Astronomy, University of Science and Technology of China, Hefei, Anhui 230026, P. R. China\\}
\affiliation{School of Astronomy and Space Science, University of Science and Technology of China, Hefei 230026, P. R. China\\}

\author{Jun-Xian Wang（王俊贤）}
\affiliation{CAS Key Laboratory for Research in Galaxies and Cosmology, Department of Astronomy, University of Science and Technology of China, Hefei, Anhui 230026, P. R. China\\}
\affiliation{School of Astronomy and Space Science, University of Science and Technology of China, Hefei 230026, P. R. China\\}

\author{Zhen-Bo Su（苏镇波）}
\affiliation{CAS Key Laboratory for Research in Galaxies and Cosmology, Department of Astronomy, University of Science and Technology of China, Hefei, Anhui 230026, P. R. China\\}
\affiliation{School of Astronomy and Space Science, University of Science and Technology of China, Hefei 230026, P. R. China\\}

\author{Guang-Cheng Xiao（肖广成）} 
\affiliation{Department of Physics, Jinggangshan University, Ji'an, 343009, P. R. China\\}



\begin{abstract} 
We present the timing analysis of 10 archived \XMM observations with an exposure of $>40$ ks of Markarian 421.
Mrk 421 is the brightest high-frequency-peaked BL Lac object (HBL) emitting in X-rays produced by electrons accelerated in 
the innermost regions of a relativistic jet pointing toward us.
For each observation, we construct averaged X-ray spectra in 0.5--10 keV band, as well as 100 s binned light curves (LCs) in various subbands.
During these observations, the source exhibited various intensity states differing by close to an order of magnitude in flux, 
with the fractional variability amplitude increasing with energy through the X-ray band.
Bayesian power spectral density analysis reveals that the X-ray variability can be characterized by a colored noise, 
with an index ranging from $\sim-1.9$ to $-3.0$.
Moreover, both the standard cross-correlation function and cross-spectral methods indicate that the amount 
of time lags increases with the energy difference between two compared LCs.
A time-dependent two-zone jet model is developed to extract physical information from the X-ray emission of Mrk 421.
In the model, we assume that the jet emission mostly comprises a quasi-stationary component and a highly variable one. 
Our results show that the two-zone model can simultaneously provide a satisfactory description for both the X-ray spectra 
and time lags observed in different epochs, 
with the model parameters constrained in a fully acceptable interval.
We suggest that shocks within the jets may be the primary energy dissipation process responsible for triggering the rapid variability, 
although magnetic reconnection cannot be excluded. 
\end{abstract}



\section{Introduction} \label{sec:intro}

The highly non-thermal continuum emission from blazars is believed to originate from a relativistic jet, directed 
at a small angle ($\lesssim10^\circ$) with respect to our line of sight \citep{Urry1995PASP}.
The broadband spectral energy distributions (SEDs) of blazars are characterized by a typical two-hump structure: 
a low-energy (LE) hump that lies in the infrared to X-ray range and a high-energy (HE) hump that appears 
in the $\gamma$-rays ranging from MeV to TeV energies \citep{Fossati1998MNRAS,Abdo2010a,Fan2016ApJS}.
BL Lacertae (BL Lac) objects are a subclass of blazars distinguished by their weakness or absence of broad emission lines in the optical band. 
Based on the peak frequency of the LE hump, BL Lacs are further subdivided into
subclasses of high-, intermediate- and low-energy peaked BL Lac objects: HBLs, IBLs, LBLs, respectively.
LBLs are defined to be objects with the LE hump in the IR-optical band, and HBLs as those with the LE hump peaking in the UV to the X-ray band. 
IBLs are distinguished by their LE hump peaking at optical to UV frequencies.

Another remarkable characteristic of blazars is the flux variation throughout the whole observable electromagnetic 
spectrum (i.e., from radio to very high-energy $\gamma$-rays) in diverse timescales ranging from a few minutes to years. 
The nature of the variability is similar to red noise, i.e., longer timescale variability has larger amplitude.
Variations in flux are most dramatic and occur on timescales as short as a few hours or less at X-ray/TeV $\gamma$-ray energies 
\citep[e.g.,][]{Ferrero2006A&A,Abeysekara2017ApJ,Yan2018,Noel2022ApJS,Devanand2022ApJ}.
Such rapid variability is commonly known as microvariability, intraday variability (IDV), or intranight variability,
and is thought to be caused by the acceleration and radiative cooling of the emitting particles in the jets 
\citep[e.g.,][]{Zhang2002MNRAS,Zhang2006ApJ,Bottcher2019ApJ,Das2023MNRAS}.

It is commonly accepted that the LE hump is attributed to the synchrotron radiation produced by highly relativistic electrons/positrons in the blazar jet.
 The HE $\gamma$-rays hump is primarily interpreted as the inverse Compton (IC) scattering of the internal synchrotron photons \cite[SSC; e.g.,][]{Bloom1996}, 
 or as the scattering of photons intercepting the jet from an external source  \citep[EC; e.g.,][]{Dermer2009ApJ}, such as the accretion disk, 
 the broad-line region, or the dusty/molecular torus.
For BL Lacs, there is a relatively simple picture that most of the non-thermal emission from the jet can arise as the synchrotron 
and SSC radiation of relativistic electrons accelerated in a single blob of plasma
which itself moves relativistically outward along the jet.
In particular for HBL Mrk 421, the picture is strongly supported by the correlated X-ray/TeV flares and the roughly 
quadratic dependence of the X-ray and $\gamma$-ray fractional variability \citep[e.g.,][]{Fossati2008ApJ,Aleksic2015,Balokovic2016}.

According to the simplified picture, various theoretical time-dependent and steady models have been developed \citep[e.g.,][]{Mastichiadis1997A&A,Kirk1998A&A,Chiaberge1999MNRAS,Kusunose2000ApJ,Li2000ApJ,Bottcher2002ApJ,Finke2008,Zacharias2013ApJ,Boula2022}.
Such theoretical models have led to significant progress in our understanding of the jet physics and the underlying particle acceleration and cooling processes in HBLs.
However, the conventional one-zone models are faced with a challenge posed by various variability and correlation studies 
\citep[e.g.,][]{Fossati2000ApJa,Fossati2000ApJb,Brinkmann2005A&A,Aharonian2009A&A,Abeysekara2020,Acciari2021A&A}.
One of the most peculiar aspects of blazar variability is the orphan flares detected in various blazars, e.g., orphan optical-UV flare observed in PKS 2155-304 \citep{Wierzcholska2019Galax}, orphan X-ray flare observed in S5 0716+714 \citep{Rani2013A&A}, and orphan TeV $\gamma-$ray flare observed in B2 1215+30 \citep{Abeysekara2017a}. In these orphan events, the absence of correlated variability in other bands poses a serious issue for the single-zone SSC model, where the emission at various bands is thought to be radiated by the same population of relativistic electrons.
Actually, there may be a need for two distinct zones to explain the observations 
\citep[e.g.,][]{Tavecchio2011A&A,MacDonald2015ApJ,Prince2019ApJ,Xue2019ApJ,Aguilar2022MNRAS,Wang2022PhRvD,Liu2023MNRAS}.
In the scenario, one zone accounts for quiescent behavior, 
while another zone may be responsible for a detected outburst triggered by a localized energetic dissipation events 
such as magnetic reconnection \citep[e.g.,][]{Giannios2009,Giannios2013}, shocks \citep[e.g.,][]{Marscher1985ApJ,Spada2001MNRAS,Larionov2013ApJ}, 
and MHD instabilities within the jet itself \citep[e.g.,][]{Giannios2006A&A,Nalewajko2012MNRAS}.

The optimal radiative window for exploring the variations in HBLs is X-ray, which is believed to be the synchrotron radiation 
arising from the high-energy tail of electrons accelerated in the innermost region of the jets.
Characterizing the X-ray variability of HBLs is therefore important for exploring the physical nature of acceleration and cooling processes in blazar jets.
 A general property observed in HBLs is the "harder-when-brighter" behavior, which refers to a trend where an increase in flux 
 is accompanied by a hardening of the spectrum as expected from a shift of the synchrotron peak to higher energies 
 \citep[e.g.,][]{Brinkmann2003A&A,Ravasio2004A&A,Abeysekara2017ApJ,Gupta2020Galax,Chandra2021ApJ,Noel2022ApJS}.
Of particular interest is the formation of loops in the hardness ratio-intensity (HR-I) diagrams, which provides a model-independent 
diagnostic of the underlying physical process responsible for the observed emission.
In theory, a clockwise loop in the HR-I diagram may indicate that the cooling process dominates the system, 
while an anti-clockwise loop may be caused by the acceleration process.
The hysteresis of X-ray spectral variability can serve as an indication on whether the acceleration or the cooling process controls the observed variations.

Instead, the X-ray time delays measured between the soft-energy and hard-energy band variations, which are accompanied by the hysteresis loop, 
can allow us to deduce the underlying physical parameters for acceleration and radiation in the jet \citep{Zhang2006ApJ,Das2023MNRAS,Gokus2024MNRAS}. 
Particularly, the X-ray Fourier time lags could consistently provide stronger constraints and offer valuable insights 
into the underlying mechanisms at work in the blazar jet \citep{Zhang2002MNRAS,Kroon2014ApJ,Finke2014ApJ,Finke2015ApJ,Lewis2016}.

In our work, the primary purpose is to explore the physics of blazar jets by modeling both the X-ray spectrum and the associated time lags.
To quantitatively diagnose the cooling and acceleration behavior of the electrons responsible for the observed X-ray emission, 
both the cross-correlation function (CCF) and cross-spectral methods are employed to determine the time lags of variability in different X-ray bands.
Then, we perform a detailed theoretical modeling of the time-dependent electron acceleration and radiation 
transport processes in the jets to extract the physical information contained in the X-ray emission.
We focus our analysis on he IDV timescale X-ray variability of the famous HBL Mrk 421, using data taken from the public archives of \XMM.

The paper is organized as follows. 
Section \ref{sec:predi} gives the theoretical time lags resulting from the synchrotron cooling and Fermi acceleration mechanisms.
In Section \ref{sec:analysis}, we describe the archived \XMM observations of Mrk 421, and data reduction, as well as the X-ray variability and time lags analysis of the public archival data.
In Section \ref{sec:model}, we present the theoretical model and the modeling results.
A discussion follows in Section \ref{sec:diss}. Finally, we summarize the results in Section \ref{sec:summary}.

\begin{deluxetable}{lcccc}
\centerwidetable
\tabletypesize{\scriptsize}
\tablecaption{The Mrk 421 observations from the EPIC-pn on broad XMM-Newton.}
\label{tab:table1}
\tablehead{
\colhead{Obs. ID}  & \colhead{Date}  & \colhead{MJD} &  \colhead{PN mode}  &  \colhead{Exp.}   \\
   & \colhead{(yyyy-mm-dd)}  & \colhead{} &  \colhead{}  &  \colhead{ks}                                        
}  
\startdata
0099280301	& 2000-11-13	&51861.93& SW-TK    & 46.6		\\	
0136540701	& 2002-11-14	&52592.02& LW-TK 	   & 69.9		\\	
0136541001	& 2002-12-01	&52609.97& TI-ME 	   & 69.9	 	\\	
0158970101	&2003-06-01	&52791.55& SW-TN 	   & 40.9		\\			
0150498701	& 2003-11-14	&52957.69& TI-TN 	   & 47.7	 \\	
0158971201	&2004-05-06	&53131.12& TI-ME 	   & 64.9		\\	
0158971301	&2005-11-09	&53683.78& TI-TN 	   & 58.7		\\	
0302180101	&2006-04-29	&53854.88& TI-TN 	   & 40.6		\\	
0411080301	&2006-05-28	&53883.09& SW-ME   & 68.4		\\	
0502030101	&2008-05-07	&54593.08& TI-TN  	   & 41.9 		 \\
\enddata 
\tablecomments{ SW, LW and TI stand for Small Window, Large Window and Timing science modes, respectively. TK, ME and TN stand for Thick, Median, and Thin filters, respectively.}
\end{deluxetable}

\section{Theoretical prediction}\label{sec:predi}
It is commonly accepted that the X-rays from blazars are attributed to the synchrotron emission of highly relativistic electrons in the jets.
The observed soft lag between the soft $E_{\rm s,keV}$ and high $E_{\rm h, keV}$ energy band can be estimated from 
\citep[e.g.,][]{Zhang2002MNRAS,Ravasio2004A&A,Zhang2006ApJ,Devanand2022ApJ,Das2023MNRAS}
\begin{eqnarray}\label{lag:cool}  
\tau_{\rm s}\simeq -53.7(1+z)^{\frac{1}{2}}\delta_{\rm D}^{-\frac{1}{2}}B'^{-\frac{3}{2}}\Delta{E}~\rm min,
\end{eqnarray}
where $\Delta{E}\equiv\left(E_{\rm s,keV}^{-1/2}-E_{\rm h,keV}^{-1/2}\right)$, $z$ is the source's redshift, $B'$ is the magnetic 
field strength and $\delta_{\rm D}$ is the Doppler factor of the emission region in the jet.
In the derivation, $E_{\rm s/h,keV}$ is the characteristic synchrotron energy of electrons with the Lorentz factor $\gamma'$.
Note that the simple estimation of the time delay doubles that in \cite{Finke2014ApJ} based on the analytical 
solution of the electron continuity equation using Fourier transformation method.

On the other hand, the acceleration process is not well understood yet and could operate in different ways.
Particularly, diffusive shock acceleration is a widely discussed mechanism responsible for electron acceleration in the blazar jets.
In the case of shock acceleration, the mean electron acceleration rate can be expressed as \citep[see][and the references therein]{Lewis2016}
\begin{equation}\label{eqs:shock}
\dot\gamma'_{\rm sh}=A_0^{\rm sh}\gamma',~A_0^{\rm sh}\equiv \frac{u_-^2}{c \ell_{\rm MHD}\chi}\frac{\chi-1}{\chi+1}
\end{equation}
where $u_-$ is the upstream velocity measured in the frame of the shock, $\chi$ is the shock compression ratio, 
$\ell_{\rm MHD}$ is the coherence length of the magnetohydrodynamic (MHD) waves.

The electrons may also experience stochastic acceleration due to resonant interactions with MHD turbulence.  
In the quasi-linear theory \citep[e.g.,][]{Schlickeiser1989ApJ,Becker2006}, 
the MHD turbulence is described by a combination of Alfv\'{e}n waves with different wave numbers. In the stochastic 
particle-wave interactions, the momentum diffusion coefficient of particles
 is determined by the wave-turbulence power spectrum as $W(k)\propto k^{-q}$, where $k$ is the wavenumber.
 Here, we consider the hard-sphere turbulent spectrum (i.e., $q=2$), leading to the acceleration timescale being independent of the electron energy. 
 It favours the production of the brightest flares due to the most efficient re-acceleration of high-energy electrons.
 Moreover, the stochastic acceleration model can reasonably reproduce the broad-band blazar spectra 
 \citep[e.g.,][]{Katarzy2006A&A,Weidinger2010ASTRA,Dmytriiev2021MNRAS}, 
 in particular, very hard spectra \citep[e.g.,][]{Stawarz2008} and curved spectra \citep[e.g.,][]{Tramacere2011}.
 Such results suggest that the hard-sphere index may be supported in the blazar jet.

In the hard sphere approximation, the momentum diffusion coefficient can be phenomenologically described as
\begin{equation}
D_p(\gamma')=D_0\gamma'^2,
\end{equation}
for ultra-relativistic electrons. Then, the mean stochastic energy gain rate is given by
\begin{equation}
\dot\gamma'_{\rm sto}=4D_0\gamma', 
\end{equation}
where $D_0\propto s^{-1}$ is the momentum diffusion rate constant.

In either case,  the hard lag is expected and the observed hard lag is predicted to be \citep{Hu2021}
\begin{equation}
\tau_{\rm h}\simeq\frac{1+z}{\delta_{\rm D}}t_{\rm acc}'\Delta{E},
\end{equation}
where $\Delta{E}\equiv \ln (E_{\rm h,keV})-\ln (E_{\rm s,keV})$, $t'_{\rm acc}$ is an energy-independent acceleration timescale. 
Assuming the synchrotron peak energy $E_{\rm pk,keV}$ is determined by balancing the acceleration and synchrotron cooling, i.e., $t'_{\rm acc}=t'_{\rm syn}(E_{\rm pk,keV})$,
the expected lag can be expressed as
\begin{equation}\label{lag:acc}  
\tau_{\rm h}\simeq53.7(1+z)^{\frac{1}{2}}\delta_{\rm D}^{-\frac{1}{2}}B'^{-\frac{3}{2}}E_{\rm pk,keV}^{-1/2} \Delta{E}~\rm min.
\end{equation}
Note that the simple estimation is valid only for $E_{\rm s,keV}<E_{\rm h,keV}\lesssim E_{\rm pk, keV}$.

Accordingly, the expected lags increase with increasing differences $\Delta{E}$ of the compared energy bands, and
 the signs of lags indicate whether the relativistic electrons responsible for the observed X-rays are dominated by the acceleration or cooling processes.

\section{DATA ANALYSIS and results}\label{sec:analysis}

\begin{figure}
\figurenum{1}
\includegraphics[width=0.23\textwidth] {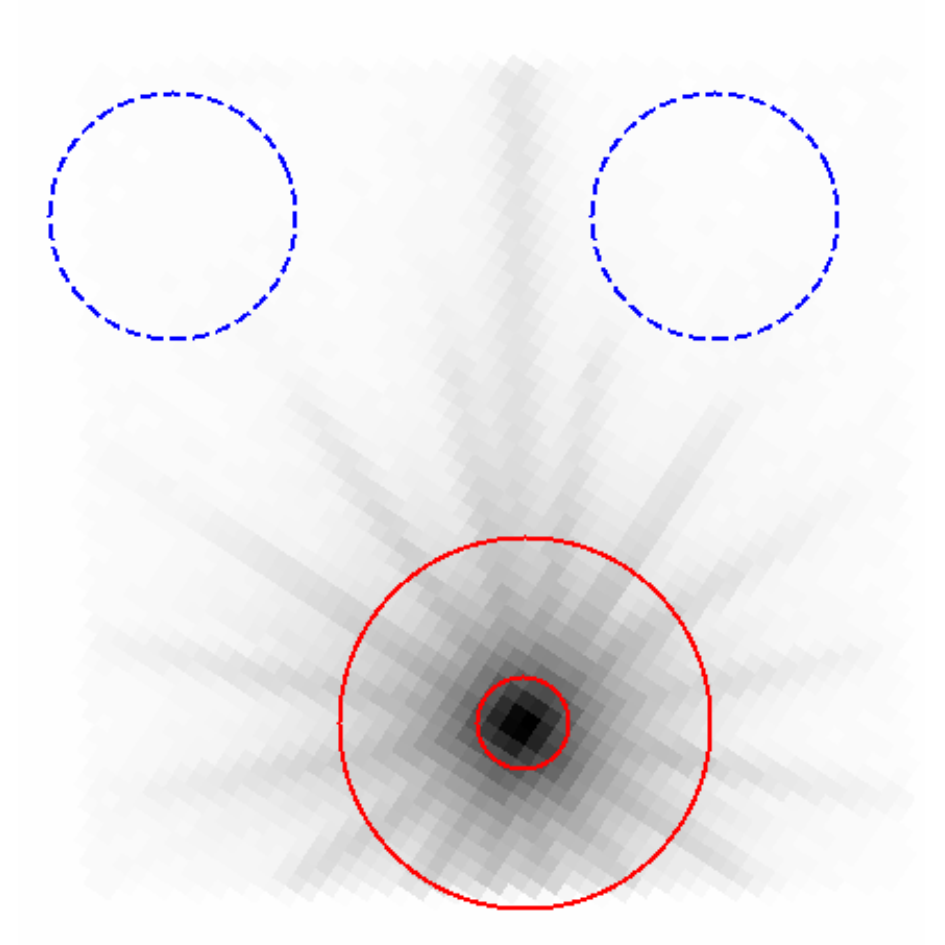}
\includegraphics[width=0.23\textwidth] {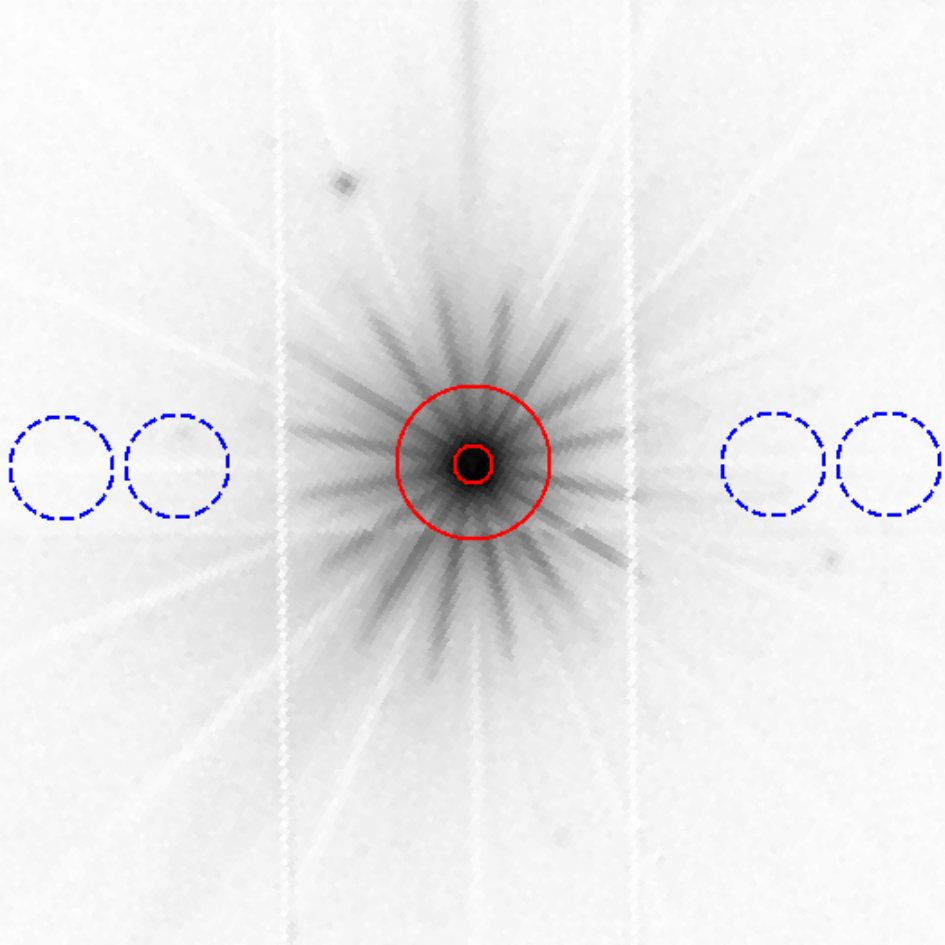} 
\caption{\label{fig:region} Examples of the source and background regions for SW (0099280301) and LW (0136540701) modes. The 0.5--10 keV images derived with \textit{evselect} are plotted using \textit{ds9}. The source regions, red annuli with inner radius of 15\arcsec and outer radius of 60\arcsec , along with the background regions, blue circles (two for SW and four LW modes) with radius of 40\arcsec , are over-plotted.  
}
\end{figure}

\begin{table*}
	\centering
	\caption{The best-fitting log-parabolic model parameters and the fractional rms amplitude for 10 \XMM observations.}
	\label{tab:table2}
	\tabletypesize{\scriptsize}
	\begin{tabular}{lccccccc} 
		\hline\hline
		Obs.ID& $E_{p} F_p$ 	& $E_p$   	   &	$b$      &  \colhead{$F_{\rm var}(\%)$}&  \colhead{$F_{\rm var}(\%)$}&  \colhead{$F_{\rm var}(\%)$}&  \colhead{$F_{\rm var}(\%)$}\\
			   &$10^{-10}~{\rm ergs/cm^2/s}$ 	& keV	  	    & 		&  \colhead{(0.5--1) keV} &  \colhead{(1--2) keV}  &  \colhead{ (2--4) keV }  &  \colhead{(4--10) keV}  	       \\
		\hline
	0099280301& $4.43\pm0.17$	& $0.30\pm0.10$ 	   & $0.21\pm0.05$		&$2.55 \pm 0.13$		&$4.59 \pm0.13$	& $ 8.98 \pm 0.23$   &  $16.82 \pm 0.36$	 \\	
	0136540701& $3.67\pm0.05$	& $0.54\pm0.25$ 	   & $0.07\pm0.03$		&$13.64 \pm 0.1$		&$18.13 \pm0.1$	& $ 24.26 \pm 0.17$ &  $32.00 \pm 0.25$	 \\	
	0136541001& $2.15\pm0.17$	& $0.15\pm0.08$ 	   & $0.16\pm0.04$		&$3.41 \pm0.05$		&$4.66 \pm0.06$	& $ 6.14 \pm 0.11$   &  $8.07 \pm 0.21$ 	 \\	
	0158970101& $2.98\pm0.46$	& $0.16\pm0.10$ 	   & $0.28\pm0.08$		&$6.15 \pm0.13$		&$7.00 \pm0.18$	& $ 8.68 \pm 0.40$   &  $11.81 \pm 0.99$	\\
	0150498701& $4.89\pm0.09$	& $0.42\pm0.07$ 	   & $0.22\pm0.03$		&$7.96 \pm0.05$		&$11.2 \pm0.06$	& $ 16.19 \pm 0.11$ &  $22.41 \pm 0.19$	\\	
	0158971201& $9.95\pm0.47$	& $\gg10$ 		   &$0.01\pm0.0$ 		&$9.17 \pm0.04$		&$13.8 \pm0.05$	& $ 20.10 \pm 0.08$ &  $27.87 \pm 0.12$	\\		
	0158971301& $5.35\pm0.21$	& $0.27\pm0.07$ 	   & $0.23\pm0.03$		&$7.57 \pm0.04$		&$9.17 \pm0.05$	& $ 11.07 \pm 0.09$ &  $13.70 \pm 0.16$	\\
	0302180101& $3.18\pm0.09$	& $0.27\pm0.16$ 	   & $0.07\pm0.02$		&$3.36 \pm0.04$		&$4.82 \pm0.05$	& $ 7.46 \pm0.09$   &  $11.80 \pm 0.13$	 \\	
	0411080301& $6.48\pm0.05$	& $1.83\pm0.19$ 	   & $0.12\pm0.03$		&$3.84 \pm0.07$		&$5.05 \pm0.08$	& $ 7.02 \pm0.14$   &  $10.70 \pm 0.20$	 \\	
	0502030101& $4.99\pm0.43$	& $0.13\pm0.07$ 	   & $0.17\pm0.04$		&$6.96 \pm0.04$		&$8.34 \pm0.05$	& $ 9.79 \pm0.10$   &  $11.38 \pm 0.19$	\\
		\hline
	\end{tabular}
\end{table*}

\begin{figure*}
\vspace{2.2mm} 
\figurenum{2}
\label{figs:xrays}
\centering
\includegraphics[width=0.48\textwidth] {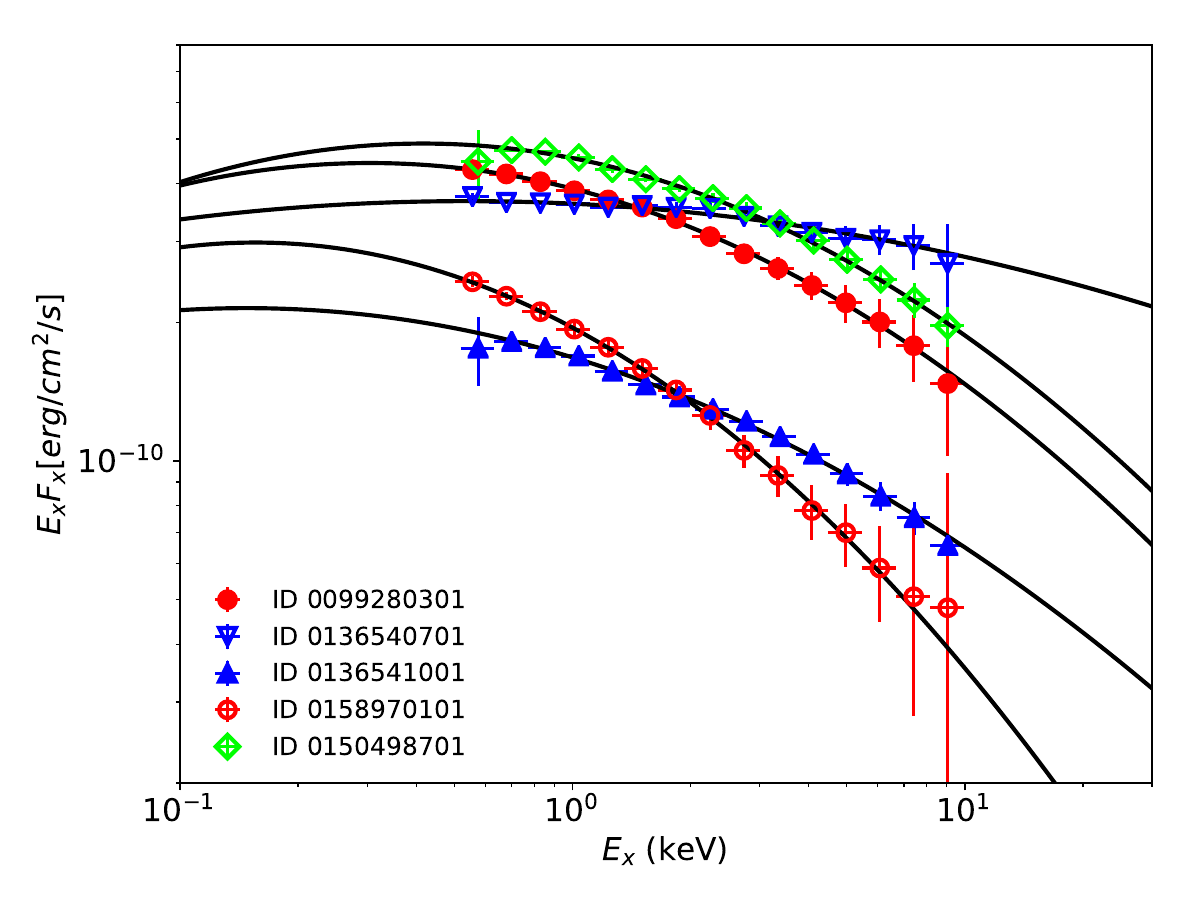}
\includegraphics[width=0.48\textwidth] {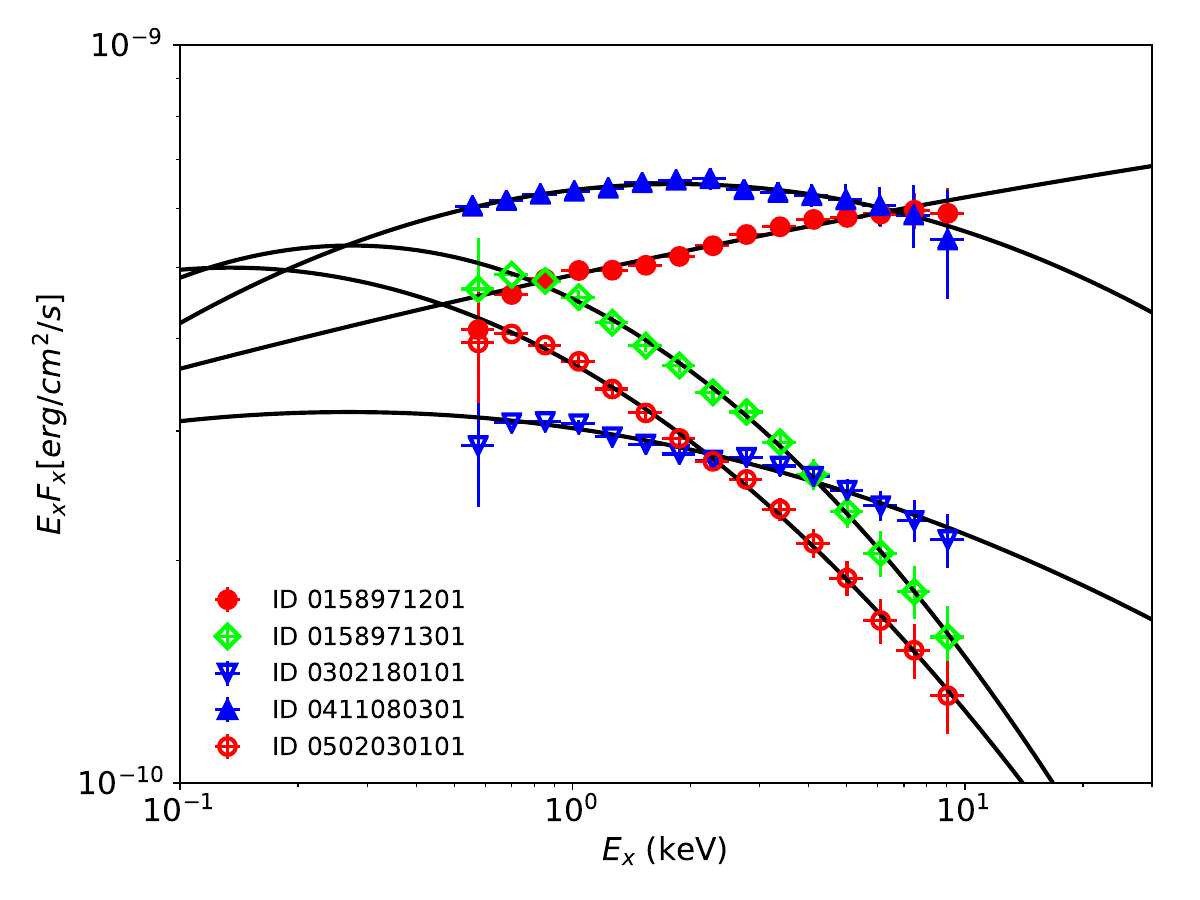}
\caption{The X-ray spectrum of Mrk 421 measured in 10 observations.
The solid black lines represent the best-fitting model obtained with a log-parabola function.
}
\vspace{2.2mm}
\end{figure*}

Mrk 421 is a prominent high-energy peaked blazar, with the synchrotron peak energy 
ranging from a fraction of a keV to several keV, even up to 10 keV \citep[e.g.,][]{Tramacere2009A&A,Ushio2010,Balokovic2016}.
It is one of the closest and brightest blazars in the X-ray bands, at redshift $z=0.031$. Because of its brightness, Mrk 421 has been studied in detail with \XMM among other instruments (e.g., NuSTAR, Chandra and Suzaku). 
\XMM with its high sensitivity, spectral resolving power and broad energy bandwidth, allows unprecedented time-resolved spectroscopy, uninterrupted by gaps because of a 48-hour-long orbital period \citep{Jansen2001A&A}.
It is therefore uniquely well-suited for studying IDV variability.
In this work we focus on the European Photon Imaging Camera (EPIC) pn data \citep{Struder2001A&A} of \XMM to investigate the variability of Mrk 421,
 as it is more sensitive and less affected by photon pile-up effects.

\subsection{Data Selection and Reduction}

So far, there are more than 150 archive \XMM observations of Mrk 421. In order to perform high-quality timing analysis 
in a wide temporal frequency range, we search the \XMM database for archival observations of Mrk 421 with an EPIC-pn 
exposure $>40$ ks, and get a sample of 12 observations. Among them, two observations (Obs. IDs: 0099280101 and 0845000901) 
have multiple (two and three respectively) EPIC-pn exposures in different modes within a single observation. 
As none of these five individual exposures meets the $>40$ ks criterion, we drop these two and obtain a sample of 10 observations, 
the detailed log of which is shown in Tab. \ref{tab:table1}. Note by coincidence, or due to unknown XMM strategy, all these observations 
were performed between 2002 and 2008. Among them, the pn camera was operated in Image mode for four and Timing mode for six observations.

Raw data are reduced using the latest XMM-Newton Science Analysis System (SAS, version 20.0.0) and the Current Calibration Files, 
following the standard pipeline in the \XMM user guide. For data of Image mode, we firstly extracted the source light curves (LCs) and spectra within
 a circular region with a radius of 60$\arcsec$, while extracting background from nearby source-free regions. However, using SAS task \textit{epatplot}, 
 we find serious pile-up effect in most observations. Therefore, we adopt annular regions to extract the source products, 
 the inner radius being chosen to range from 15$\arcsec$ and 25$\arcsec$, depending on the severity of the pile-up effect. 
 We present examples of how the source and background regions are adopted in Fig. \ref{fig:region}. 
For data of Timing mode, we extract the source products with RAWX in [27:47], while background in [3:5],
following the ``XMM ABC Guide''\footnote{\url{https://heasarc.gsfc.nasa.gov/docs/xmm/abc/}}. Basically, no source-free region 
is available for the Timing mode observations where the PSF is larger than the field-of-view, and the background region is selected 
to minimize the contamination of the source \citep[see][for a detailed analysis]{Ng_2010}. Such a small background region leads 
to small number of background counts (0.5--10 keV count rate $\sim$ 0.2 cts/s), and thus biased estimation of the background level. 
However, since the source is very bright and the background only contributes to $<$ 1 per cent within the source region, this bias should be negligible to the analysis in this work.
We find no significant pile-up effect for the data of Timing mode.
To study variability on a timescale of minutes, LCs, with a time bin of 100 s, 
are further reprocessed with the task \emph{epiclccorr}, applying corrections and background subtraction.


In Fig. \ref{figs:xrays}, we present the unfolded X-ray spectrum within the 0.5--10 keV energy range, averaged over each of the 10 observations.
The spectra are corrected for absorption with a neutral hydrogen column density fixed to the Galactic value of $N_{\rm H}=1.3\times10^{20} \rm ~cm^2$.
To describe the X-ray spectral curvature of Mrk 421,
a log-parabolic model has been used to fit the observed data:
\begin{equation}
E_x F_x = E_{pk} F_{pk} \left(\frac{E_{x}}{E_{pk}}\right)^{-b\log(E_{x}/E_{pk})}
\end{equation}
where $E_{pk}$ and $E_{pk} F_{pk}$ are the peak energy and flux, respectively, and $b$ measures the spectral curvature around $E_{pk}$.
All the spectra can be satisfactorily described by the model, with the best-fit parameters reported in Tab. \ref{tab:table2}. 
Note that the X-ray spectrum in Obs. ID 0158971201 may be represented by a power law with a photon spectral index of $\sim-1.9$.

We extracted the EPIC-pn LCs in two main energy bands, namely soft (0.5--2 keV) and hard (2--10 keV) X-rays, 
with the central energies being 1.25 and 6 keV, respectively.
Additionally, we divided the soft X-ray band into two narrow bands: 0.5--1 keV and 1--2 keV, 
while the hard X-ray band was split into 2--4 keV and 4--10 keV.
The corresponding central energies are 0.75, 1.5,  3 and 7 keV, respectively.

 To quantify the strength of X-ray variability at different bands, we calculate the fractional root mean square (rms) variability amplitude $F_{\rm var}$\footnote{
The fractional rms variability amplitude is defined as $F_{\rm var}=\sqrt{\sigma^2_{\rm XS}/\overline{x}^2}$,
with $\sigma^2_{\rm XS}=S^2-\overline{\sigma_{\rm err}^2}$ and $\overline{x}$ being the excess variance and the arithmetic mean of a light curve $x_{\rm j}$, respectively.
Here, the sample variance and mean square error are $S^2=\frac{1}{N-1}\sum^N_{\rm j=1}(x_{\rm j}-\overline{x})^2$ and $\overline{\sigma_{\rm err}^2}=\frac{1}{N}\sum^N_{\rm j=1}\sigma_{\rm err,j}^2$, respectively.
}.
The uncertainty on $F_{\rm var}$ is calculated following the method described in \cite{Vaughan2003MNRAS}.
We calculate $F_{\rm var}$ and its associated error $(F_{\rm var})_{\rm err}$ in all four narrow energy bands for the 10 observations.
The results of the variability analysis are summarized in Tab.  \ref{tab:table2}.
It can be seen that during the 10 observations the source shows significant flux variation based on two conditions: $S^2>\overline{\sigma^2_{\rm err}}$ and $F_{\rm var}>3\times (F_{\rm var})_{\rm err}$ \citep{Dhiman2021MNRAS,Devanand2022ApJ}. 
Moreover, $F_{\rm var}$ tends to larger toward higher energy.

For illustration, the normalized LCs in the 1--2, 2--4 and 4--10 keV bands are shown in Fig. \ref{figs:lcs} in Appendix \ref{append:lc}.
In the figure, we also show power spectral density (PSD) that is also often used to characterize the variability of blazar spectra.
Due to limitations of observation durations, the PSDs are derived only within the range of approximately $10^{-5}$ to $5\times10^{-3}$ Hz,
and the PSDs are well described by a power-law component combined with a constant component.
The spectral index of the PSD at lower frequencies falls within $\sim$1.9--3, indicating a rapid decrease in variability amplitude towards shorter timescales. 
On the other hand, the constant component at higher frequencies is attributed to uncertainties in the measured count rate, often referred to as white noise. 
Our findings demonstrate that the PSDs in the majority of observations are well above the white noise level up to $\sim2\times10^{-4}$--$10^{-3}$ Hz, 
which corresponds to variability timescales on hour to sub-hour timescales.
Notably, in special cases (e.g., Obs. ID 0502030101), the variability is still present, reaching frequencies of $\sim (2-3)\times10^{-3}$ Hz, 
which translates to timescales shorter than 10 minutes.

In the following analysis, we investigate the time delays associated with the interband variability in different X-ray energy bands.
To maximize the utility of the available data, the interband time delays include time lags for 4--10 keV with respect to 0.5--1, 1--2, 0.5--2 and 2--4 keV, 
delays for 2--4 keV with respect to 0.5--1, 1--2, and 0.5--2 keV, as well as delays for 2--10 keV with respect to 0.5--1, 1--2, and 0.5--2 keV, and delays between 0.5--1 and 1--2 keV.
To comprehensively assess the possible time delays between the low- and high-energy variations, 
we employ two distinct techniques: the cross-correlation function (CCF) method in the time domain
and the cross-spectral method in the frequency domain.

\begin{figure*}
\vspace{2.2mm} 
\figurenum{3}
\centering
\includegraphics[width=0.33\textwidth] {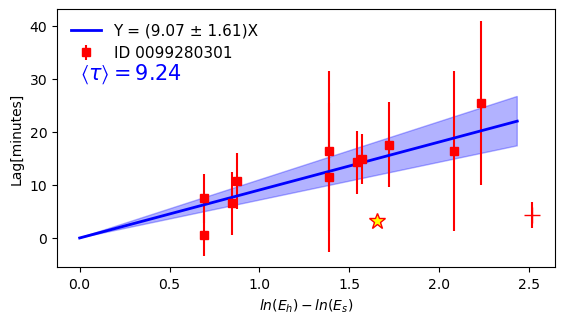} 
\includegraphics[width=0.33\textwidth] {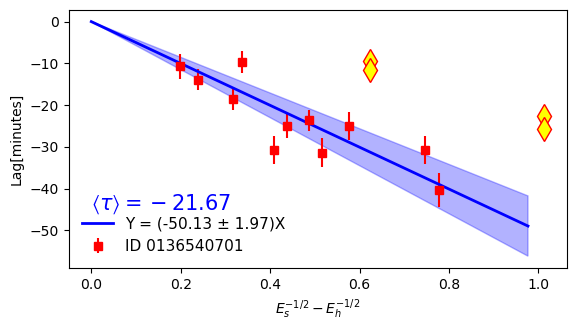} 
\includegraphics[width=0.33\textwidth] {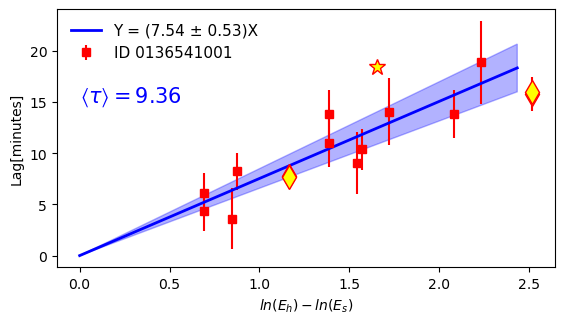} 
\includegraphics[width=0.33\textwidth] {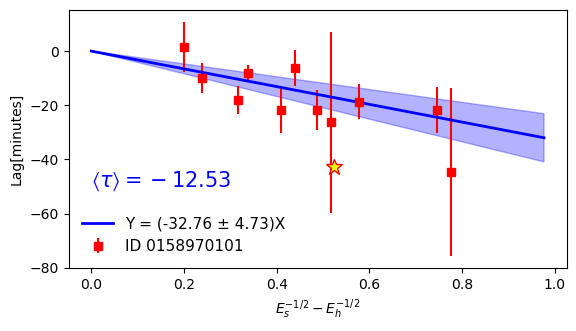} 
\includegraphics[width=0.33\textwidth] {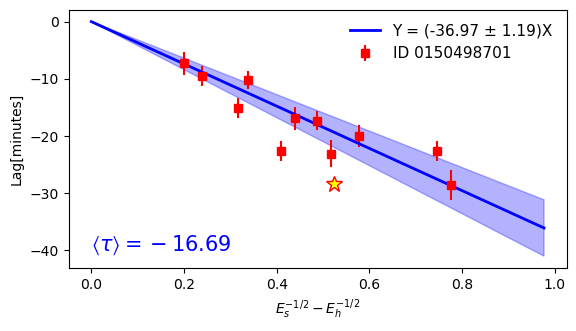} 
\includegraphics[width=0.33\textwidth] {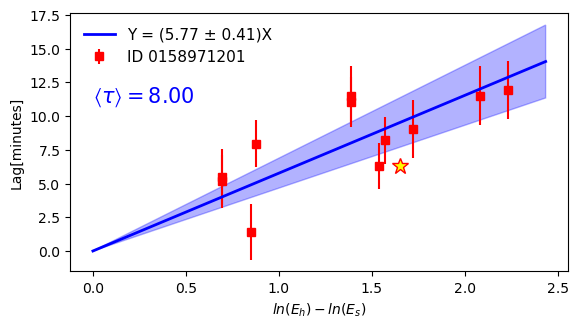} 
\includegraphics[width=0.33\textwidth] {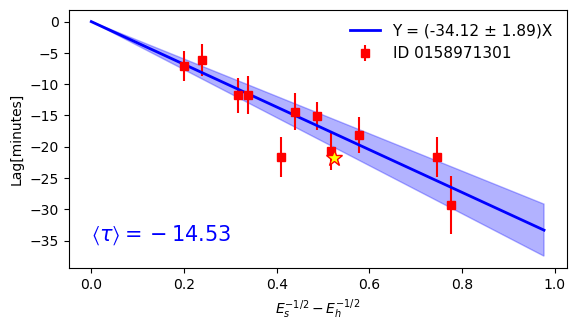} 
\includegraphics[width=0.33\textwidth] {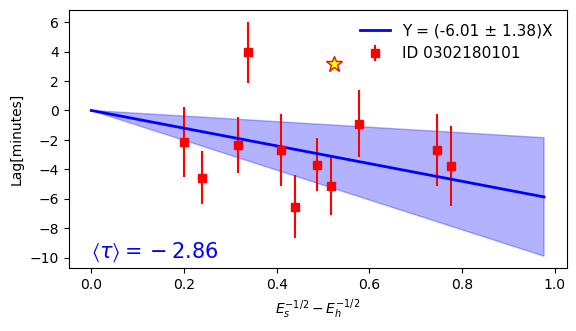} 
\includegraphics[width=0.33\textwidth] {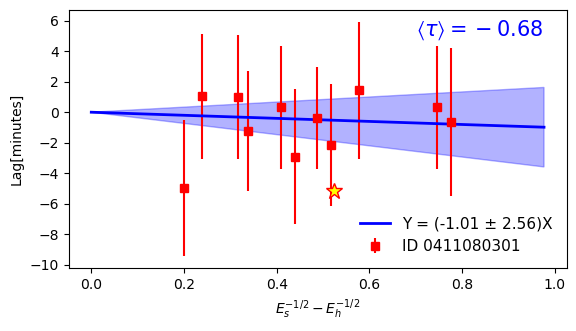} 
\includegraphics[width=0.33\textwidth] {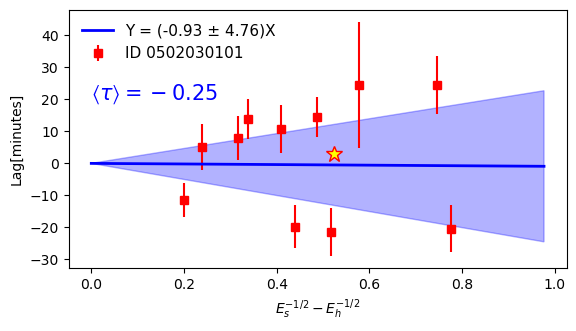}		
\caption{The  CCF time lags as a function of the energy differences $\Delta E$. 
Which $\Delta E$ is used depends on whether the weighted mean lag is negative or positive lag. 
A linear function is fitted to each of the trends, and the uncertainty region is shaded in blue.
The stars denote the results from \cite{Noel2022ApJS}.The diamonds in Obs. ID 0136540701 and 0136541001 denote the results from \cite{Ravasio2004A&A}. 
For Obs. ID 0099280301, the lag reported in \cite{Brinkmann2003A&A} is denoted by the symbol `+' .
\label{figs:fig3}
}
\vspace{2.2mm}
\end{figure*}

\subsection{Cross-correlation analysis in the time domain} 
The lags between different X-ray bands are calculated using the popular interpolated cross-correlation functions (CCFs) method described in \cite{Peterson1998PASP}.  
As suggested in the previous works \citep[e.g.,][]{Peterson1998PASP,Zhang2006ApJ}, the centroid of the CCF $\tau_{\rm cent}$ is used to quantify 
the time lags between two highly variable LCs, representing the intrinsic complexities of the underlying variability processes. 
The primary reason is that $\tau_{\rm cent}$ may account for the CCF asymmetries that are usually seen in AGN variability. 
Uncertainties on the lags are estimated using the flux randomization and random subset selection (FR/RSS) simulation.
To estimate the uncertainties, the distribution of the CCF centroid is built up based on all points with correlation coefficients in excess of 80\% of the CCF peak value (i.e., the maximum correlation coefficient). 
The calculations were done using a publicly available code {\it{PyMCCF}} \footnote{https://bitbucket.org/oknyansky/python\_mccf/src/master/}.
For each pair of LCs, the time lag $\tau$ and its uncertainty $\sigma$ are estimated by fitting a Gaussian model to the cross-correlation centroid distribution (CCCD),
and the weighted average $\langle\tau_{\rm c}\rangle$ of these values is computed \footnote{
To the lag $\tau$ obtained for each pair of LCs, we assigned as weight $1/\sigma^2$, where $\sigma$ represents 68\% confidence limits around the lag corresponding to the peak of the Gaussian function.
} .
The results are shown in Fig. \ref{figs:fig3}.  
For illustrative purposes, the CCCDs from the selected 10 observations are shown in Fig. \ref{figs:cccd} in Appendix \ref{append:cccd}.

In Fig. \ref{figs:fig3}, we display the CCF lags computed from the original LCs as a function of the differences $\Delta{E}$ of the compared energy bands, 
with $\Delta E\equiv E_{\rm s,keV}^{-1/2}-E_{\rm h,keV}^{-1/2}$ for $\langle\tau_{\rm c}\rangle<0$, and $\Delta E\equiv\ln (E_{\rm h,keV})-\ln (E_{\rm s,keV})$ otherwise.
For comparison, we also show the measured lags in several previous works, where the time lags are estimated with different techniques: 
(1) fitting the CCF with an asymmetric Gaussian profile to find the peak position \citep{Brinkmann2003A&A,Ravasio2004A&A}; 
(2) fitting the central peak of the Discrete correlation function (DCF) with a Gaussian function and taking the peak position of the function as the delay value \citep{Noel2022ApJS}.

It is clear from Fig.\ref{figs:fig3} that in the first eight observations the interband time lags exhibit a systemic trend, i.e., the amount 
of lag $|\tau_{\rm c}|$ tend to become larger with increasing cross-correlated energy difference $\Delta E$.
For each case, linear regression analysis of the scatter plot indicates that the $\tau_{\rm c}-\Delta E$ trend can be adequately described 
by the expected linear relation, and most of the data points are within 95\% confidence ranges.
Among the eight observations, there are three cases exhibiting the systematic hard lag and five cases exhibiting the systematic soft lag. 
Except for Obs. ID 0302180101, the time lags predicted in the $\tau_{\rm c}-\Delta E$ trends are compatible with previous results measured with different techniques. 
Based on the DCF maximum fitting, a minor hard lag of $\sim3$ minutes between the 0.3--2.0 keV and 2.0--10 keV X-ray bands was reported in \cite{Noel2022ApJS} for Obs. ID 0302180101
\footnote{In the work, a positive value for the lag means that  the soft channel signal is delayed with respect to the hard signal, and it is known as a soft lag.
Conversely, a negative value indicates a hard lag, with the hard channel signal lagging behind the soft signal.}.
On the contrary, there may exist a clear soft-lag trend in the scatter plot of $\tau_{\rm c}-\Delta E$, 
although a slight hard lag of $4.0\pm2.0$ minutes may be present between 0.5--1 and 1--2 keV bands,
and this measurement lies outside the 95\% confidence intervals of the best fitting. 
Using the best-fit relation for $\tau_{\rm c}-\Delta E$ trend, a soft lag of $\sim3$ minutes between 0.3--2 and 2--10 keV bands can be predicted.
However, it is worth noting that any time lags comparable to or less than bin size (100 s) adopted should be taken with caution.

In the last two observations: Obs. IDs 0411080301 and 0502030101, there are no systematic trends.
In Obs. ID 0411080301, no evidence for time leads or lags is present. Our results show that almost all of the lags are consistent with zero lag within the $1\sigma$ intervals.
Moreover, we notice no clear trend in the plot of $\tau-\Delta E$, where the best-fitting slope is statistically consistent with zero within 95\% confidence intervals.
As claimed by previous works \citep[e.g.,][]{Ravasio2004A&A,Ferrero2006A&A,Noel2022ApJS}, absence of a non-zero time lag may be due to a superposition of the various lags in individual features.
In Obs. ID 0502030101, one can see both the significant positive and negative lags. 
The results show that the flux variation at the 4-10 keV reference band leads those at the other four lower energy bands, 
whereas the remaining 7 pairs of LCs exhibit clear hard lags.
Throughout the observation, \XMM  may sample only a decaying phase of a possible major flare (see Fig.\ref{figs:lcs}). 
The presence of non-zero lag may be mainly caused by the differences in the decay slopes of the compared LCs, 
producing, on average, soft lags or hard lags \citep[see detailed discussion in][]{Ravasio2004A&A}.
Actually, previous work showed a minor hard lag between 0.3--2 keV and 2--10 keV, with the variation of soft photons preceding that of hard photons by 0.17 ks. 
It appears to be compatible with the lag predicted by our best-fit relation.

It is worth noting that the CCF time lags measured from the structured or asymmetrical LCs as a whole may roughly represent 
a weighted average value combining the delays of the various substructures or an index of the LC asymmetry \citep{Ravasio2004A&A}. 
Thus, one should perform a careful check of the behaviour of the single subcomponent before using the results to test the blazar models.
However, it is a tedious task to check the results based on the CCF techniques \citep[e.g.,][]{Brinkmann2003A&A,Ravasio2004A&A,Brinkmann2005A&A}.
An alternative approach for checking and measuring the lags is based on Fourier techniques, 
which can provide the frequency-dependent time lag of the average cross-power spectrum between two different time series.
The cross-spectral method may be superior to the commonly used  CCF method for measuring the time lag from the complex or irregular LCs \citep{Zhang2002MNRAS}.

\begin{figure*}
\vspace{2.2mm} 
\figurenum{4}
\centering
\includegraphics[width=0.33\textwidth] {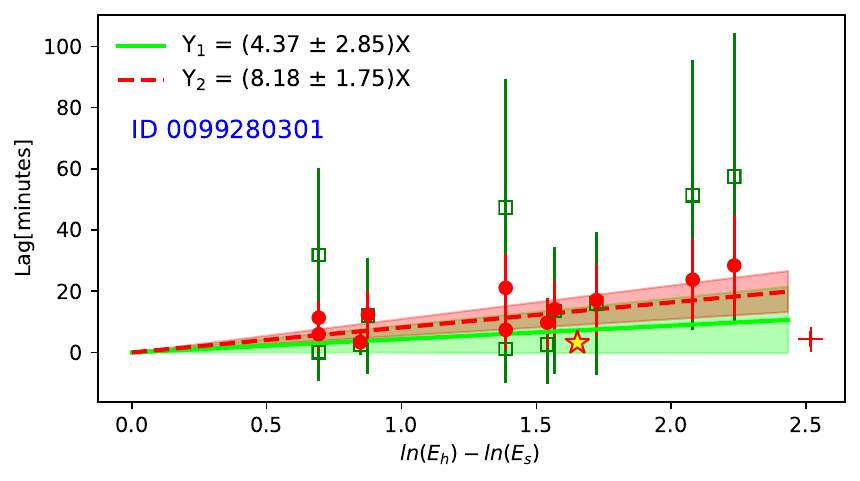} 
\includegraphics[width=0.33\textwidth] {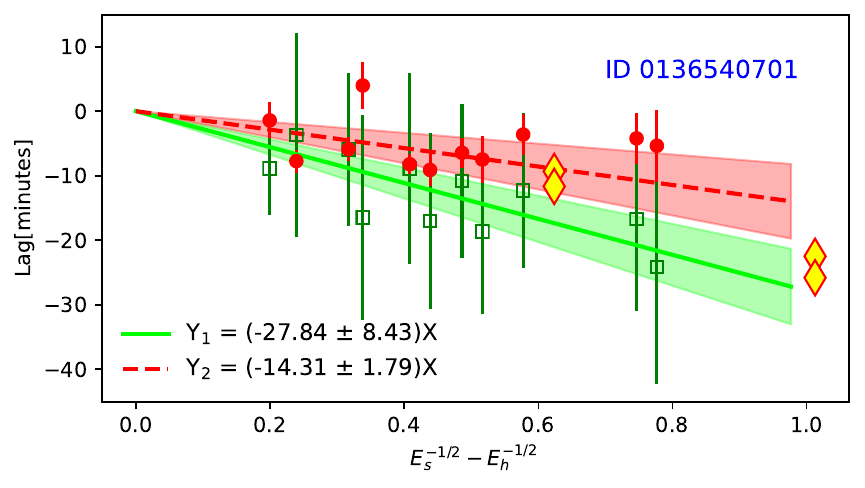} 
\includegraphics[width=0.33\textwidth] {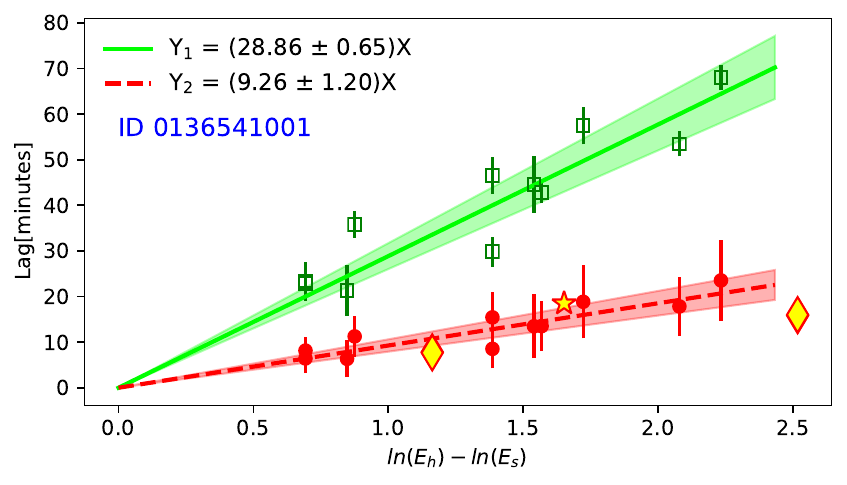} 
\includegraphics[width=0.33\textwidth] {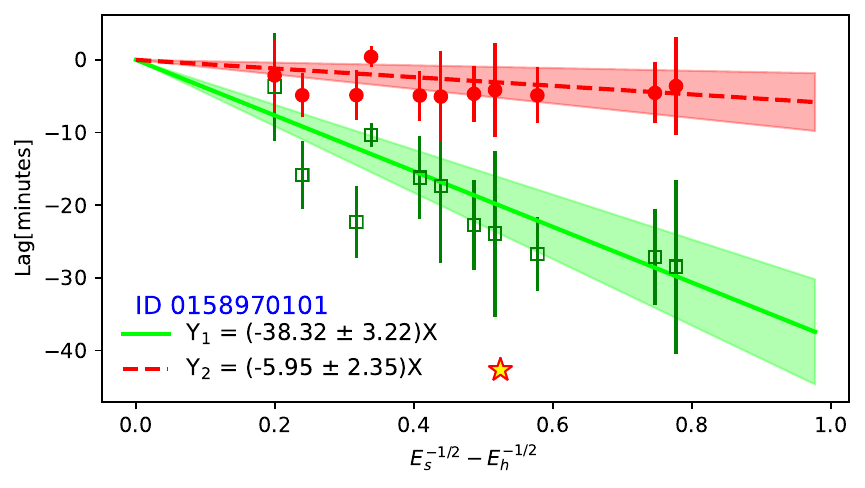} 
\includegraphics[width=0.33\textwidth] {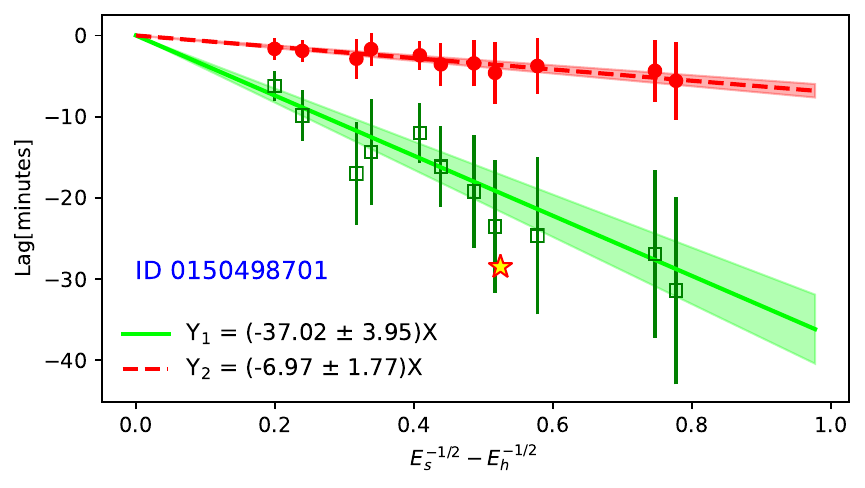} 
\includegraphics[width=0.33\textwidth] {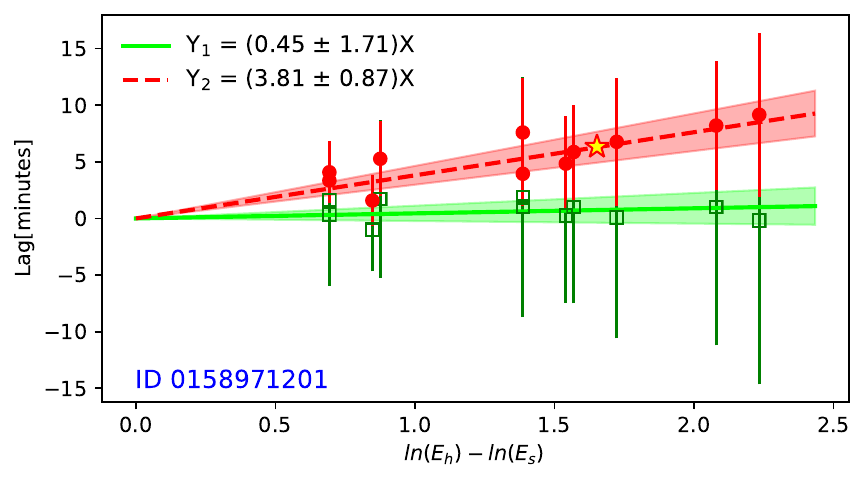} 
\includegraphics[width=0.33\textwidth] {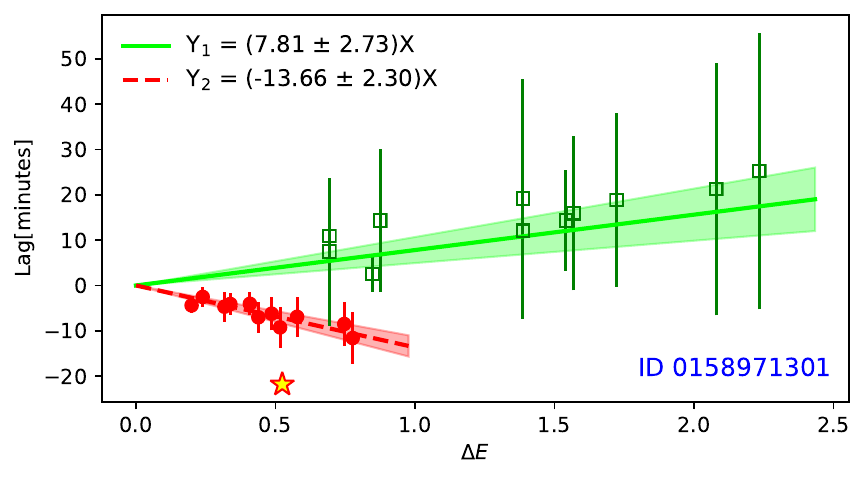} 
\includegraphics[width=0.33\textwidth] {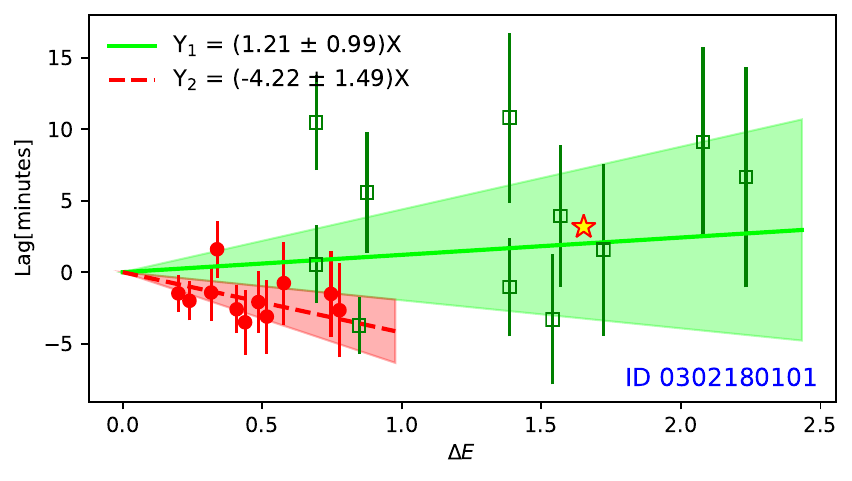} 
\includegraphics[width=0.33\textwidth] {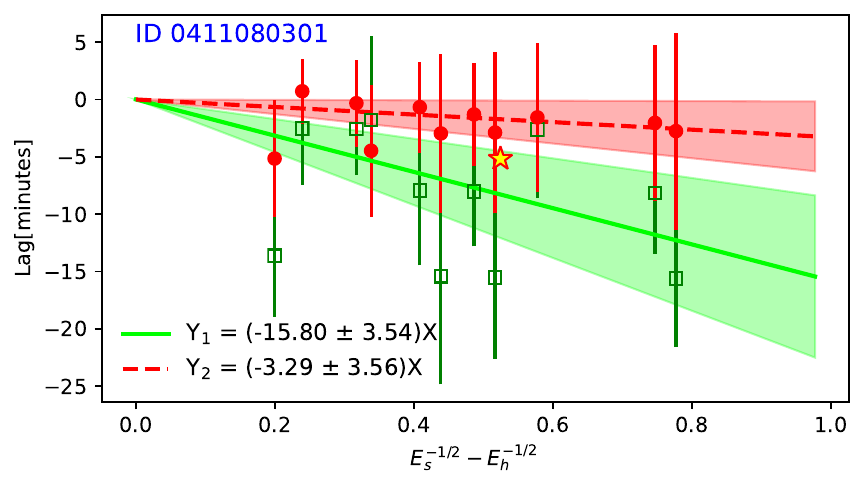} 
\includegraphics[width=0.33\textwidth] {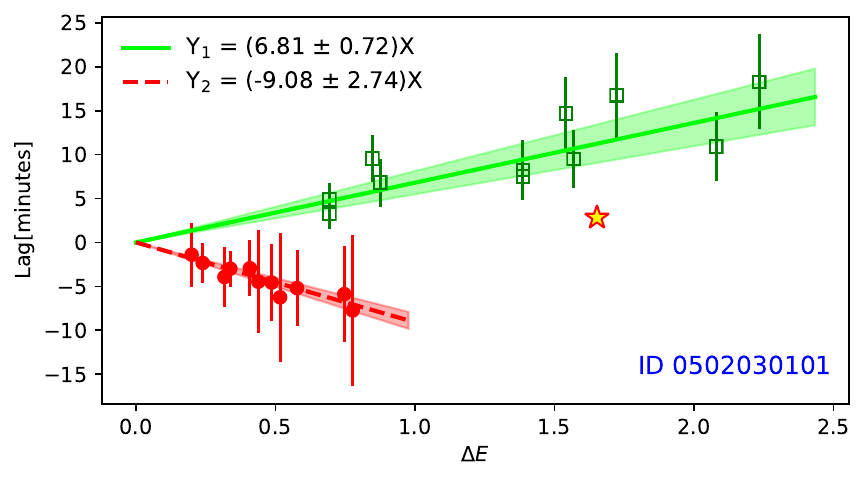}		
\caption{The lowest-frequency time lags as a function of the differences $\Delta{E}$. 
The open green squares denotes the lowest-frequency lag in $\tau_1$, and the filled red circles denotes the lowest-frequency lag in $\tau_2$. 
\label{figs:fig4}
}
\vspace{2.2mm}
\end{figure*}

\subsection{Cross-spectral analysis in the frequency domain} 
In the section, we present the analysis of the X-ray time lags in the Fourier domain.
We calculate the frequency-resolved time lags using a standard Fourier approach reviewed in detail by \cite{Uttley2014A&ARv}.
Briefly, estimation of the lag consists of three major steps.
Firstly, LCs are split into segments of identical length in two bands, and the periodograms and cross-spectrum are 
calculated using the DFT for each segment of the resulting LCs (see Appendix \ref{append:dft}).
To reduce the effects of noise, the periodograms and the cross-spectrum are then averaged over Fourier frequencies 
in a given frequency bin centered at $\nu_f$, as well as over the number of segments,
 allowing the raw coherence \textbf{to} be calculated from the binned periodograms and cross-spectrum \citep{Vaughan1997ApJ}.
 Finally, one measures the phase lags from the argument of the cross-spectrum in each bin and divides them by $2\pi\nu_f$ to 
 give a lag-frequency spectrum between a pair of LCs at two bands,
 and their uncertainties are given by the coherence in each frequency bin.
The procedure was carried out using the X-ray timing analysis package {\it{pyLag}}\footnote{http://github.com/wilkinsdr/pylag}.

For each pair of LCs, we obtain the lag-frequency spectra in five logarithmically spaced bins, 
ranging from the lowest frequency $T^{-1}$ (where $T$ is the time duration of each observation) 
to the Nyquist frequency of $5\times10^{-3}$ Hz.
The resulting lag-frequency spectra are denoted as $\tau_1$.
To explore the variability at different timescales in more detail, 
we then divide each LC into two segments of equal duration and recalculate the lag-frequency spectra, referred to as $\tau_2$. 
The minimum frequency bin in $\tau_2$ is about twice that in $\tau_1$. It helps us to reveal the potential profile of the lag-frequency spectra.
The results reveal that, for most observations, lags predominantly appear in the low Fourier-frequency bin, where the color noise dominates (see Appendix \ref{append:lc}).
We show the resulting lags at the lowest frequencies in $\tau_1$ and $\tau_2$ in Fig.\ref{figs:fig4}, as denoted by the green open squares and red filled circles, respectively. 
To show the dependence of the lags on the frequency, the lag-frequency spectra for 0.5--2 keV/1--2 keV bands, with respect to 2--10 keV band, are plotted in Fig. \ref{figs:fig6},
as denoted by the red and green symbols, respectively.
Moreover, a direct comparison of the CCF lags with the Fourier time lags is also shown in the figure.
It is clear that the CCF lags in different epochs are in good qualitative agreement with the Fourier lags at the lowest frequency bin in $\tau_1$ or $\tau_2$.

Similar to the results quantified with the CCF method, the lowest-frequency Fourier lags in the plot of $\tau-\Delta E$ follow the expected linear relation.
The results further confirm the presence of hard lags, as measured by the CCF method in Obs. IDs 0099280301, 0136541001 and 0158971201, 
along with soft lags in Obs. IDs 0136540701, 0158970101 and 0150498701. 
Comparing the lags measured by two different methods, we note that the lowest-frequency hard lags in $\tau_2$ are more closely aligned with the CCF lags compared to those in $\tau_1$.
Conversely, the lowest-frequency soft lags in $\tau_1$ are more consistent with the CCF lags than those in $\tau_2$.

For Obs. IDs 0158971301, 0302180101 and 0502030101, 
the cross-spectral analyses indicate that the lowest-frequency lags in $\tau_2$ exhibit systematic soft lags, while those in $\tau_1$ exhibit an opposite trend. 
In Obs. ID 0158971301, the lowest-frequency soft lags in $\tau_2$ are comparable with the lags measured using CCF method,
while those in $\tau_1$ exhibit a systematic hard-lag trend, although these results are associated with large errors.
In Obs. ID 0302180101, the lowest-frequency lags in $\tau_1$ are highly scattered, and no clear trend can be observed. 
However, the lag between $0.3-2$ and $2-10$ keV band predicted by the best-fit result is in good qualitative agreement with that estimated in \cite{Noel2022ApJS}.

For Obs. ID 0411080301, all points in the plot of $\tau-\Delta E$ are consistent with zero lags, which is in agreement with the measured CCF lags.
However, it seems that a soft-lag trend may exist.

\section{Model description}\label{sec:model}
Generally, the non-thermal emission of blazars is thought to be the combination of radiation from various regions of the jets with electron spectra at different stages of evolution \citep[e.g.,][]{Blandford1979,Konigl1981}.
Given the X-ray observations of Mrk 421, the LCs in most epochs can be characterized by long-term trends lasting for time scales comparable to or longer than the observations' length,
 with superimposed small features occurring on short timescales of hours or even a few minutes.
This may suggest that more than one emission region may be in operation in the system \citep{Brinkmann2005A&A}. 
Since we mainly focus on the X-ray radiation emerging from the innermost regions of the jets,
we employ the simplest version of a multi-zone jet model to investigate the jet's physical properties contained in the broadband X-ray spectrum and the interband time lags measured in different epochs.
In the model, the blazar emission predominantly consists of a quasi-stationary component that changes only on long timescales and a non-stationary component 
responsible for short-timescale variability.

For convenience, the compact dissipation zone responsible for the rapid variation is usually treated as a spherical blob of size $R'$ 
containing a randomly tangled magnetic field of strength $B'$. 
As it propagates outward along the jet, a population of ultra-relativistic electrons (and positrons) is injected uniformly throughout the blob.
The injected electrons experience further acceleration due to interactions with shocks or MHD waves, and radiative losses until they escape from the blob.
The most important radiation mechanisms for electrons are synchrotron and SSC radiation.
Due to the relativistic beaming effects, the intrinsic radiation produced in the comoving frame of the blob is Doppler boosted by a factor of $\delta_{\rm D}^4$ in the observer's frame, 
while the variability timescale is reduced by a factor of $\delta_{\rm D}^{-1}$.
Then, rapid variability can be observed.

The quasi-stationary emission may arise from an extended inhomogeneous or stratified region responsible for generating X-ray emission. 
For the modeling of the quasi-stationary emission, 
we simplify the spatial geometry as a spherical plasmoid with different physical properties including the magnetic field strength, 
Doppler factor, spatial size, and the emitting electrons.
Thus, the blob may represent an average over the entire region. 
To facilitate parameter comparison between different epochs, the quasi-stationary emission remains constant and is treated as the background emission.

In the model, the population of the injected electrons is assumed to be provided by a significantly smaller co-spatial non-radiative acceleration region,
where seed electrons undergo significant energization by shocks \citep[e.g.,][]{Kirk1998A&A,Bottcher2019ApJ,Chandra2021ApJ} or magnetic reconnection events \citep[e.g.,][]{Sironi2014ApJ,Sironi2015MNRAS}, 
but contribute only a negligible amount of radiation due to a low magnetic field.
Thus, it is reasonable to consider that the injected electrons exhibit a time-dependent nature.
For simplicity, we assume that the time-dependent injection process of the energetic electrons is described by the time-dependent function $Q'_{\rm e}(\gamma',t')$, which
can be decoupled into the product of $f_{\rm e}'(t)$ and $q'_{\rm e}(\gamma')$, 
with $f'_{\rm e}(t')$ and $q'_{\rm e}(\gamma')$ being the time-dependent profile and time-independent energy distribution, respectively.

In these two zones, we assume that the electron injection spectrum $q_e'(\gamma')$ follows the same power-law distribution form characterized by four parameters:  the normalization constant $q_0'$,
the minimum $\gamma'_{\rm i,min}$ and maximum $\gamma'_{\rm i,max}$ Lorentz factors and the power-law index $p$, represented by:
$q'_{\rm e}(\gamma',t')=q_{\rm 0}'\gamma'^{-p} H(\gamma';\gamma'_{\rm i, min},\gamma_{\rm i, max}')$, 
where $H(\cdot)$ is the Heaviside function defined by $H=1$ if $\gamma'_{\rm i, min}\le\gamma'\le\gamma_{\rm i, max}'$ and $H = 0$ otherwise.

\subsection{Electron Transport Equation}

For simplicity, we assume that the two components are decoupled from each other.
Then, the evolution of the emitting electron distribution (EED) in both zones can be independently described by the transport equation \citep[e.g.][]{Schlickeiser1984,Stawarz2008,Diltz2014JHEAp,Hu2023ApJ}
\begin{eqnarray}\label{acc-transport1}
\frac{\partial N'_{\rm e}(\gamma',t')}{\partial t'} &=& \frac{\partial}{\partial \gamma'} \left\{\gamma'^2D_p(\gamma')  \frac{\partial}{\partial \gamma'}\left[\frac{N'_{\rm e}(\gamma',t')}{\gamma'^2}\right]  \right\} \nonumber\\
&-& \frac{\partial}{\partial \gamma'}\left[\left(A_0^{\rm sh}\gamma'+\dot\gamma'_{\rm rad}\right) N'_{\rm e}(\gamma',t') \right] \nonumber\\
&-& \frac{N'_{\rm e} (\gamma',t')}{t'_{\rm esc}} + Q'_{\rm e}(\gamma',t').
\end{eqnarray}
where  $D_p (\gamma')$ is the momentum diffusion coefficient for interactions with MHD waves, 
$A_0^{\rm sh}$ accounts for first-order Fermi acceleration used in Eq. \ref{eqs:shock}, 
and $t'_{\rm esc}$ is the escape timescale which describes spatial transport of the electrons in the blob. 
The term $\dot\gamma'_{\rm rad}$ represents the radiative losses due to synchrotron and SSC processes, and is calculated using standard formulas in the literatures\citep[e.g.,][]{Jones1968,Moderski2005,Finke2008}.

In brief,  we assume that the observed X-ray and TeV $\gamma$-rays are the sum of two standard blobs that radiate simultaneously and characterized by different physical conditions.
For each zone, there are at least 10 physical quantities, i.e., $B'$, $\delta_{\rm D}$, $R'$, $A_0^{\rm sh}$, $D_0$, $t'_{\rm esc}$, $q_0'$, $\gamma_{\rm i,min}'$,  $\gamma_{\rm i,max}'$ and $p$.
Among those, the first three quantities describe the global properties of the zone, while the injection electron spectrum is described by the last four quantities.
The quantities $A_0^{\rm sh}$, $D_0$ and $t'_{\rm esc}$ determine the acceleration and escape of the injected electrons.
The EED can be obtained by solving the diffusion equation numerically, 
and the expected synchrotron and SSC radiation can be calculated following the methods described in the literatures \citep[e.g.,][]{Rybicki1979book,Crusius1986A&A,Finke2008,Deng2021}.

To explore the parameter space more effectively, we use the observed synchrotron peak frequency $\nu_{\rm pk}$ and electron escape efficiency $\eta_{\rm esc}$ 
as the input parameters instead of $D_0$ and $t'_{\rm esc}$.
Since the peak Lorentz factor $\gamma'_{\rm pk}$ corresponding to $\nu_{\rm pk}$ is approximately equal to 
the equilibrium Lorentz factor $\gamma'_{\rm eq}$ determined by the balance of the synchrotron cooling and acceleration,
the energy-independent acceleration timescale $t'_{\rm acc}$ can be evaluated by the relationship \citep{Hu2021}
\begin{equation}\label{eqs:tacc}
t'_{\rm acc}=\frac{6\pi m_{\rm e}c}{\sigma_{T}B'^{3/2}}\sqrt{\frac{\nu_0 \delta_{\rm D}}{\nu_{\rm pk}(1+z)} },
\end{equation}
where $\nu_0=4m_{\rm e}c^2/3hB_{\rm cr}(1+z)$, with $m_{\rm e}$, $c$, $h$ and $B_{\rm cr}$ being the rest mass of electrons, 
light speed, Planck's constant and critical magnetic field strength, respectively.

Then, the momentum diffusion coefficient can be deduced through the following relationship: 
\begin{eqnarray}
D_0&=&\frac{1}{(4+a_{\rm sh})t_{\rm acc}'}
\end{eqnarray}
where $a_{\rm sh}\equiv A_0^{\rm sh}/D_0$.
In particular, we assume that stochastic acceleration plays a dominant role in energization of the injected electrons. 
This is motivated by the observations that the X-ray spectral shape of Mrk 421 is better represented 
by a log-parabola model rather than by a single power law or broken power-law models \citep[e.g.,][]{Brinkmann2003A&A,Ravasio2004A&A}, 
and the curvature parameter decreases as the synchrotron peak energy increases \citep[e.g.,][]{Tramacere2007}.
Without loss of generality, we set $a_{\rm sh}=0$ in our modelling.

\begin{deluxetable*}{ccccccccccccc}
\centerwidetable
\label{tab:table3}
\renewcommand\tabcolsep{8.0pt}
\tablecaption{Input model parameters and derived quantities for the simultaneous SEDs collected in 2009 and 2013 MWL campaigns.}
\tablehead{
MWL camps &\colhead{$\delta_{\rm D}$}	&	\colhead{$B'$}      & \colhead{$R'$}  	& \colhead{$\nu_{\rm pk} (E_{\rm pk})$}	& \colhead{$q_0'$}    & \colhead{$\gamma_{\rm i,min}'$} & \colhead{$\gamma'_{\rm i,max}$}  & \colhead{$p$} &   \colhead{$\eta_{\rm esc}$} & \colhead{$t'_{\rm acc}$}   & \colhead{$D_0$}  & \colhead{$U_{\rm e}'/U_{\rm B}'$} \\
	&\colhead{}	 &\colhead{G}	  &\colhead{$10^{15}$cm}	 	&\colhead{$10^{17}$ Hz	(keV)} &\colhead{$\rm s^{-1}$} 	     &	\colhead{$10^3$}	 &\colhead{$10^5$}	&  &	 &\colhead{$R'/c$}   &\colhead{$10^{-6}\rm s^{-1}$}    &\colhead{}
}  
\decimalcolnumbers
\startdata
	2009 	&27		&0.08    	 &12.0   	&$2.5 (1.03)$ &$4.03\times10^{47}$ 	&$2.0$	&$35.0$ 	&$2.5$	& $1.26$&$1.69$	&$0.37$	& 8.6	 \\
	2013 	&25		&0.1    	 &18.5   	&$0.5 (0.21)$ &$2.38\times10^{48}$ 	&$2.0$	&$12.0$ 	&$2.7$	& $1.26$&$1.69$	&$0.24$	& 2.0	 \\
\enddata 
\tablecomments{Column(1): Multi-wavelength campaigns. Column (2): The Doppler factor. Column (3): The magnetic field strength. Column (4): The synchrotron peak frequency (in units of keV). Column (5): The injection rate.  Column (6) and (7): The minimum and maximum Lorentz factor of the injected electrons. Column (8): The spectral index of the injected electron spectrum. Column (9): The electron escape efficiency. 
Column (10): The acceleration timescale in units of the dynamical timescale. Column (11) and (12): The momentum diffusion coefficient and equipartition parameter derived from the SED modeling.}
\end{deluxetable*}

On the other hand, the escape timescale associated with spatial diffusion can be written as $t'_{\rm esc}= t'_{\rm acc}/\eta_{\rm esc}$,
using the well-known relation between the momentum and spatial diffusion coefficient given by
$D_p(\gamma')D_x(\gamma')\simeq\gamma'^2\beta_A^2c^2/9$ with $\beta_A$ denoting the dimensionless Alfv\'{e}n velocity.

The full time-dependent and steady-state EED are computed numerically by solving Eq. (\ref{acc-transport1}).
The implicit Crank-Nicolson scheme is used to numerically solve the transport equations.
We tested our numerical code with the analytical solution for the steady-state and time-dependent EED in Appendix \ref{append:solution}.

\subsection{Modeling the X-ray and TeV $\gamma$-ray background emission}

For a better understanding of the observed X-ray spectrum and time lags in different epochs, 
we first need to model the quasi-stationary background emission
from the jet that generates the synchrotron X-rays and TeV $\gamma$-rays emitted via the SSC process.

\cite{Abdo2011} and \cite{Balokovic2016} organized two coordinated MWL observations of Mrk 421 
taken in 2009 January-June and 2013 January-March, respectively. 
The 4.5-month-long MWL campaign in 2009 collected an extensive multifrequency data set that is simultaneous and 
a good representation of the typical SED from Mrk 421 in a relatively low flux state.
During the campaign in 2013, the source was in a historically low-activity state, and four SED snapshots were constructed.
In particular, Mrk 421 exhibited exceptionally low X-ray and TeV flux on 2013 January 10 (MJD 56302) and 20 (MJD 56312).

\begin{figure*}
\vspace{2.2mm} 
\figurenum{5}
\centering
\includegraphics[width=\textwidth] {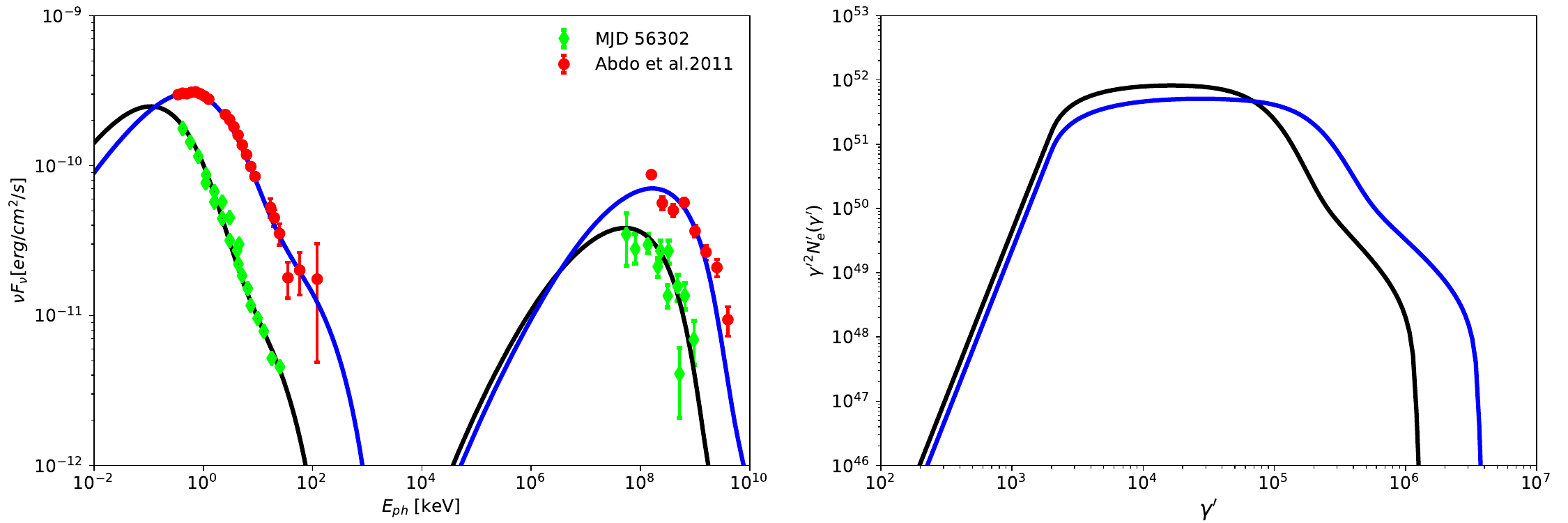} 
\caption{Modeling of the simultaneous X-ray and TeV $\gamma$-ray spectrum constructed in 2009 and 2013 MWL campaigns.
\label{figs:lowest}
}
\vspace{2.2mm}
\end{figure*}

In Fig. \ref{figs:lowest}, we show the simultaneous X-ray and TeV $\gamma$-ray spectra on 2013 January 10 
and those from the 2009 campaign, along with the theoretical SEDs reproduced by our model for comparison.
The theoretical X-ray spectra are obtained using a quasi-equilibrium EED resulting from continual electron injection, 
and the resulting EEDs for two SEDs are shown in the right panel of Fig. \ref{figs:lowest}.
In our modeling, the three parameters $\nu_{\rm pk}$, $\gamma'_{\rm i,max}$ and $p$ are adjusted to reproduce the shape of the observed X-ray spectra,
and the other four parameters $\delta_{\rm D}$, $B'$, $R'$ and $q_0'$ are mainly determined 
on the basis of the X-ray and TeV $\gamma$-ray flux, as well as to avoid superluminal escape of electrons.
Since we have not taken into account simultaneous radio and optical-UV data, which may be contaminated by contributions from larger-scale jet radiation,
the values of $\eta_{\rm esc}$ and $\gamma'_{\rm i,min}$ cannot be effectively constrained.
In the current SED modeling, $\eta_{\rm esc}=1.26$ is chosen to yield a standard spectral index of $\sim2.2$,
resulting from continuous injection of mono-energetic electrons with Lorentz factor well below the equilibrium Lorentz factor \citep[e.g.,][]{Hu2021}.
The continuously injected electrons with power-law spectra lead to a harder EED with a power-law index of $\sim2.0$ in the low-energy part compared to the mono-energetic injection.
Meanwhile, we set $\gamma'_{\rm i,min}$ to $2\times10^3$, which is roughly equivalent to the ratio of proton mass to electron mass, $m_{\rm p}/m_{\rm e}$.

The input model parameters and our interested physical quantities inferred from the observed SEDs are reported in Tab. \ref{tab:table2}.
It can be seen that the global properties of the quasi-stationary zone and the diffusion coefficient $D_0$ (or $t'_{\rm acc}$) 
inferred from the two SEDs do not differ substantially, 
although the value of $\nu_{\rm pk}$ derived from the daily SED on MJD 56302 is a factor of five smaller in comparison 
with the typical SED accumulated over 4.5 months (also see Fig.\ref{figs:lowest}).
The inferred values of $D_0$ are similar to those obtained in previous works for Mrk 421 \citep{Lewis2016,Hu2023ApJ}, and for other HBLs \citep[e.g.,][]{Hu2021}.

Moreover, our results show that the deduced Doppler factor, magnetic field and the blob size are comparable to previous estimations 
of Mrk 421 from SED modeling \citep{Mankuzhiyil2011,Abdo2011,Balokovic2016}.
We also note that the modeling of the daily SED on MJD 56302 yields the equipartition parameter $U_{\rm e}'/U_{\rm B}'$ that is much closer to unity,
and it differs substantially from that resulting from the typical SED in 2009.
An attractive feature is that the equipartition can minimize the jet power \citep[e.g.,][]{Finke2008}.
The difference in the physical quantity is driven mostly by the significantly larger electron energy density $U'_{\rm e}$ required by the SED modeling of Mrk 421 in 2009.
This may be consistent with the fact that, besides quiescent episodes, the 2009 observations could contain 
a large number of similarly short flaring events that may be potentially linked to the electron injections.
As can be seen from Fig. \ref{figs:xrays},  the levels of observed X-ray fluxes in 6 out of the 10 selected epochs are very similar to those measured in the 2009 MWL campaign.

Additionally, we note that the observed X-ray fluxes in all 10 epochs are higher than those observed in the extremely low-activity state in 2013.
Therefore, the theoretical SED that corresponds to the low-activity state may serve as an estimate for the background radiation of Mrk 421, and
is used to decompose the X-ray spectrum observed in the 10 epochs.

\begin{deluxetable*}{ccccccccccc}
\centerwidetable
\label{tab:table4}
\renewcommand\tabcolsep{8.0pt}
\tablecaption{Input model parameters and derived quantities for the daily X-ray spectrum from 10 observations.}
\tablehead{
Obs. ID &\colhead{$\delta_{\rm D}$}	&	\colhead{$B'$}      & \colhead{$R'$}  	& \colhead{$\nu_{\rm pk} (E_{\rm pk})$}	& \colhead{$q_0'$}    & \colhead{$\gamma'_{\rm i,max}$}  & \colhead{$p$}  & \colhead{$t'_{\rm acc}$}   & \colhead{$D_0$}  & \colhead{$U_{\rm e}'/U_{\rm B}'$} \\
	&\colhead{}	 &\colhead{G}	  &\colhead{$10^{15}$cm}	 	&\colhead{$10^{17}$ Hz	(keV)} &\colhead{$\rm s^{-1}$} 	    	 &\colhead{$10^5$}	&  	 &\colhead{$R'/c$}   &\colhead{$10^{-6}\rm s^{-1}$}    &\colhead{}
}  
\decimalcolnumbers
\startdata
	0099280301	&42		&0.23    	 &1.7   	&$7.0 (2.89)$ &$1.06\times10^{46}$ 		&$0.9$ 	&$2.30$	&$1.82$	&$2.42$	&$8.0$ 	\\
	0136540701	&25		&0.17 	 &3.2   	&$3.0 (1.24)$ &$2.18\times10^{44}$			&$13.0$ 	&$1.78$	&$1.80$	&$1.30$	&$13.9$	\\
	0136541001	&28		&0.11	 &3.2		&$11.5(4.76)$&$2.54\times10^{46}$			&$2.0$	&$2.35$	&$1.86$	&$1.26$	&$18.8$    \\
	0158970101	&30		&0.23    	 &2.3   	&$2.0 (0.83)$ &$6.30\times10^{45}$			&$5.0$ 	&$2.25$	&$2.13$	&$1.53$	&$4.6$ 	\\
	0150498701	&40		&0.20 	 &2.5   	&$2.5 (1.03)$ &$1.46\times10^{45}$			&$7.0$ 	&$2.10$	&$2.50$	&$1.20$	&$5.7$ 	\\
	0158971201	&47		&0.14   	 &1.7   	&$38.0 (15.7)$&$8.44\times10^{45}$ 		&$3.0$ 	&$2.25$	&$1.74$	&$2.53$	&$33.4$ 	\\
	0158971301	&43		&0.15   	 &3.0   	&$5.0 (2.07)$ &$2.63\times10^{46}$			&$2.0$ 	&$2.40$	&$2.35$	&$1.06$	&$8.2$ 	\\
	0302180101	&38		&0.22 	 &1.6	 	&$8.0 (3.31)$ &$3.55\times10^{45}$			&$10.0$ 	&$2.20$	&$1.84$	&$2.54$	&$9.4$ 	\\
	0411080301	&53		&0.10	 &2.0  	&$9.0 (3.72)$ &$1.92\times10^{44}$			&$5.5$ 	&$1.95$	&$5.36$	&$0.70$	&$51.4$ 	\\
	0502030101	&36		&0.17  	 &3.0   	&$5.5 (2.27)$ &$2.62\times10^{46}$			&$2.0$ 	&$2.35$	&$1.70$	&$1.47$	&$7.3$ 	\\
	\hline
$\langle\cdot\rangle$&38		&0.17	&2.4		&9.2(3.8)	      &	$1.09\times10^{46}$			&5.0		&2.19		&2.31	&1.6		&16.1	\\
\enddata 
\tablecomments{
Column (1) : Observation ID. Here, we fixed $\gamma'_{\rm i,min}=2\times10^3$ and $\eta_{\rm esc}=1.26$.  
}
\end{deluxetable*}

\subsection{Modeling X-ray spectrum and  lag-frequency}

\begin{figure*}
\vspace{2.2mm} 
\figurenum{6}
\centering
\includegraphics[width=0.4\textwidth] {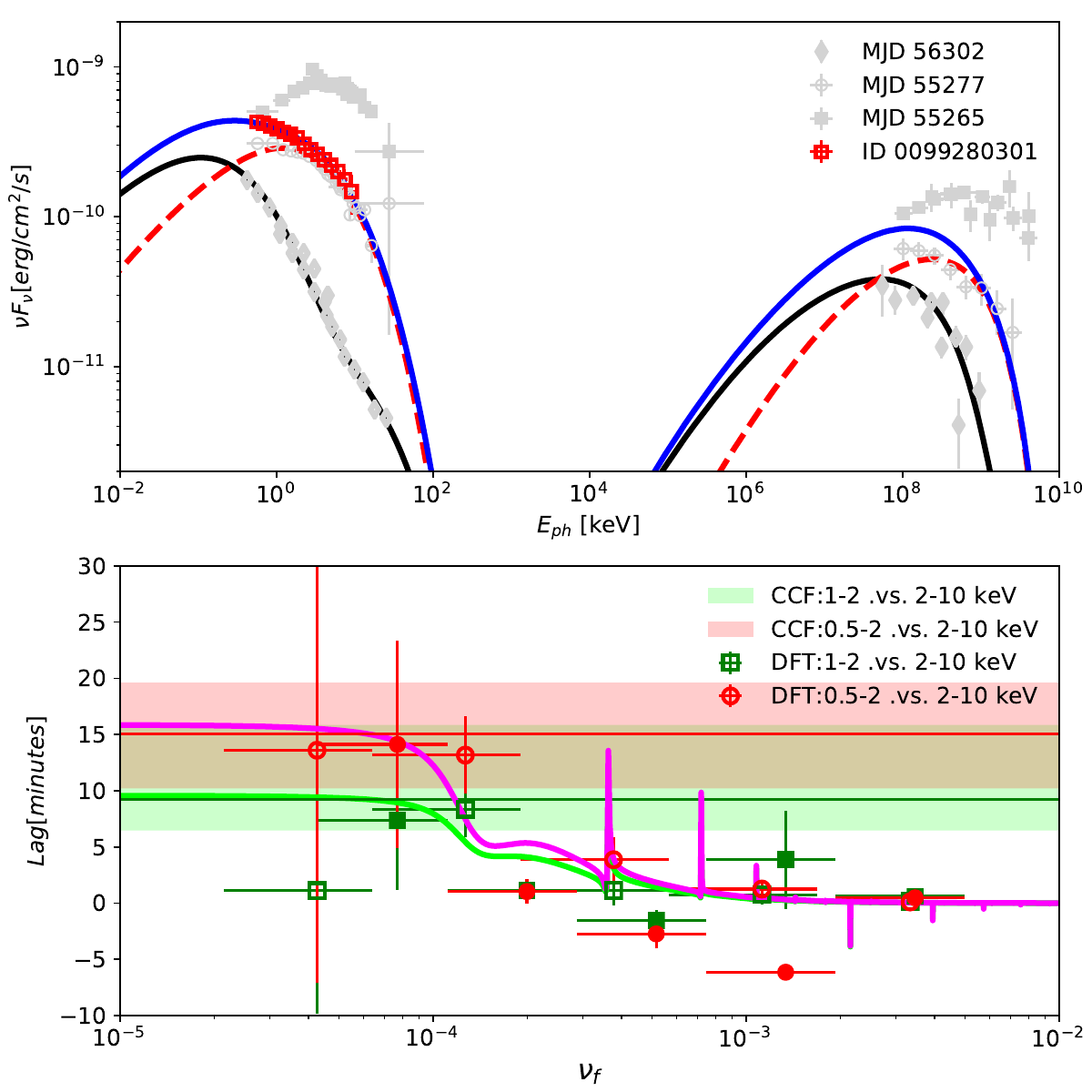} 
\includegraphics[width=0.4\textwidth] {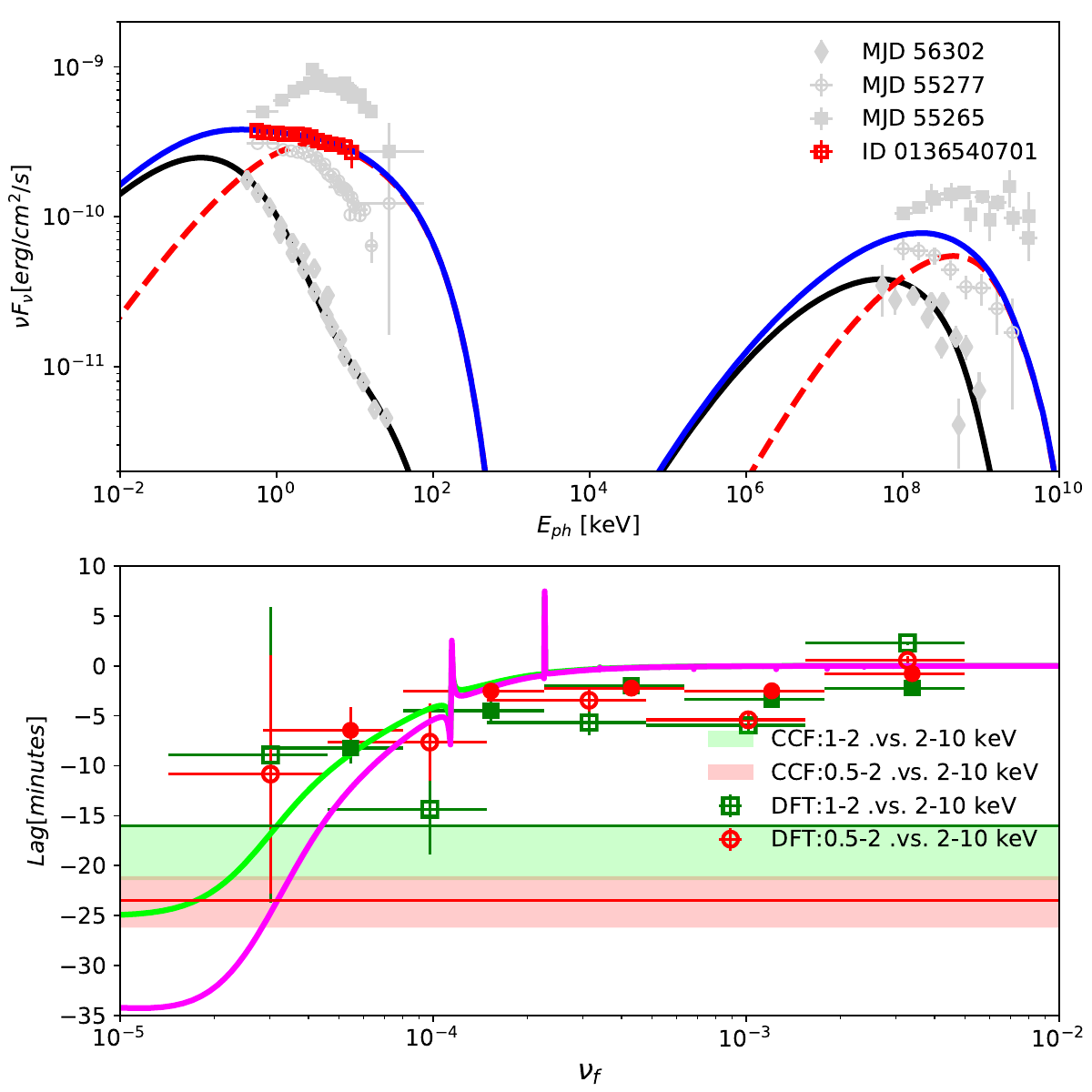} 
\includegraphics[width=0.4\textwidth] {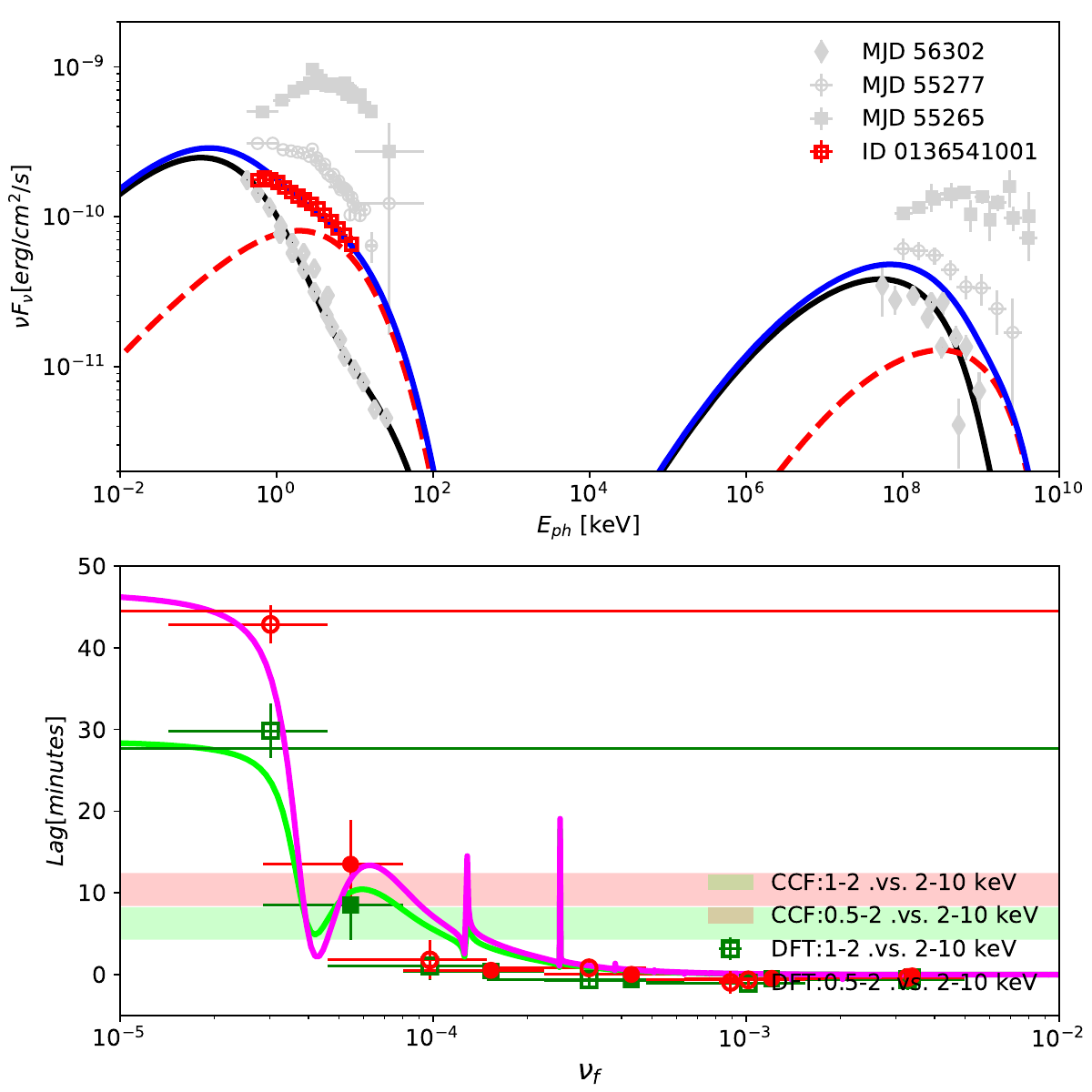} 
\includegraphics[width=0.4\textwidth] {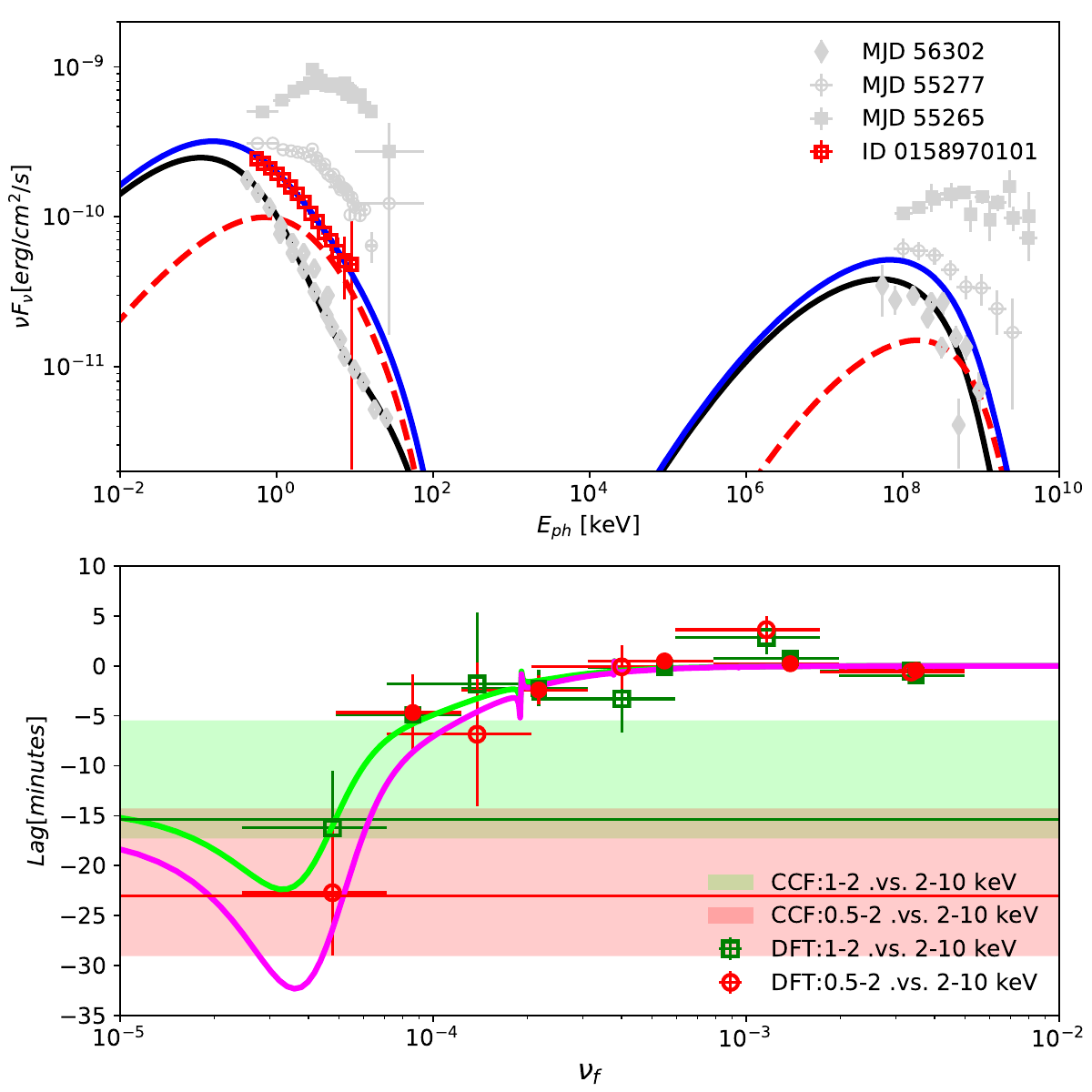} 
\caption{Modeling of the 0.5--10 keV X-ray spectrum and the Fourier time lags observed in different epoch.  
In the upper part of each panel, filled gray diamonds represent the simultaneous X-ray and TeV $\gamma$-ray data points on 2013 January 10, as reported in \cite{Balokovic2016}.
The black squares and dark gray circles show the high- and low-state SEDs during flare activity in March 2010, respectively. The data points are taken from \cite{Aleksic2015}.
The red dashed lines and gray solid lines denote the SEDs from the non-stationary and quasi-stationary zone, respectively.
The blue solid lines denote the total SED from two zones.
In the lower part of each panel, the magenta and green solid lines respectively represent the theoretically predicted lag-frequency time lags for the 2--10 keV band with respect to the 0.5--2 and 1--2 keV bands,
while the corresponding CCF time lags resulting from our model are denoted by the horizontal red and green lines, respectively.
\label{figs:fig6}
}
\vspace{2.2mm}
\end{figure*}

\begin{figure*}
\vspace{2.2mm} 
\figurenum{6}
\centering
\includegraphics[width=0.4\textwidth] {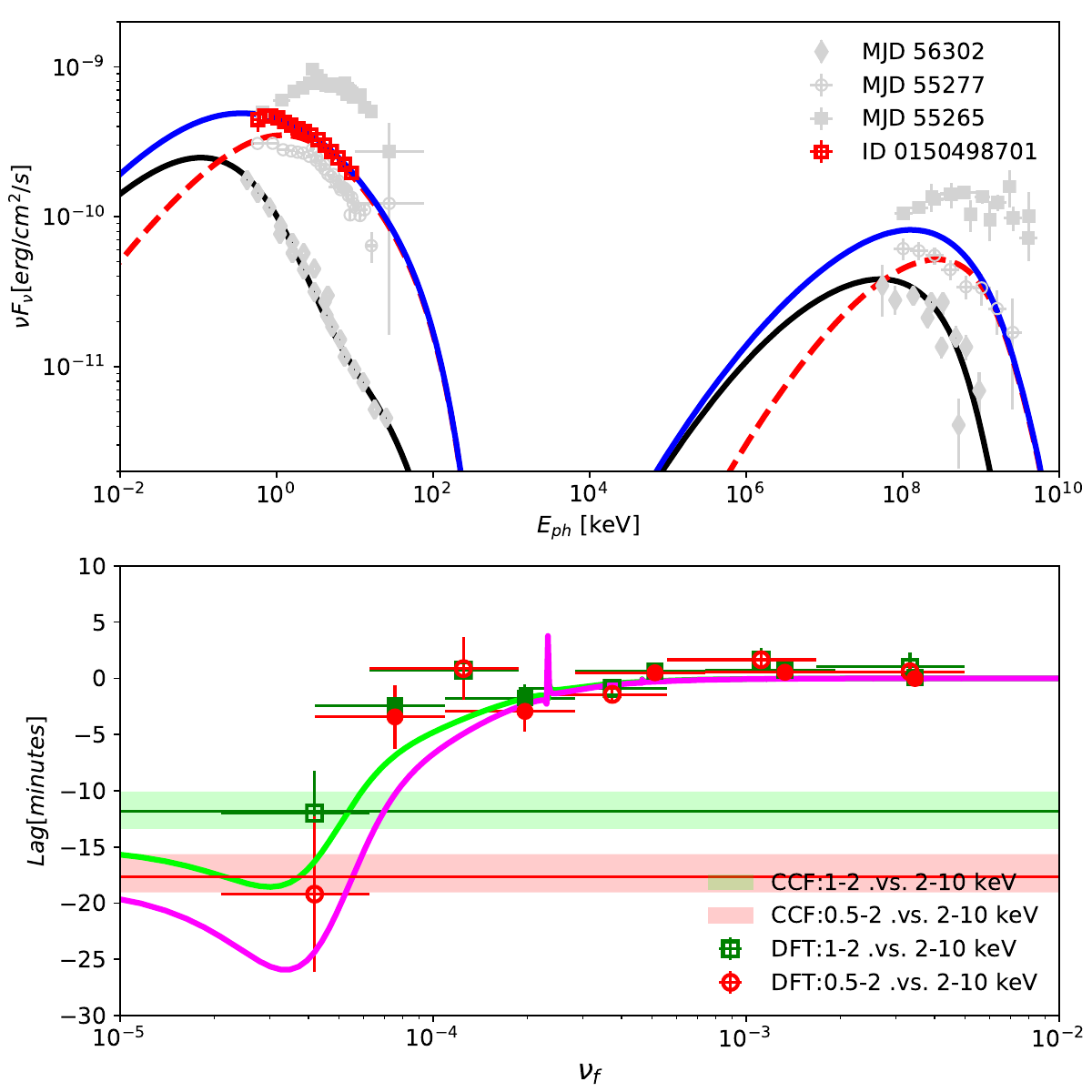} 
\includegraphics[width=0.4\textwidth] {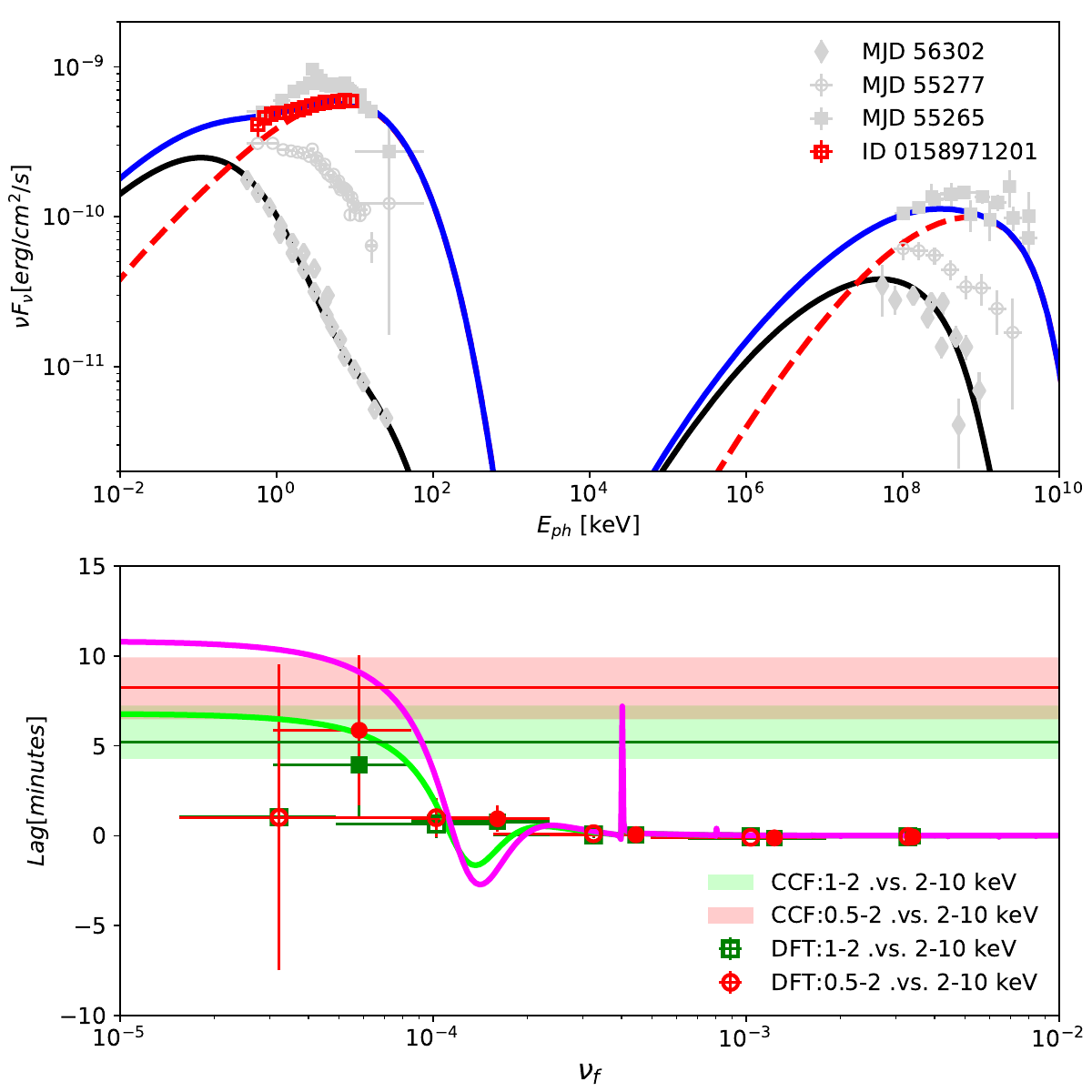} 
\includegraphics[width=0.4\textwidth] {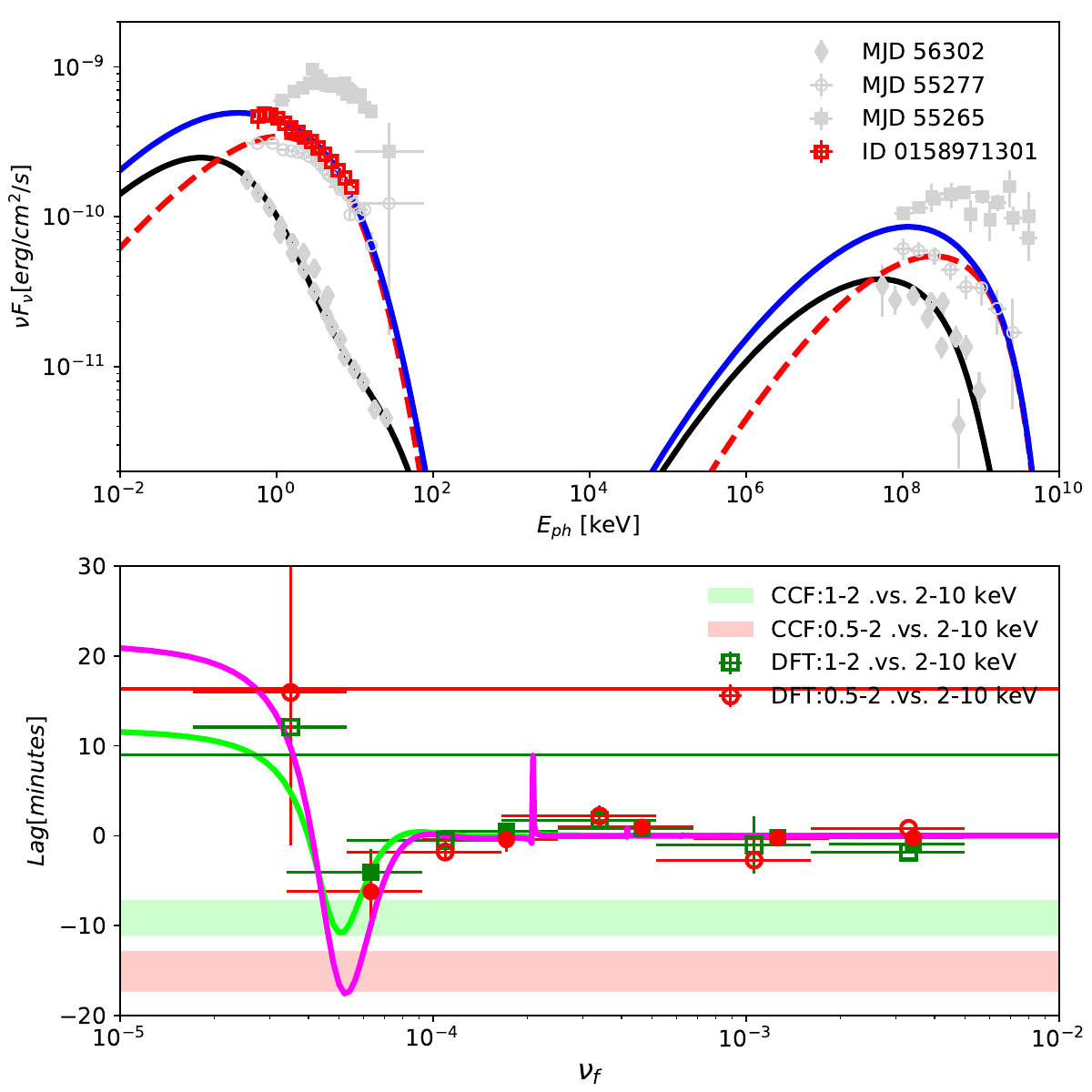} 
\includegraphics[width=0.4\textwidth] {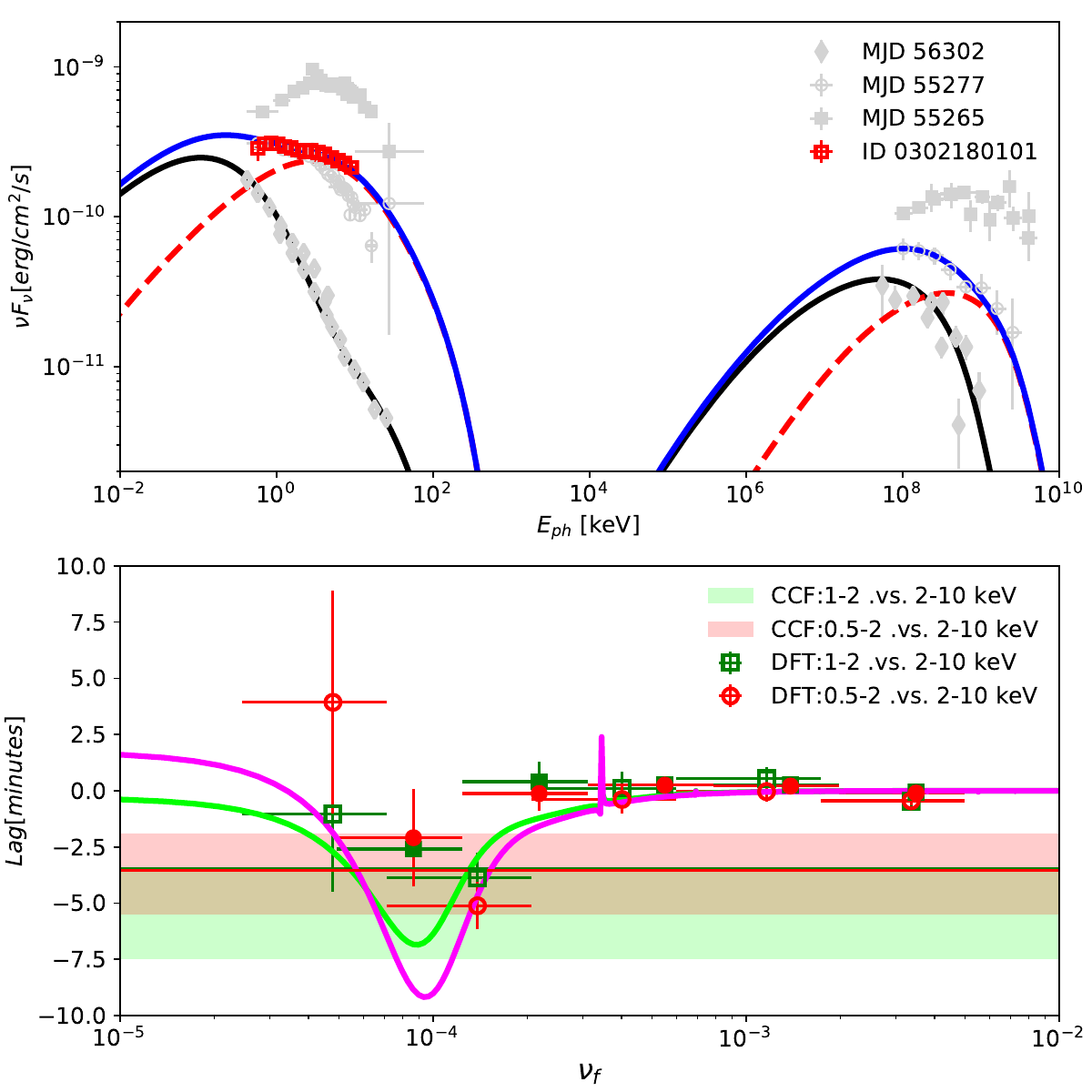} 
\includegraphics[width=0.4\textwidth] {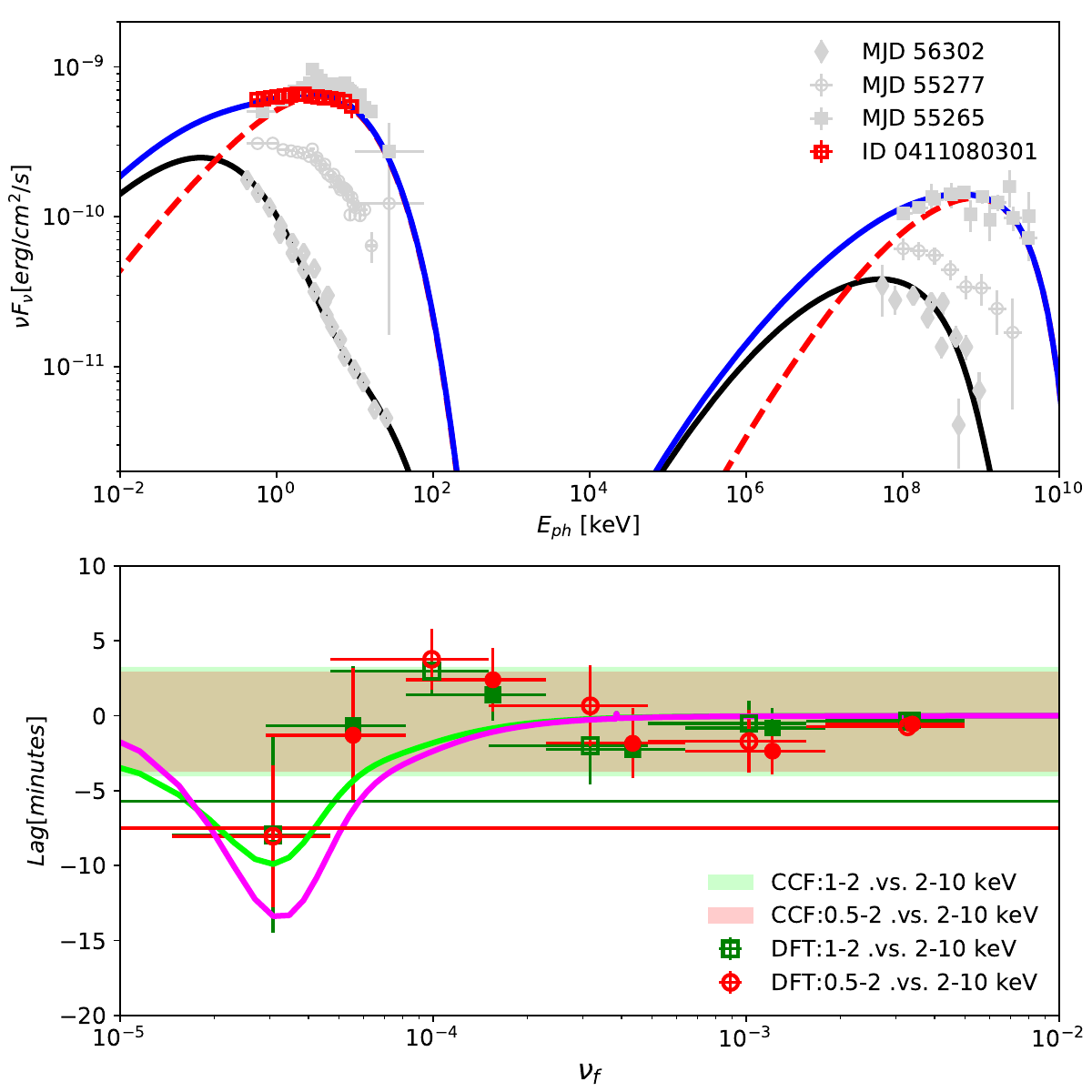} 
\includegraphics[width=0.4\textwidth] {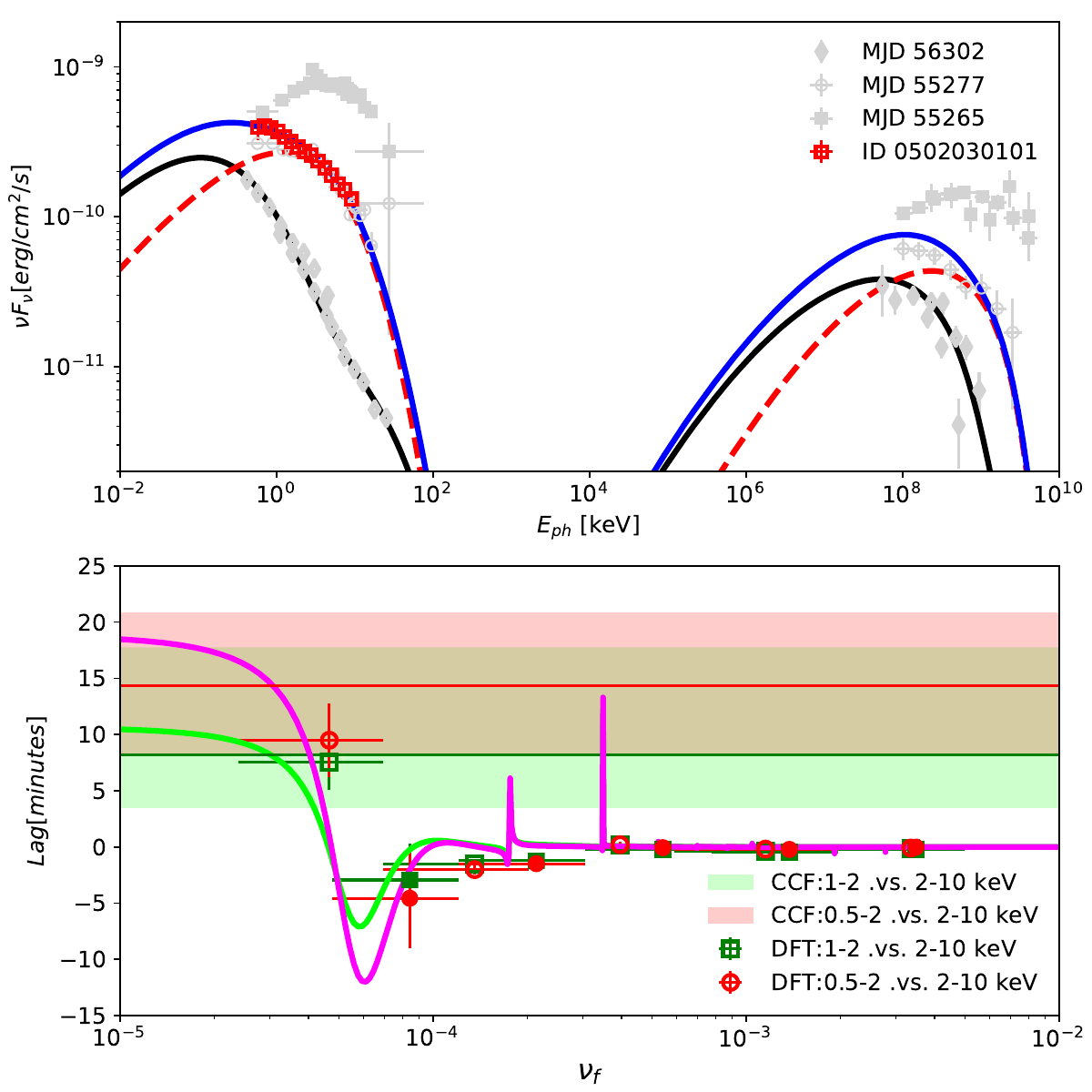}
\caption{continued.
}
\vspace{2.2mm}
\end{figure*}

In this section, we aim to simultaneously account for both the broadband X-ray spectrum and the time lags observed during 10 observations. 
For each observation, a comparison of the observed results and the simulated results is shown in Fig. \ref{figs:fig6}. 
In the course of the fitting, $\gamma'_{\rm i,min}=2\times10^3$ and $\eta_{\rm esc}=1.26$ are always held fixed,
and the remaining seven parameters listed in Tab. \ref{tab:table4} are allowed to vary.

Since the lack of simultaneous TeV $\gamma$-ray observations hinders our ability to effectively constrain the jet properties,
we refer to the simultaneous X-ray and TeV $\gamma$-ray data from the flaring activity in March 2010 \citep{Aleksic2015} 
as a benchmark for adjusting the parameters used to describe the X-ray spectrum in our study.
To illustrate the spectral and flux range of variability, we present the daily SEDs on MJD 55265 and MJD 55277,
which represent the highest and lowest flux levels during the flaring activity, respectively. 
The simultaneous data are shown in all plots of Fig. \ref{figs:fig6} as denoted by gray solid squares and empty circles, respectively.

In the modeling, we qualitatively fit the observed X-ray spectrum from 10 observations, 
using the steady-state solution generated by electrons that are continuously injected into the jet.
The steady-state SSC model approach allows us to observe basic physical parameters 
averaged over the observation in the variable blob \cite[e.g.,][]{Lewis2016,Lewis2018,Diltz2015,Diltz2016,Hu2021}.

As indicated by the fractional variability amplitude reported in Tab. \ref{tab:table2}, which increases with energy, 
one may argue that the X-ray flux variability is not dominated by changes in the global properties ($B'$, $R'$ and $\delta_{\rm D}$), 
but rather produced by variations in the emitting electron spectrum with the strongest variations occurring in the highest-energy electrons. 
Besides, the diffusion coefficient $D_0$ is not likely to vary on the same timescales as the observed X-ray emission \citep{Lewis2016,Diltz2015}.
For simplicity, the theoretical LCs are produced with the assumption of an impulsive injection of electrons,
 and then the standard DFTs for the resulting LCs are performed to give the theoretical lag-frequency spectra (Appendix \ref{append:solution}).
 To intuitively understand the results, the time lags are measured using the CCF centroid, which is very reliable for a single smooth flare event.
In order to be consistent with the basic assumption of spatial homogeneity, we require the injection time $t'_{\rm inj}$ to be longer than $t'_{\rm dyn}=R'/c$ \citep[e.g.,][]{Li2000ApJ}. 
Without loss of generality, we apply a constant injection with duration of $2R'/c$ to the injection rate of the electrons to produce the theoretical LCs in different bandpasses.

Both the observed CCF and Fourier time lags may provide important clues about the underlying physical processes.
Since the former may represent mixed results for a pair of the complex and structured LCs,
it would be inappropriate to attempt to model those results with the CCF centroid measured from a pair of simulated LCs generated by a single impulsive injection.
However, the Fourier time lags caused by acceleration or cooling processes do not depend on the time envelope of the electron injection
\footnote{
It is worth noting that the LC observed in a single observation is only a random realization of the intrinsic variability that may be characterized by a PSD with a power-law form (i.e., colored noise) \citep{Timmer1995A&A,Emmanoulopoulos2013MNRAS}.
Thus, it may be more reasonable to simulate theoretical LCs numerically with the assumption of an injection with a random time envelope, resulting in a colored noise component. 
However, this does not affect the calculation of time lag in the frequency domain (see Appendix \ref{appdex:lag}).
Moreover, this will introduce a variety of additional parameters that are not well known, and the simulated LC is typically not directly compared with the observed LC. 
}.
We intend to mainly reproduce the lag-frequency spectra measured between the main hard (2--10 keV) and soft (0.5--2 keV) bands, 
which is the most frequently used in the spectral and temporal analysis of the X-ray variability.
To impose a better constraint on the model parameters, we further reproduce the lag-frequency spectra measured between 2--10 keV and 1--2 keV.
In Fig. \ref{figs:fig6}, we also show the corresponding CCF lags measured from the simulated and observed LCs.

From Fig. \ref{figs:fig6}, it can be seen that the theoretical X-ray spectrum agrees fairly well with the observational data, while
our model satisfactorily describes the lag-frequency spectra combined with $\tau_1$ and $\tau_2$.
Moreover, we note that the CCF lags resulting from the simulated LCs are consistent with those resulting from the observed LCs, except for Obs. IDs: 0136541001, 0158971301 and 0411080301.

As can be seen from Tab. \ref{tab:table4}, the successful fits for the 10 epochs indicate that the inferred Doppler factor $\delta_{\rm D}$ lies in a reasonable range of 25--53.
According to \cite{Hervet2019}, a perturbation with a high $\delta_{\rm D}>30$ crossing multiple recollimation shocks 
is required to interpret the X-ray variability pattern of Mrk 421 observed by Swift-XRT during 13 years of observation.
Such a high value of $\delta_{\rm D}$ is also required in the study performed by \cite{Banerjee2019MNRAS} on the time-dependent modeling of Mrk 421 in an internal shock scenario.
Alternatively, the emitting plasma blobs with a high $\delta_{\rm D}$ are easily produced in the plasmoid-driven reconnection framework \citep[e.g.,][]{Giannios2009,Giannios2013,Petropoulou2016}. 
When the magnetization is high enough, the Lorentz factor of these plasma blobs in the rest frame of the jet can be as high as 50.

Meanwhile, we note that the size of the active zones $R'$ lies in the range $(1.6-3.2)\times10^{15}$ cm, which is smaller by a factor of about 6--12 compared to the quasi-stationary zone.
Combined with the large $\delta_{\rm D}$, the significantly smaller $R'$ gives rise to the rapid variability with characteristic time scale of the order of a few thousand seconds, 
which are consistent with the observations.

Moreover, the successful model parameters yield $U'_{\rm e}/U'_{\rm B}$ that significantly depart from equipartition in the range $\sim$5--51.
This suggests that the high-energy emission from Mrk 421 may be produced in a kinetic energy flux-dominated jet.
Such a departure from equipartition has often been reported in various studies for Mrk 421 during the different active 
states changing on the sub-hour-to-day timescales \citep[e.g.,][]{Shukla2012A&A,Aleksic2012A&A,Aleksic2015,Sinha2016A&A}.
Particularly, the departure was revealed by the SED modeling with an internal shock model \citep{Banerjee2019MNRAS}.
In the context of our model, this is mostly related to the increase in the electron injection rate ($\propto q_0'/R'^3$), which is more than two to three orders of magnitude larger than in the quasi-stationary zone.
Meanwhile, the power-law index of the injected electron spectrum is smaller than 2.5--2.7 inferred from the modeling of accurately measured X-ray spectra during the low-activity states in 2009 and 2013.
This suggests that the energetic electrons injected into the active and quasi-stationary zones may be associated with different acceleration mechanisms (see discussion in Section \ref{diss:acc}).

Another significant difference is the diffusion coefficient. 
The modeling indicates that $D_0$ in the active zones ranges from $(0.7-2.54)\times10^{-6}$, which is about a factor of 3--10 larger than the quasi-stationary zone.

The differences in the physical properties of the two zones may support the idea that the active zones and the quasi-stationary zone correspond to physically distinct regions within the jet.

On the other hand, comparing the physical properties of the active zones responsible for the similar shape and flux level of X-ray emission reveals 
that the main differences may be related to changes in the diffusion coefficient $D_0$ and/or the physical quantities related to the electron injection spectrum (i.e.,$\gamma'_{\rm i,max}$ and $p$).
These quantities are key in determining the presence of soft or hard time lags.
The global properties of the blobs ($B'$, $\delta_{\rm D}$ and $R'$) vary within a factor of about two, which is associated with the observed time lags.

\section{Discussion}\label{sec:diss}

\subsection{dependence of time lags on the model parameters}\label{diss:lag}

The time lags measured between LCs at different bands are an extraordinary opportunity to break degeneracies of the physical quantities involved in the SED modeling.
Based on the analytic solutions to the electron transport equation in the Fourier domain, \cite{Finke2014ApJ} explored the behaviors of Fourier time lags associated with variability in blazars.
However, the authors do not take into account electron acceleration in the model, and therefore it was only able to produce soft time lags.
To explain the formation of the hard time lags in blazar jets, \cite{Lewis2016} further extended the approach and developed an analytical physical 
model incorporating stochastic acceleration, shock acceleration, synchrotron cooling, adiabatic expansion, and electron escape regulated by Bohm diffusion. 
The model is able to qualitatively reproduce the hard time lags observed in the 1998 April 21 flare from Mrk 421, as obtained 
using \emph{BeppoSAX}, as well as the X-ray spectrum observed at the peak of the flare.
The authors indicated that the hard time lags and the peak X-ray spectrum
require two populations of mono-energetic seed electrons , which have very different origins and undergo very different levels of shock acceleration and adiabatic losses.

In the context of our model, the observed X-ray spectrum is considered to be a sum of the flux of the quiescent background component and the active component,
while the X-ray flux variation and the associated lags are mainly caused by the active component.
In our model, the use of a power-law distribution for the electron injection spectra is motivated by two physically well-known dissipation processes: magnetic reconnection events and shocks.
Compared with previous works, our model can simultaneously produce the observed X-ray spectrum and both the soft and hard time lags using the same set of parameters.
It is necessary to show how the Fourier time lag profiles and the CCF lags vary, when the simultaneous X-ray and TeV $\gamma$-ray 
spectrum are kept relatively constant by changing multiple physical quantities over a physically appropriate range at once.

We perform the study by varying model parameters with respect to two baseline models, referred to as Cases A and B. 
In Case A, the electrons responsible for the hard X-ray emission are dominated by the radiative cooling process, while in Case B, they are dominated by the acceleration process.
Our baseline parameter values for Cases A and B are selected to reproduce the strictly simultaneous X-ray and TeV $\gamma$-ray data 
of Mrk 421 on MJD 55277, which represents the typical shape of the X-ray spectrum observed in Mrk 421.
The detailed discussions are given in Appendix \ref{app:est}, where the two baseline models are plotted in Fig. \ref{figs:fig11}, with baseline parameters reported in Tab. \ref{tab:para1}.
In Appendix \ref{app:est}, we also show the effects of changing the parameters associated with the properties of the turbulence wave and the injected electron spectra.
The results indicate that the X-ray time lags can greatly constrain the jet physics and offer valuable insights into the underlying mechanisms at work in the jet.

Given the theoretical prediction in Section \ref{sec:predi}, the magnetic field strength and Doppler factor are two critical parameters for determining the measured time lags. 
On the other hand, it is worth noting that the time lags may be related to the light-crossing timescale determined by the size of the active zone.
Meanwhile, the global properties of the active zone are important to determine the cutoff of the TeV $\gamma$-ray spectrum.
Here, we mainly illustrate the effects of variations in $B'$, $\delta_{\rm D}$ and $R'$.

\begin{figure*}
\vspace{2.2mm} 
\figurenum{7}
\label{figs:fig7}
\centering
\includegraphics[width=0.45\textwidth] {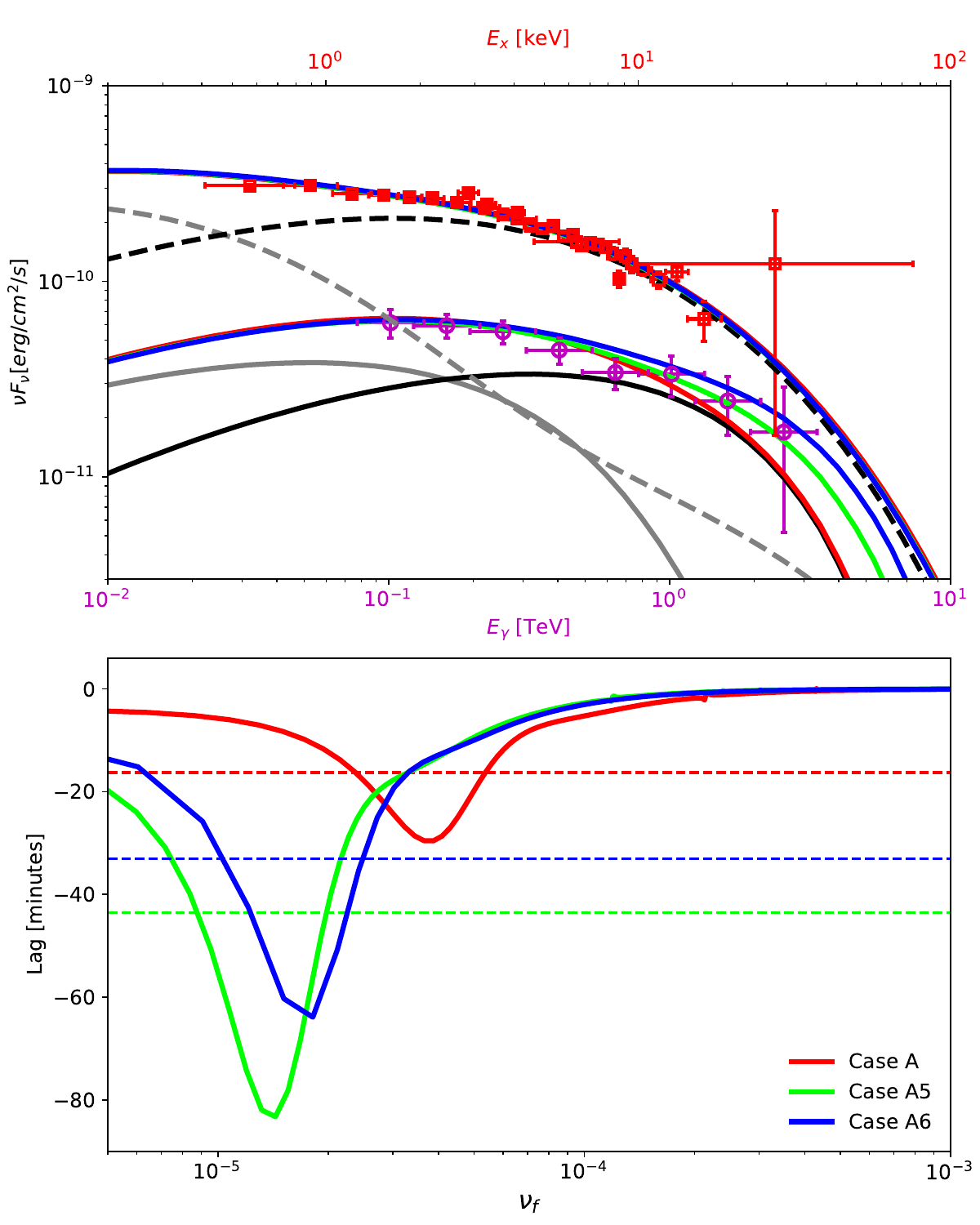} 
\includegraphics[width=0.45\textwidth] {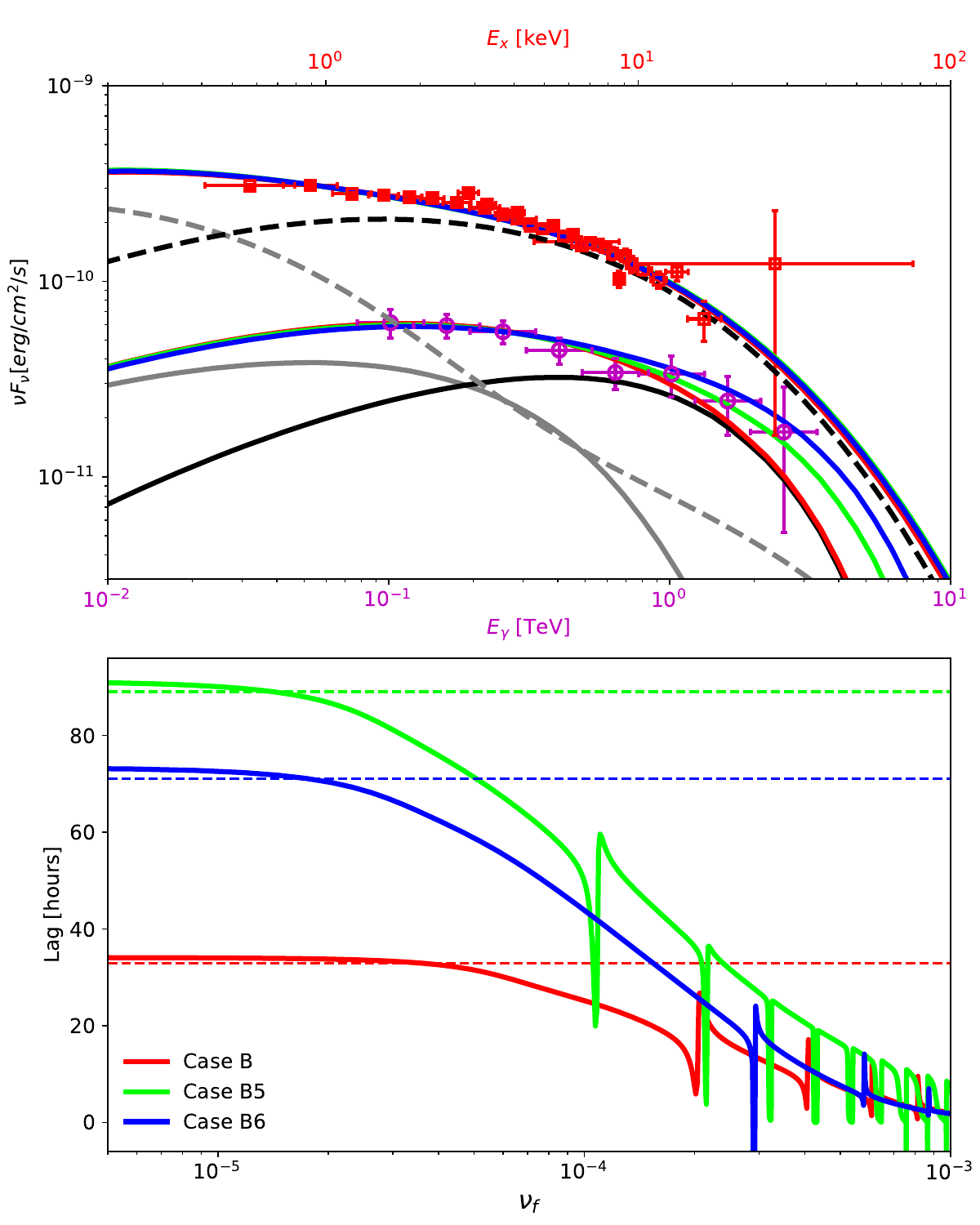} 
\caption{
The theoretical SEDs (upper) and time lags (lower) resulting from variations in $B'$, $\delta_{\rm D}$ and $R'$. 
In the lower panels, the Fourier time lag profiles for different cases are indicated in the legend, and the corresponding CCF lags are marked as horizontal lines.
In the upper panels, the empty red squares and magenta circles represent the simultaneous X-ray and TeV $\gamma$-ray data on 2010 March 22 (MJD 55277), reported in \cite{Aleksic2015}, respectively.
The gray dashed and solid lines are the synchrotron and SSC spectrum from the (quasi-)stationary zone, respectively.
The synchrotron and SSC spectrum from the non-stationary zone for Case A (left) and B (right) are denoted by the black dashed and solid lines, respectively.
The red, green and blue solid lines are the SEDs from non-stationary zone that correspond to the three cases indicated by the legend of the lower panels, 
superimposed with the SED from the (quasi-)stationary zone.
}
\vspace{2.2mm}
\end{figure*}

Fig.\ref{figs:fig7} is designed to depict the dependence of the resulting SED and time lags on the parameters involved.
We decrease the strength of the magnetic field from $B' = 0.15$ G to $B' = 0.08$ G.
To keep the Compton dominance approximately constant, we counterbalance the decrease in $B'$ by increasing $R'$ (Case A5 and B5) or $\delta_{\rm D}$ (Case A6 and B6).
The adjustments of $q_0'$ and $\gamma'_{\rm i,max}$ (or $\nu_{\rm pk}$) are required to ensure that the synchrotron spectrum remains relatively constant.
From the view of the produced SEDs, the difference in the resultant TeV $\gamma$-ray spectrum is visible.
Specifically, the SSC spectrum tends to shift toward higher energies as $B'$ decreases, and
further increases in $\delta_{\rm D}$ results in a further shift of the component toward the high-energy end.

On the other hand, the measured  $|\tau_{\rm c}|$ increases with the decrease of $B'$ (Case A5 and B5), and then decreases with the increase of $\delta_{\rm D}$ (Case A6 and B6).
The ratios of the CCF lags in Case A5 and A6 with respect to that in Case A are respectively $\tau_{\rm c,A5}/\tau_{\rm c,A}\simeq2.7$ and $\tau_{\rm c,A6}/\tau_{\rm c,A}\simeq2.0$, 
approximately equal to $\sqrt{B'^3_{\rm A}/B'^3_{\rm A5}}\simeq2.6$ and $\sqrt{\delta_{\rm D,A}B'^3_{\rm A}/\delta_{\rm D,A5}B'^3_{\rm A6}}\simeq2.2$.
The ratios $\tau_{\rm c,B5}/\tau_{\rm c,B}\simeq 2.7$ and $\tau_{\rm c,B6}/\tau_{\rm c,B}\simeq 2.2$ are in accord with the theoretical predication
$\sqrt{\nu_{\rm pk, B}B'^3_{\rm B}/\nu_{\rm pk, B5}B'^3_{\rm B5}}\simeq2.5$ and $\sqrt{\nu_{\rm pk, B}\delta_{\rm D,B}B'^3_{\rm B}/\nu_{\rm pk, B5}\delta_{\rm D,B5}B'^3_{\rm B6}}\simeq2.1$.

Meanwhile, the time lags measured in the Fourier domain also exhibit a similar variation.
The maximum time lags in the frequency-dependent spectra approximately double the CCF lags in Case A5 and A6. 
In Case B5 and B6, the CCF lags are comparable to the Fourier time lags at low frequencies.

The results demonstrate that the values of our interested physical quantities extracted using both X-ray spectra and the associated 
interband time lags may be in a well-constrained parameter space.
Moreover, we conclude that the Fourier time lag spectra may be a more promising diagnostic of the nature of the underlying physical processes in blazar jets.

\subsection{Variability and associated energy dissipation process}\label{diss:acc}

The rapidly variable emission may be related to impulsive magnetic reconnection events \citep{Giannios2009,Giannios2013} or shocks \citep{Spada2001MNRAS,Bottcher2019ApJ,Chandra2021ApJ} occurring in the relativistic jets.
The two well-known energy dissipation processes can efficiently generate a population of electrons following the form of a power-law distribution.
 However, an important difference between these two mechanisms is the magnetization of the dissipation region in the jets. 
When the jet is dominated by kinetic energy flux at the dissipation distance, the most natural candidate for powering the jet emission is shocks 
that transfer a significant fraction of the kinetic energy to electrons.
On the other hand, in a magnetically dominated dissipation region, magnetic reconnection is a more likely mechanism for depositing magnetic energy into the radiating electrons.

Using all public \XMM X-ray observations for Mrk 421, \cite{Yan2018} found that the distributions of the flare-profile parameters obey a power-law form that is a signature of a self-organized criticality model.
Magnetic reconnection processes may therefore be supported.
On the other hand, the detailed particle-in-cell (PIC) simulations show that the power-law index of the electron spectra formed in magnetic reconnection depends heavily on the plasma magnetization \citep[e.g.,][]{Cerutti2012ApJ,Guo2014PhRvL,Sironi2014ApJ,Petropoulou2019ApJ}. 
For mild magnetization $1<\sigma<10$, the electron spectra are characterized by a soft power-law index with $p\gtrsim2$, whereas $p<2$ for high magnetization $\sigma\gtrsim10$.
However, a direct result of this scenario is that the electron kinetic energy density and magnetic field energy density in the dissipation region are found in substantial equipartition,
with an upper limit of $U'_{\rm e}/U'_{\rm B} \sim3$ \citep{Sironi2015MNRAS}.
This is to be compared against our results that in all cases the inferred electron energy density substantially exceeds the magnetic energy density, by more than one order of magnitude.
Thus, the magnetic reconnection process may be not expected to contribute significantly to the energetic electrons injected into the active blobs.

Shocks are a common phenomenon thought to occur in AGN jets. 
It is therefore natural to consider the possibility that the violent variability could be triggered by shock propagation inside the jet.

\cite{Brinkmann2005A&A} compared the observed variability pattern with relativistic hydrodynamic simulations of a commonly accepted internal shock model for Mrk 421 emission.
The authors indicated that LCs that looked similar to the observed ones could be represented by the numerically estimated flux superimposed with a substantial background emission of $\gtrsim$60\%.
Meanwhile, the results suggested that the observed emission of Mrk 421 may consist of at least two components: a quasi-stationary component and a flaring component.

Recently, \cite{Banerjee2019MNRAS} carried out extensive simulations to study the SEDs of Mrk 421 during a giant outburst in February 2010.
The authors showed that a time-dependent leptonic jet model in the internal shock scenario can successfully reproduce the SEDs in three spectral states, 
which represent the temporal evolution of Mrk 421 over a period of 5 days from a low- to high-flux state during the flaring event. 
Moreover, this scenario can produce significant spectral hardening from X-rays to TeV $\gamma$-rays, as well as the harder-when-brighter behavior in these energy bands.
It is also interesting to note that the MWL spectral evolution during the giant outburst could be satisfactorily described by considering the emission from two connected regions, proposed by \cite{Dmytriiev2021MNRAS}.
In the flare scenario, the steady-state emission arises from pre-accelerated electrons due to a stationary relativistic shock located in front of the emission region, referred to as the quiescent blob.
The flaring emission is generated by transient turbulent regions where the pre-accelerated electrons escaping from the quiescent blob are injected and further accelerated via interaction with turbulent plasma waves excited by turbulent motion.

Alternatively, a realistic scenario could be a standing shock generated by the interaction of the upstream extended jet plasma with a turbulent ambient medium, or by recollimation in the jet.
Such a scenario was proposed, based on the successful interpretation of the simultaneous MWL SED of Mrk 421 in the quiescent state \citep{Abdo2011} and the flaring state \citep{Aleksic2015}.
In particular, \cite{Hervet2019} probed the X-ray signature of recollimation shocks in the jet of Mrk 421. 
Using the \emph{Swift}-XRT dataset spanning 13 years, they confirmed that the X-ray variability pattern of Mrk 421 
is consistent with the scenario of the propagation of perturbations crossing the multiple stationary VLBI radio knots observed in the jet.
It is in favor of the interpretation of slow/stationary radio knots as multiple recollimation shocks.

Indeed, the high X-ray polarization observed by Imaging X-ray Polarimetry Explorer provides important evidence for the shocks operating in the jet of Mrk 421 \citep{Laura2023NatAs}.
This is because the shock front partially aligns a previously disordered magnetic field.

The power-law form of electron distribution is a key feature of the shock mechanism. 
In the non-relativistic shock regime, both the linear theory and numerical simulations predict the power-law indices $p = (r+2)/(r-1)$ which are only determined by the compression ratio $r$ \citep[e.g.,][]{Drury1983RPPh,Bell1978MNRAS,Ellison2004APh}. 
The canonical index $p=2$ is highlighted for strong shocks with $r=4$. 
On the other hand, in the relativistic shock case, the power-law index depends critically on the nature and magnitude of turbulence, the shock speed, and the shock field obliquity.
A nearly universal value of the index, $2.2 < p < 2.3$, is produced in the (semi-)analytic \citep{Kirk1989MNRAS,Kirk2000ApJ} and numerical studies of relativistic (quasi-)parallel shocks \citep[e.g.,][]{Bednarz1998PhRvL,Achterberg2001MNRAS,Ellison2004APh,Niemiec2004ApJ,Crumley2019MNRAS}.
A steeper spectrum with $p \simeq 2.5$ can be found under considering the modification of shock by the back-reaction of the accelerated particles\citep{Kirk1996A&A}.
Such a soft spectrum can also emerge from PIC kinetic plasma simulations where the Weibel instability enhances the turbulence \citep{Sironi2009ApJ,Sironi2013ApJ}.
Indeed, relativistic oblique shocks can result in a wide variety of power-law indices that vary dramatically with obliquity \citep{Summerlin2012ApJ,Baring2017MNRAS}. 
Generally, subluminal shocks possess indices in the range $1 < p < 2.5$, 
whereas superluminal shocks that possess higher obliquities have soft spectrum with $p > 2.5$.

Thus, the inferred spectral indices with $\langle p\rangle\simeq2.2$ are compatible with the theoretical predictions of the shocks scenario.
As expected from theoretical arguments, the energy dissipation at shocks does introduce a threshold energy, 
below which the electrons are not freely able to cross the shock front and then no shock acceleration is in principle allowed \citep[e.g.,][]{Begelman1990ApJ}.
For $p>2$, the threshold energy is determined as 
\begin{equation}
\gamma'_{\rm min}\simeq
\Gamma_{\rm sh}\frac{\epsilon_{\rm e}}{q}\frac{m_{\rm p}}{m_{\rm e}}\frac{p-2}{p-1},  
\end{equation}
where $\epsilon_{\rm e}$ is the fraction of shock energy that goes into the electrons, 
$\Gamma_{\rm sh}$ is the bulk Lorentz factor of the shock in the upstream rest frame, and $q$ is the number of pairs per proton \citep{Piran1999PhR,Inoue2016ApJ,Bottcher2010ApJ}.
For $q=1$ and $\langle p \rangle=2.2$, a relatively high energy, $\gamma'_{\rm min}\sim 0.17\epsilon_{\rm e}\Gamma_{\rm sh}m_{\rm p}/m_{\rm e}$,
is required to cross the shock with the thickness that is expected to be of the order of the gyroradius of protons.
Using $\Gamma_{\rm sh}\sim\langle \Gamma \rangle=\langle \delta_{\rm D} \rangle\sim38$,
gives $\gamma'_{\rm min}\sim m_{\rm p}/m_{\rm e}$ for a typical value of $\epsilon_{\rm e}\sim0.15$.
According to the modeling of broadband SED from radio to TeV $\gamma$-rays, \cite{Abdo2011} proposed that the lower limit of $\gamma'_{\rm min}$ 
can be from $2\times10^2$ up to $10^3$ in order to reproduce the SMA and VLBA radio data. 
The higher value used should be reasonable for interpreting the X-ray spectrum and associated lags.
In principle, the high minimum electron energy and the spectral index of injected electrons might be both reconciled with the relativistic proton-mediated shocks.

In conclusion, we suggest that shocks inside the jet may be more reasonable for powering the emission and triggering the rapid variability. 
Therefore, we could propose a picture that seed electrons are accelerated by randomly occurring shocks,
which could provide a fully developed power-law spectrum \citep[see discussion in][]{Tammi2009MNRAS}.
However, this process may be difficult to create any observable time lag signal because only a small amount of electrons in the upstream plasma, 
which is filled with a relatively low magnetic field, can pass through the shock, undergoing Fermi-I acceleration.
Then, the accelerated electrons are injected downstream into the radiation zone, in which most of the observed radiation is being produced due to synchrotron and IC processes, 
while the energetic electrons may be further re-accelerated by turbulent plasma waves.
Thus, a soft lag will be detected, when the stochastic acceleration in the radiation zone is less efficient than the acceleration in the vicinity of the shock, 
and the electron cooling processes dominate over the stochastic acceleration of electrons responsible for the emission at two compared bands. 
Conversely, a hard lag is expected to be observed.

\section{Summary}\label{sec:summary}

X-ray observations are a powerful diagnostic tool for understanding the physical processes taking place in the jets.
In the present paper, we reanalyzed 10 observations of the X-ray emission from HBL Mrk 421, each with an exposure of $\gtrsim40$ ks, 
using the EPIC-pn instrument on board XMM-Newton.
For each of the 10 observations, we obtained the averaged X-ray spectra in the $0.5-10$ keV band 
and extracted 100s-binned LCs in six subbands: $0.5-1, 1-2, 2-4$, $4-10$ keV, as well as $0.5-2$ and $2-10$ keV bands.
We derived the fractional rms variability amplitude $F_{\rm var}$ for the first four subbands and performed PSD analyses.
The results of the analysis indicate that: (1) During the 10 observations, Mrk 421 exhibited significant 
variability in the X-ray band, and $F_{\rm var}$ increased with increasing energy. 
(2)The PSD of the variability follows a power law spectra above the white noise level up to $\sim2\times10^{-4}-2\times10^{-3}$ Hz, 
which corresponds to variability on timescales from minutes to an hour.
Below the frequencies dominated by the white noise, the spectral index of the measured PSDs lies 
within the range of $\sim1.9-3.0$, indicating that the variability amplitude rapidly decreases toward short timescales.

Particularly interesting information is the estimation of interband time lags associated with the variability in different subbands.
These time lags can shed light on the acceleration as well as on the cooling mechanisms,
and provide a powerful tool to constrain the physical properties of the jets.
Then, the standard CCF and cross-spectral methods are employed to verify the existenc and estimate the amount of the temporal lags between two LCs.
Our results show that the CCF lags are comparable to the Fourier time lags at the lowest frequency bin.
Moreover, we confirmed that the amount of the lags $|\tau|$ increases with the energy difference $\Delta E$ between the compared LCs.
The trend in the plot of $\tau-\Delta E$ is in agreement with the linear relation expected by the synchrotron cooling or Fermi-type acceleration.

To extract the physical information of the jet contained in the X-ray spectra and the interband time lags measured by both the CCF and cross-spectral methods, 
we employed a self-consistent two-zone model, with a large emission region responsible for a quasi-stationary emission and a smaller region for highly variable emission. 
For each zone, we focused on the injection of electrons with a power-law energy distribution.
We numerically solved a time-dependent transport equation that include synchrotron and Compton energy losses, hard-sphere stochastic acceleration, and escape.
The model provides a very satisfactory description of the X-ray spectra and time delays for all of the observations.
The physical quantities involved in the model can be well constrained by the CCF lag and the frequency-dependent lag spectra.
We suggest that the main energy dissipation process triggering rapid variability may be associated with shocks in the jets.
However, magnetic reconnection should not be excluded in the scenario involving a weakly magnetized shock.


\begin{acknowledgments}
We thank the anonymous referee for constructive comments that helped improve and clarify the paper.
This work is supported by the National Natural Science Foundation of China (NSFC-12263003, 12033006 and 12373016).
\end{acknowledgments}

%

\vspace{5mm}





\appendix

\section{Light curves and power spectral density} \label{append:lc}

We show the X-ray LCs in the 1--2, 2--4, and 4--10 keV energy bands in the upper panels of all the plots in Fig.\ref{figs:lcs}.
To provide a visual method for comparing the variability amplitudes and possible time lags of the variability in the different energy bands,
the LCs have been mean-subtracted and normalized by their standard deviation \citep{Marco2011MNRAS}.
In most observations of Mrk 421, the LCs can be characterized by long-term trends lasting for timescales comparable to or longer 
than the duration of the observations, with short-term substructures superimposed on it.
In Obs. ID 0502030101, LCs exhibit a long decaying phase, while LCs observed in Obs. ID 0158970101 display a decaying phase at the beginning followed by a rising phase.

The intrinsic variability power of the source can be represented by the power spectral density (PSD).
For evaluating a PSD, the standard procedure is by calculating the periodogram, which is the modulus-squared of the Discrete Fourier Transform (DFT) of the observed X-ray LC.
To give the same units as the PSD (i.e., $(rms/mean)^2$), the periodogram is further expressed as \citep[e.g.,][]{Vaughan2003MNRAS,Uttley2014A&ARv}
\begin{equation}
I_j = \frac{2\Delta t}{\overline{x}^2N}|\mathcal{X}_j|^2
\end{equation}
where $\overline{x}$ is the mean flux of the LC, $N$ the number of samples and $\Delta t$ is the time bin
\footnote{Note that integrating the normalized periodogram over a certain frequency range and taking the square 
root gives the fractional rms variability (i.e., $F_{\rm var}$) contributed by variations over that frequency range.}.

The normalized periodogram of the raw LC is an inconsistent estimator of the PSD, 
since its standard deviation at a frequency $\nu_f$ is 100\%, leading to the large scatter in the periodogram \citep[e.g.,][]{Vaughan2003MNRAS}.
The underlying PSD is obtained by fitting a given model to the normalized periodogram 
using the Markov Chain Monte Carlo (MCMC) method and maximum likelihood estimation \citep{Vaughan2010MNRAS}.
For the fitting, we adopt a single power law (PL) model for the red noise continuum, plus a constant $C$ to account for Poisson noise: 
\begin{equation}
P(\nu_{f})=A\left(\frac{\nu_{f}}{\nu_{\rm ref}}\right)^{-\alpha} + C
\end{equation}
where $A$ is the normalization, $\alpha$ is the power-law slope index, and $\nu_{\rm ref}$ is the reference frequency.
In this approach, the fitting procedure for the considered model is equivalent to minimizing the following function
\begin{equation}
D=2\sum^{N/2}_{j=1}\left(\frac{I_j}{P_j}+\log P_{j}\right),
\end{equation}
with $I_j$ and $P_j$ being the periodogram and the model PSD at frequency $\nu_j$, respectively.

For this purpose, {\it{emcee}} sampler developed by \cite{ForemanMackey2013PASP} is employed to obtain the posterior distribution of the model parameters.
The median value of the posterior distribution is used as the best-fitting result of each parameter, while 16\% and 84\% quantiles give their uncertainties. 
The best-fitting results are shown in Fig. \ref{figs:lcs}, and the power-law index $\alpha$, logarithm of normalization $\log_{10}A$, 
and constant $\log_{10}C$ are tabulated in Tab. \ref{tab:tab6}. 
Here, $\nu_{\rm ref}$ is fixed to be the lowest frequency of the periodograms.


In the figure, we also present the frequency-binned periodograms obtained using the standard tool {\it{pylag}} for time series analysis \citep{Uttley2014A&ARv}.
In the calculations, the raw periodograms are binned into five logarithmically spaced bins, 
spanning from the lowest frequency $T^{-1}$ (where $T$ represents the total length of the LC used for variability analysis from each observation)
up to the Nyquist frequency at $5\times10^{-3}$ Hz.
To obtain a more accurate PSD estimate, the LC is divided into two equal-length segments, and then a raw periodogram is calculated for each segment.
Finally, the binned periodograms are obtained by binning the raw periodograms over frequency in a given frequency bin, as well as over both segments.
From Fig. \ref{figs:lcs}, it can be seen that the two rebinned PSDs are consistent with each other, and in agreement with the result of the Bayesian PSD analysis.
This suggests that the binned periodograms offer reliable estimates of the underlying PSD, and the choice of bin size may be suitable.
Thus, the binned method may be reasonable for variability analysis in this work.


\begin{deluxetable*}{lCccccccccccc}
\centerwidetable
\tabletypesize{\scriptsize}
\label{tab:tab6}
\renewcommand\tabcolsep{4.5pt}
\tablecaption{The best-fitting parameters of a PL plus Poisson noise model fitted to the normalized periodograms of Mrk 421 in three energy bands: 1--2, 2--4 and 4--10 keV.}
\tablehead{
\colhead{Obs ID}  & \multicolumn4c{ 1--2 keV} &\multicolumn4c{ 2--4 keV} & \multicolumn4c{4--10 keV} \\
\colhead{}  & \colhead{$\log_{10}(A)$}  & \colhead{$\log_{10}C$} &  \colhead{$\alpha$}  &  \colhead{$t_{\rm v}$}  &  \colhead{$\log_{10}(A)$}  & \colhead{$\log_{10}C$} &  \colhead{$\alpha$}  & \colhead{$t_{\rm v}$}  &  \colhead{$\log_{10}(A)$}  & \colhead{$\log_{10}C$} &  \colhead{$\alpha$}  & \colhead{$t_{\rm v}$} 
}  
\decimalcolnumbers
\startdata
0099280301	& $2.38_{-0.47}^{+0.60}$	&$-0.74_{-0.05}^{+0.04}$ & $2.78_{-0.57}^{+0.75}$  & $58.7$  & $2.91_{-0.47}^{+0.60}$& $-0.19_{-0.04}^{+0.04}$ & $2.89_{-0.56}^{+0.72}$  &$65.8$  & $3.18_{-0.40}^{+0.50}$	& $0.19_{-0.04}^{+0.04}$ & $2.61_{-0.45}^{+0.57}$  & $55.7$ 	 \\
0136540701	& $3.17_{-0.27}^{+0.31}$	&$0.49_{-0.04}^{+0.04}$ & $2.28_{-0.22}^{+0.50}$ 	& $77.8$  & $3.60_{-0.21}^{+0.26}$& $0.54_{-0.04}^{+0.04}$ & $2.16_{-0.12}^{+0.27}$   &$44.6$  & $3.89_{-0.32}^{+0.37}$	& $0.16_{-0.04}^{+0.04}$ & $2.33_{-0.25}^{+0.30}$  & $29.2$    \\
0136541001	& $2.87_{-0.42}^{+0.48}$	&$-1.30_{-0.04}^{+0.04}$ & $2.91_{-0.39}^{+0.43}$ 	& $43.0$  & $3.07_{-0.44}^{+0.49}$& $-0.80_{-0.04}^{+0.04}$ &$2.83_{-0.42}^{+0.48}$   &$50.0$  & $2.93_{-0.40}^{+0.47}$	& $-0.27_{-0.04}^{+0.04}$ &$2.34_{-0.36}^{+0.44}$  & $50.0$  	 \\
0158970101	&$2.04_{-0.32}^{+0.33}$	&$-0.65_{-0.05}^{+0.05}$ & $2.36_{-0.42}^{+0.40}$ 	& $49.4$  & $2.26_{-0.32}^{+0.32}$&$0.02_{-0.04}^{+0.05}$ & $2.48_{-0.44}^{+0.36}$   &$85.2$   & $2.26_{-0.34}^{+0.39}$	&$0.71_{-0.05}^{+0.05}$ & $2.23_{-0.70}^{+0.54}$   & $137.6$	 \\
0150498701	& $2.72_{-0.30}^{+0.35}$	&$-1.46_{-0.06}^{+0.06}$ & $2.49_{-0.24}^{+0.28}$ 	& $16.6$  & $3.28_{-0.38}^{+0.47}$& $-0.98_{-0.04}^{+0.05}$ & $3.04_{-0.35}^{+0.43}$  &$31.6$   & $3.49_{-0.39}^{+0.48}$	& $-0.46_{-0.04}^{+0.04}$ & $3.06_{-0.38}^{+0.47}$ & $40.7$     \\
0158971201	&$3.29_{-0.26}^{+0.29}$	&$-1.32_{-0.08}^{+0.07}$& $2.19_{-0.17}^{+0.18}$ 	& $8.5$    & $3.60_{-0.27}^{+0.30}$& $-0.92_{-0.09}^{+0.08}$ & $2.15_{-0.17}^{+0.19}$ &$8.5$     & $3.90_{-0.27}^{+0.32}$	& $-0.56_{-0.08}^{+0.07}$ & $2.16_{-0.18}^{+0.20}$	& $9.3$       \\
0158971301	&$2.97_{-0.39}^{+0.45}$	&$-1.54_{-0.04}^{+0.04}$	& $3.16_{-0.36}^{+0.43}$ 	& $36.6$  & $3.43_{-0.45}^{+0.55}$& $-0.99_{-0.04}^{+0.04}$ & $3.42_{-0.46}^{+0.55}$ &$49.9$   & $3.60_{-0.45}^{+0.55}$	& $-0.50_{-0.04}^{+0.04}$ & $3.22_{-0.47}^{+0.58}$	& $52.1$     \\
0302180101	&$1.84_{-0.32}^{+0.38}$	&$-1.71_{-0.07}^{+0.06}$ & $2.35_{-0.31}^{+0.37}$ 	& $20.9$  & $2.25_{-0.32}^{+0.38}$&$-1.19_{-0.06}^{+0.06}$ &$2.33_{-0.29}^{+0.36}$   &$22.6$   & $2.89_{-0.39}^{+0.48}$	&$-0.73_{-0.06}^{+0.05}$ &$2.61_{-0.39}^{+0.48}$   & $27.8$  	  \\
0411080301	&$2.70_{-0.42}^{+0.52}$	&$-1.08_{-0.04}^{+0.04}$ & $3.06_{-0.46}^{+0.57}$  & $66.3$   & $2.92_{-0.43}^{+0.54}$& $-0.64_{-0.03}^{+0.03}$ & $2.91_{-0.46}^{+0.55}$ &$68.2$  & $3.24_{-0.44}^{+0.54}$	& $-0.30_{-0.03}^{+0.04}$ & $2.88_{-0.46}^{+0.57}$ & $67.2$  	  \\
0502030101	&$2.30_{-0.26}^{+0.32}$	&$-1.60_{-0.19}^{+0.14}$& $1.91_{-0.18}^{+0.24}$  	& $6.4$    & $2.57_{-0.30}^{+0.35}$&$-1.12_{-0.12}^{+0.09}$& $1.98_{-0.23}^{+0.26}$    &$9.6$    & $2.70_{-0.34}^{+0.40}$	&$-0.49_{-0.09}^{+0.07}$& $1.94_{-0.28}^{+0.23}$   & $15.9$ 	  \\
\enddata 
\tablecomments{$t_{\rm v}$ is the shortest variability time scale (in units of minutes) derived from the PSD analysis.}
\end{deluxetable*}

\begin{figure*}
\vspace{2.2mm} 
\figurenum{8}
\centering
\includegraphics[width=0.48\textwidth] {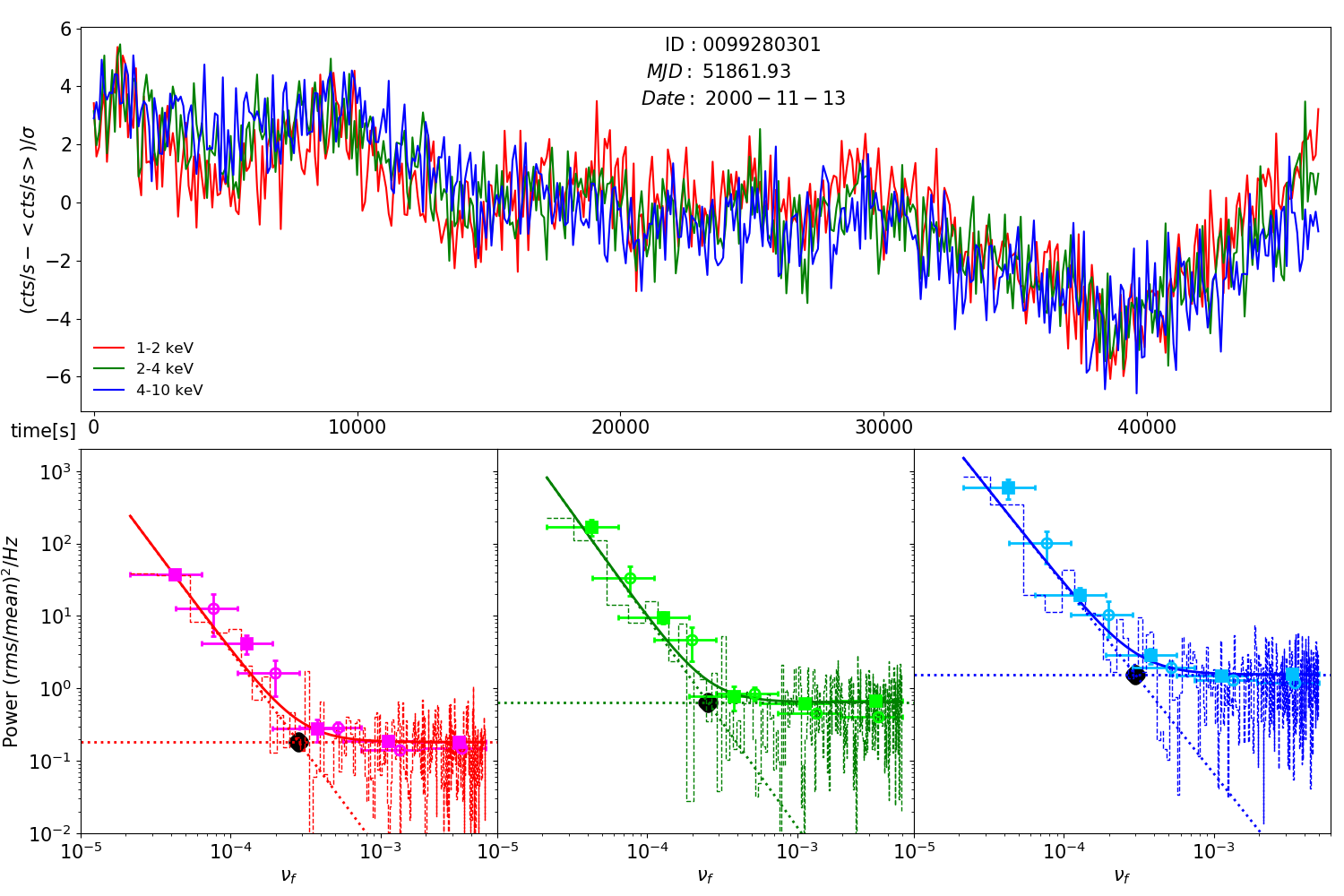} 
\includegraphics[width=0.48\textwidth] {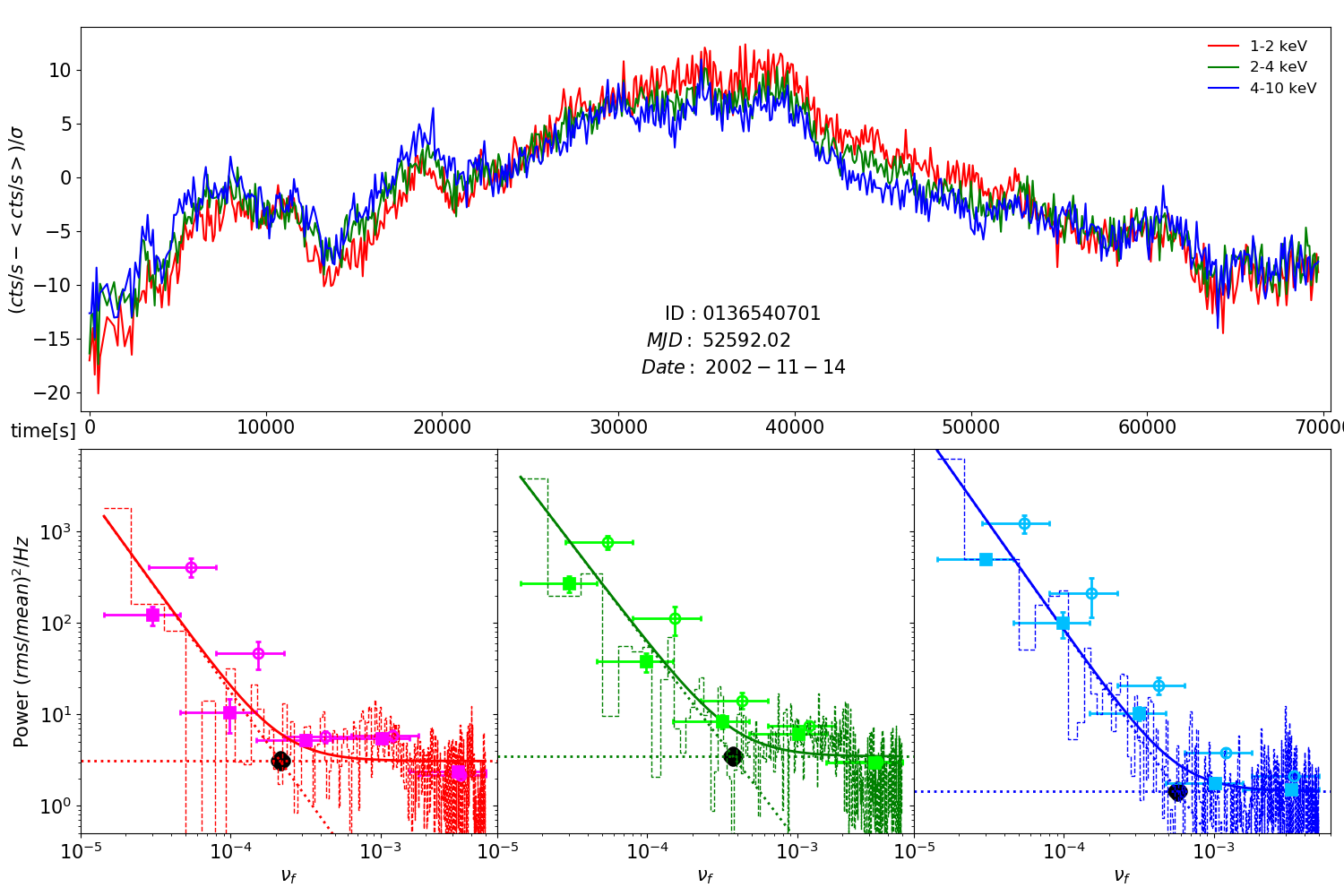} 
\includegraphics[width=0.48\textwidth] {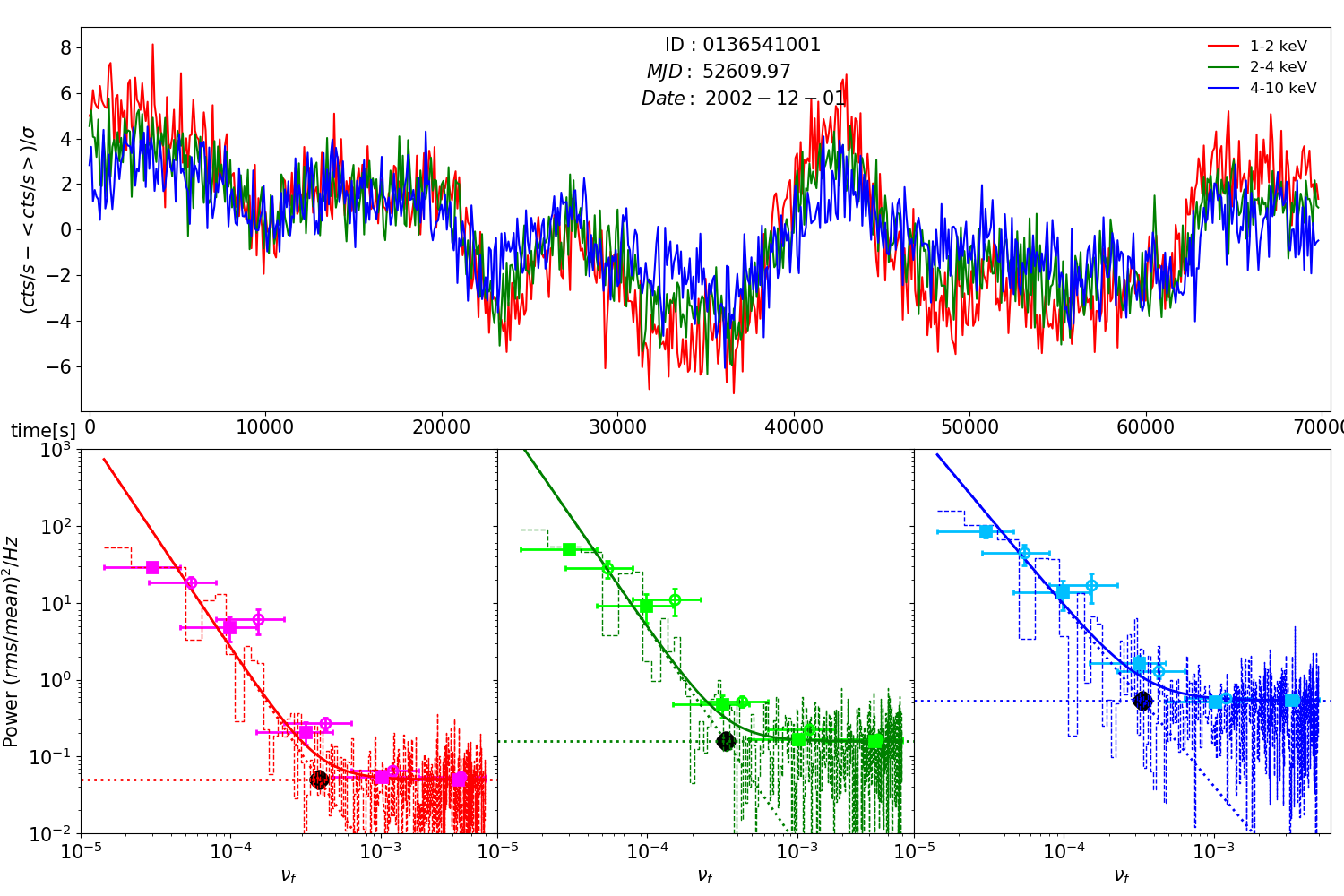} 
\includegraphics[width=0.48\textwidth] {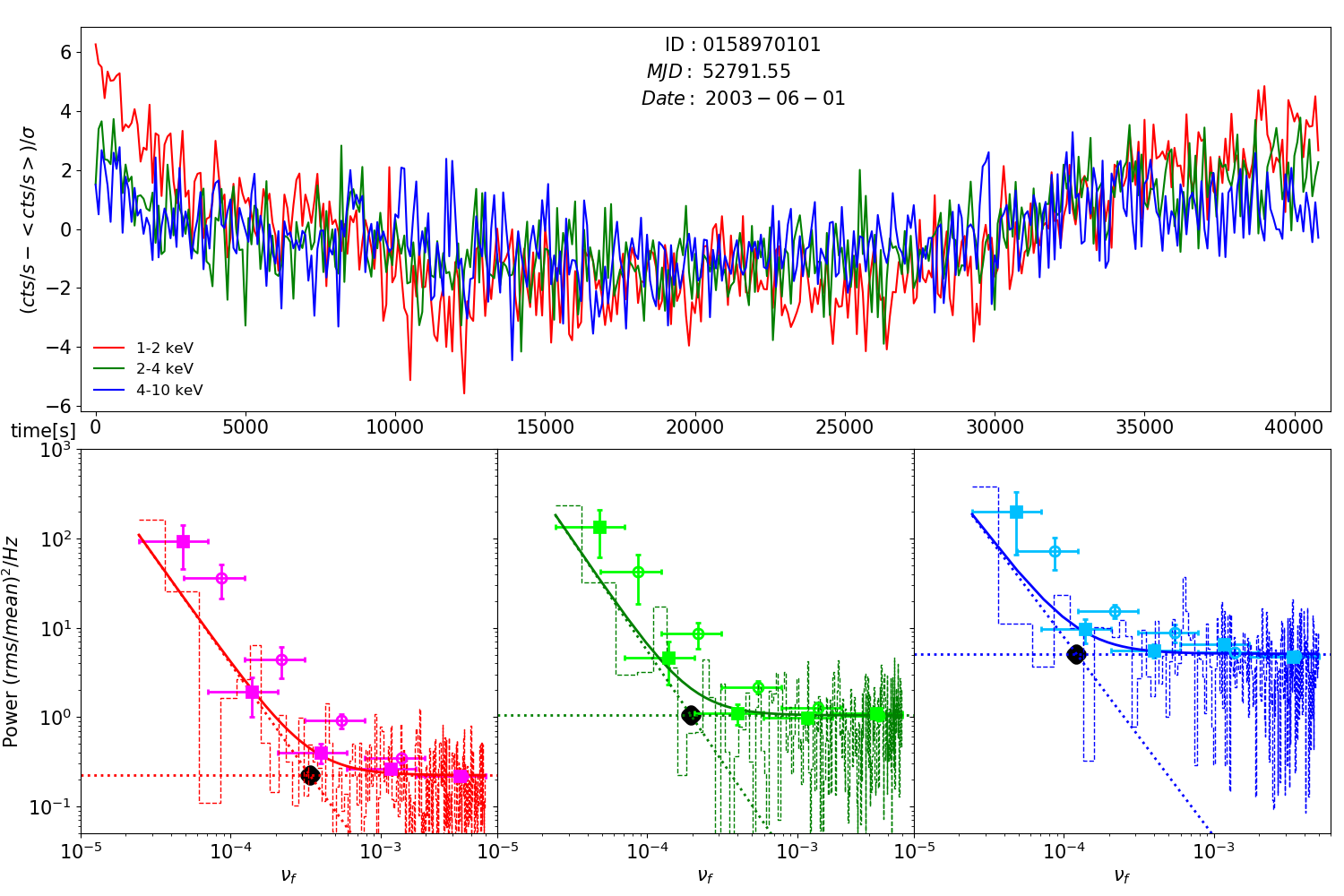} 
\caption{XMM-Newton EPIC-pn LCs and estimated power spectra of Mrk 421 obtained from 10 observations. 
In each panel, the upper part displays the LCs, while the lower part shows the corresponding PSDs.
The RMS-squared normalization is applied such that the periodogram will be calculated in units consistent with the PSD.
The solid lines represent the PSDs estimated using the Bayesian method, 
whereas the solid squares and empty circles correspond to two logarithmically rebinned versions of the PSDs.
The black filled circle marks the frequency threshold above which the variability is dominated by Poisson noise.
The dotted horizontal line marks the Poisson noise level for each band. 
\label{figs:lcs}
}
\vspace{2.2mm}
\end{figure*}

\begin{figure*}
\vspace{2.2mm} 
\figurenum{8}
\centering
\includegraphics[width=0.48\textwidth] {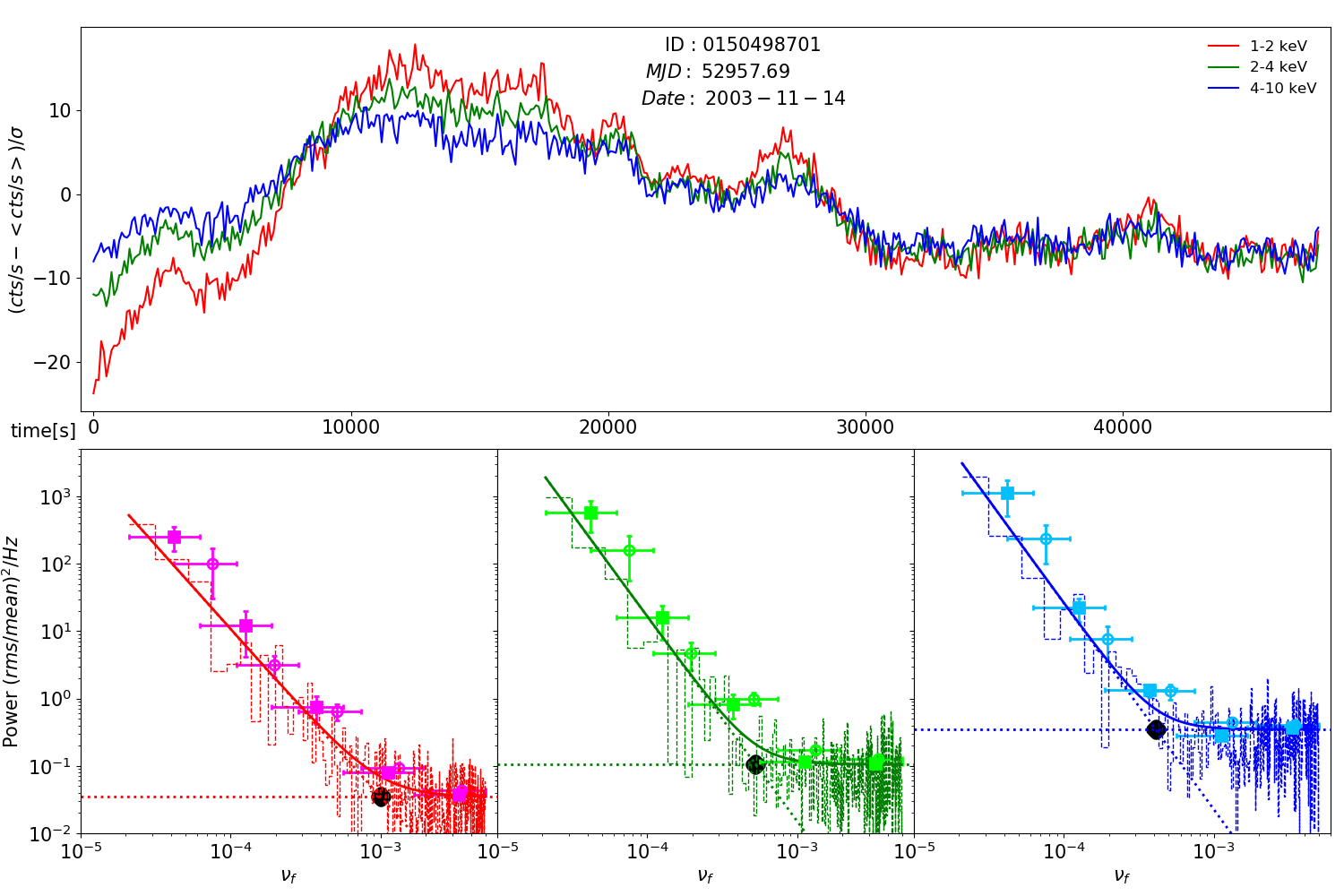} 
\includegraphics[width=0.48\textwidth] {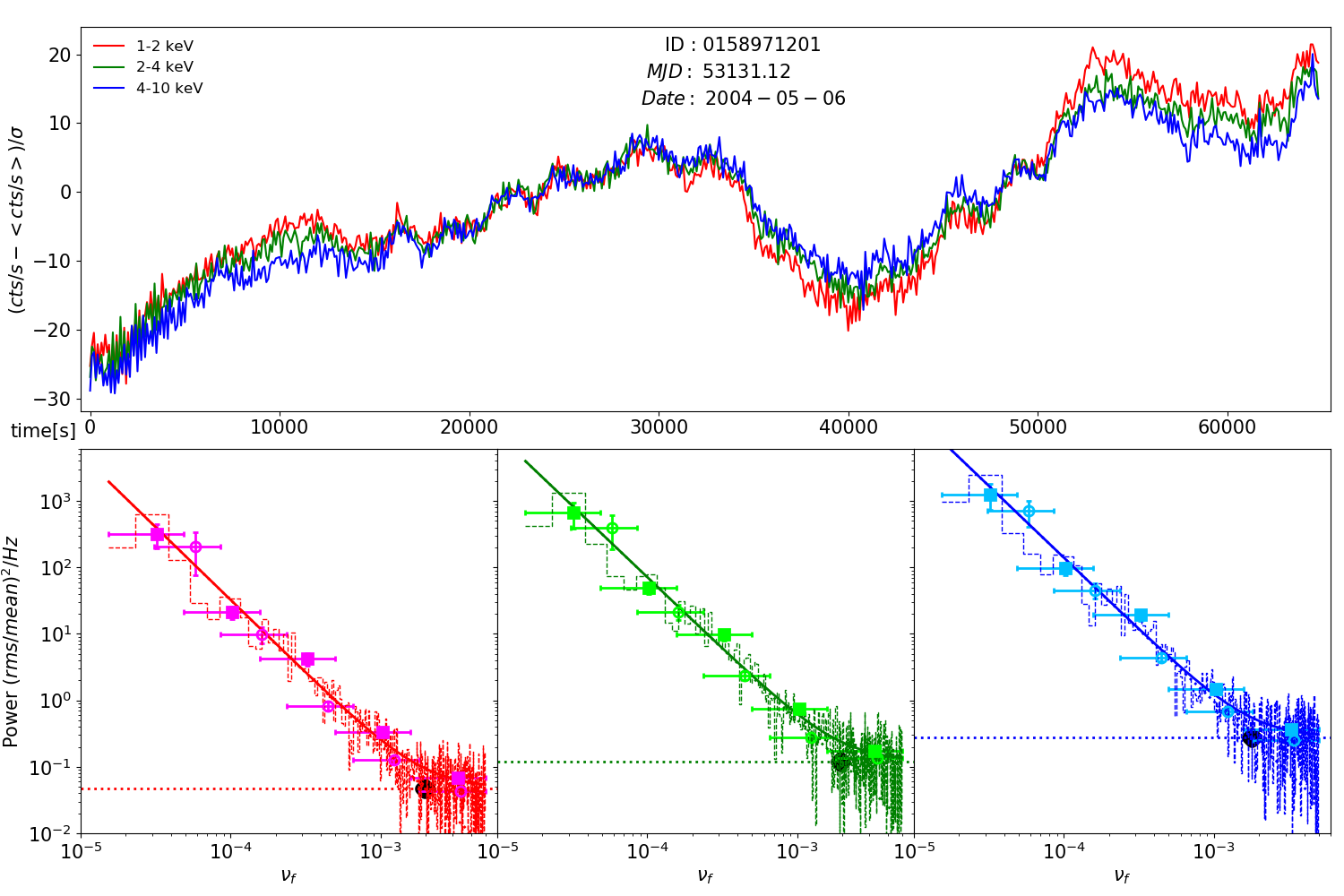} 
\includegraphics[width=0.48\textwidth] {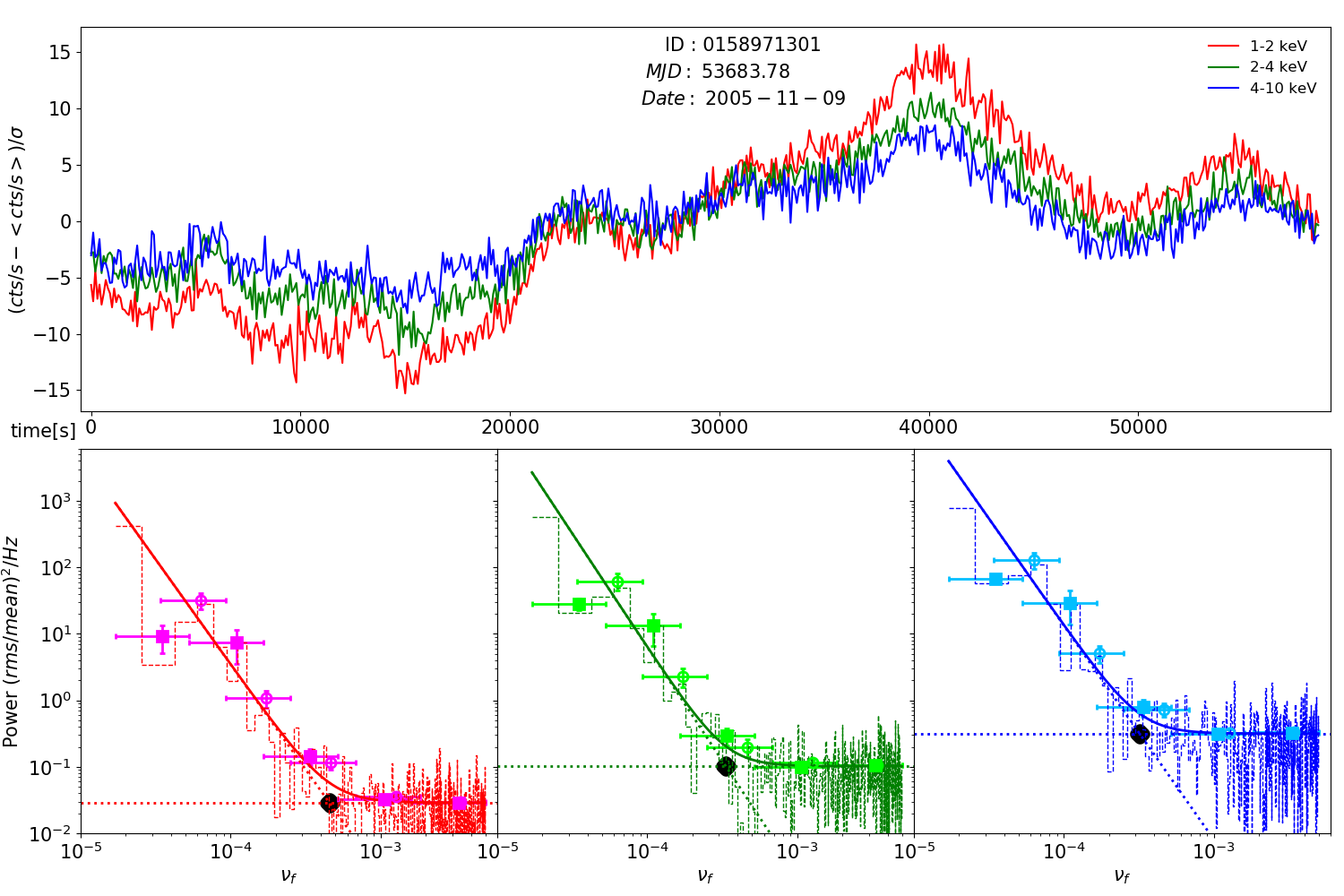} 
\includegraphics[width=0.48\textwidth] {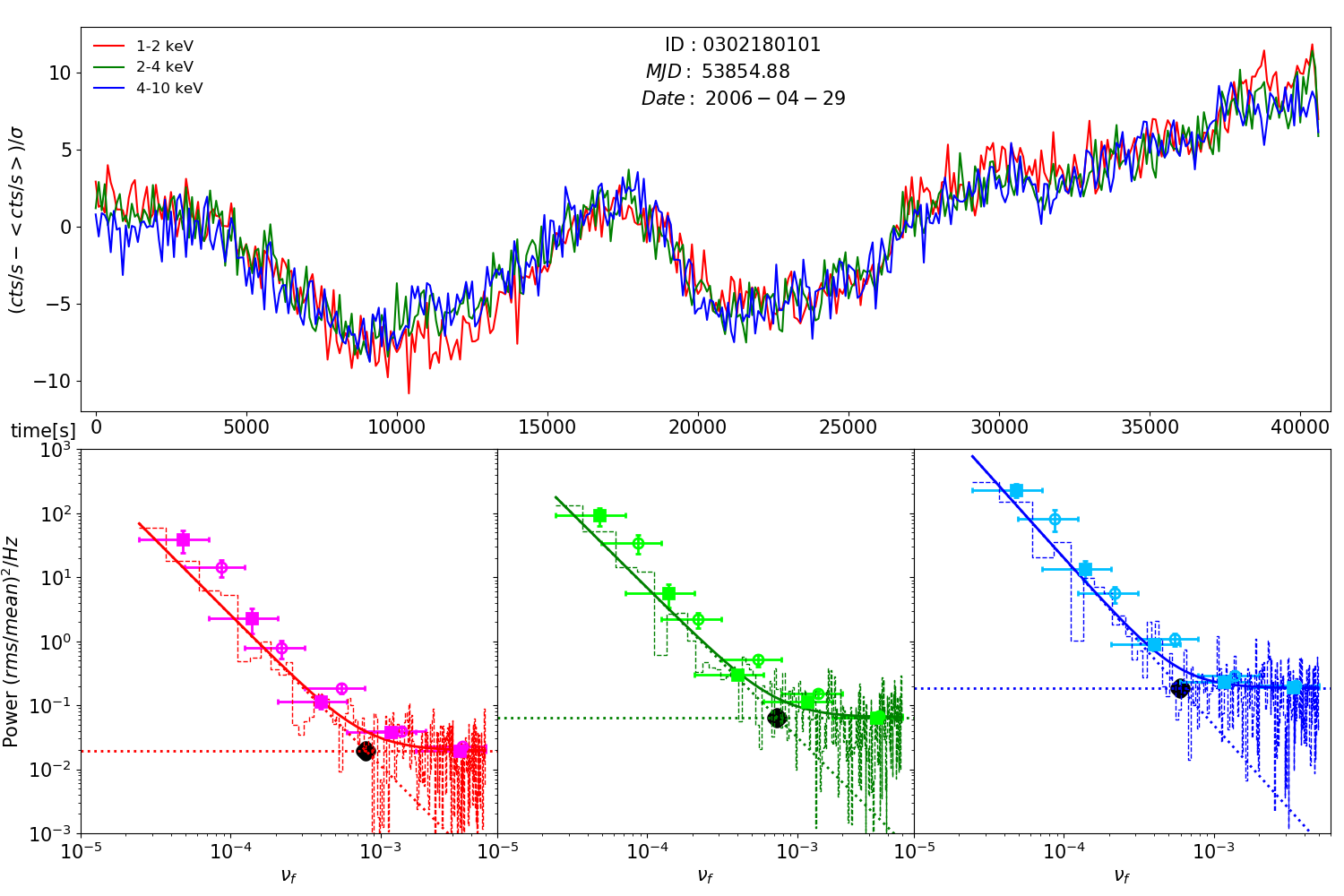} 
\includegraphics[width=0.48\textwidth] {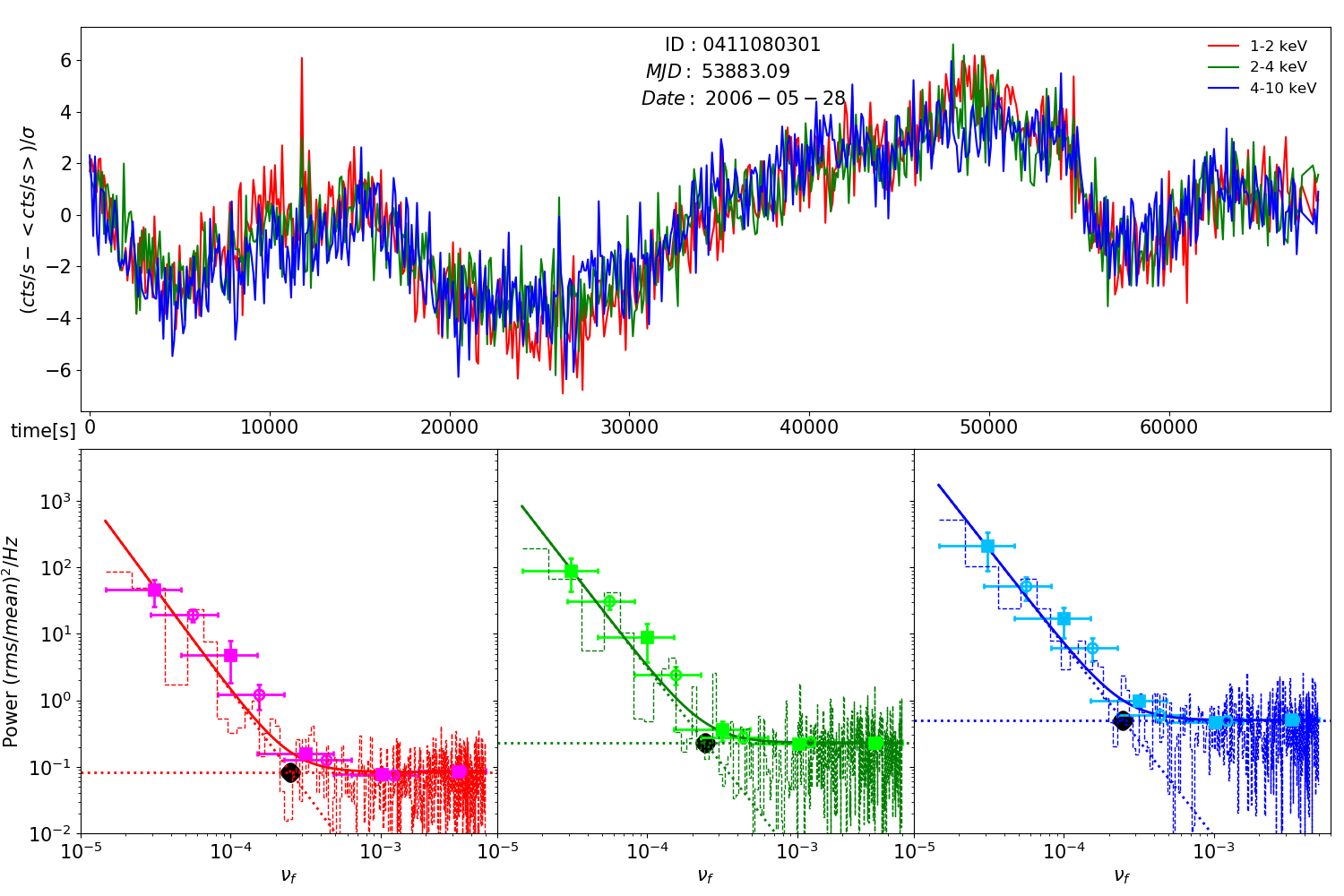} 
\includegraphics[width=0.48\textwidth] {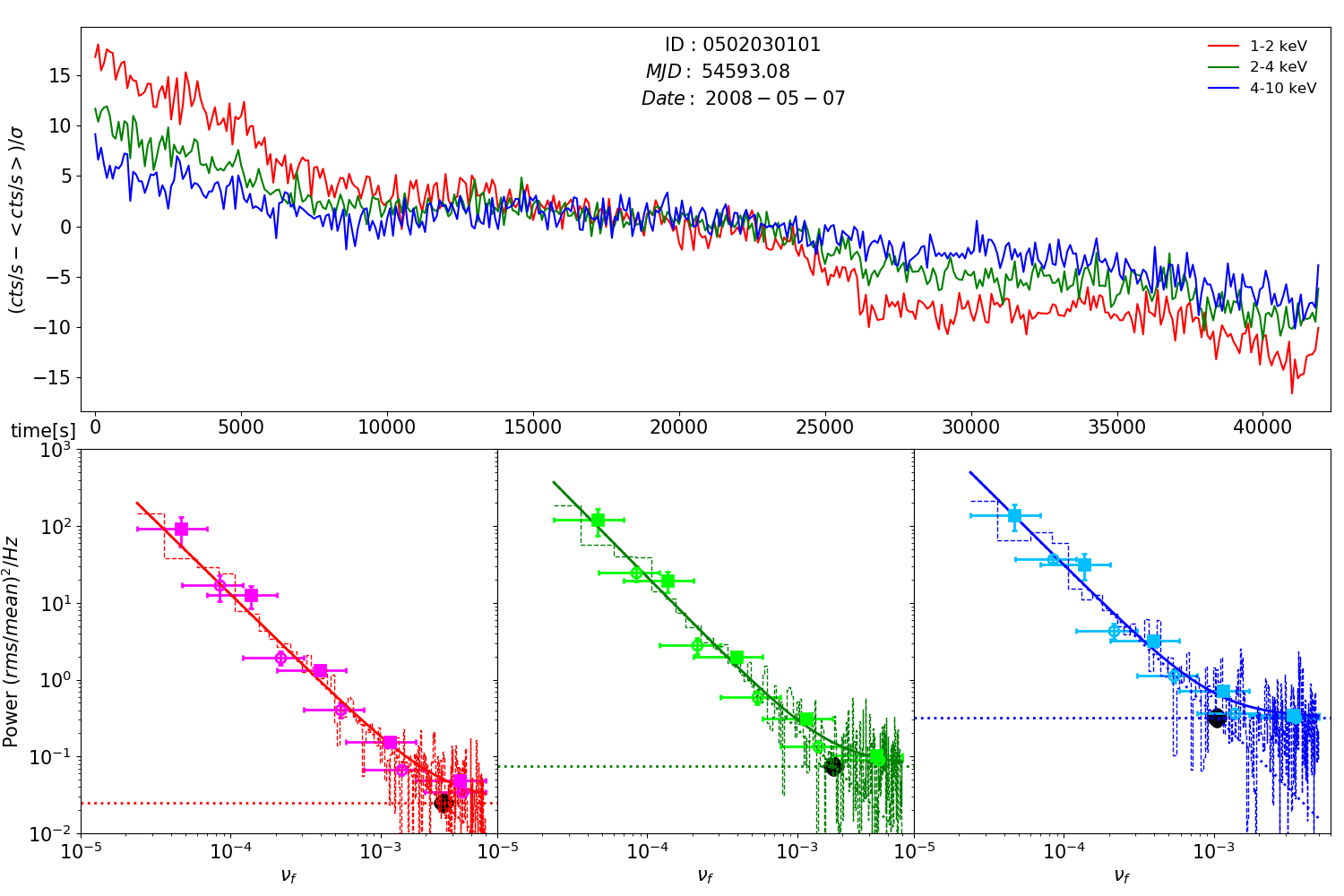}
\caption{Continued.
}
\vspace{2.2mm}
\end{figure*}

\section{Cross-correlation centroid distributions}\label{append:cccd}
\begin{figure*}
\vspace{2.2mm} 
\figurenum{9}
\centering
\includegraphics[width=0.3\textwidth] {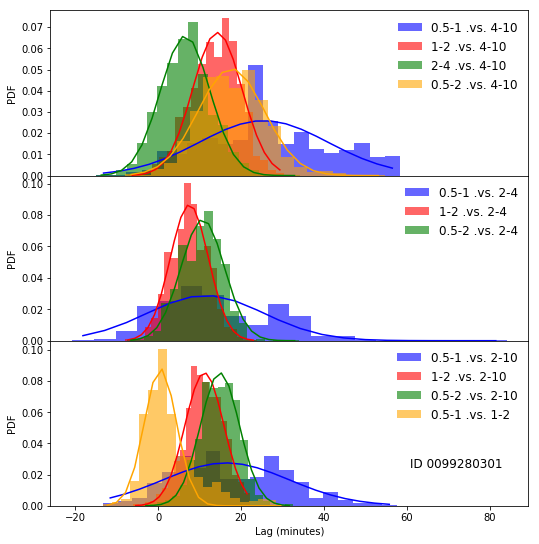} 
\includegraphics[width=0.3\textwidth] {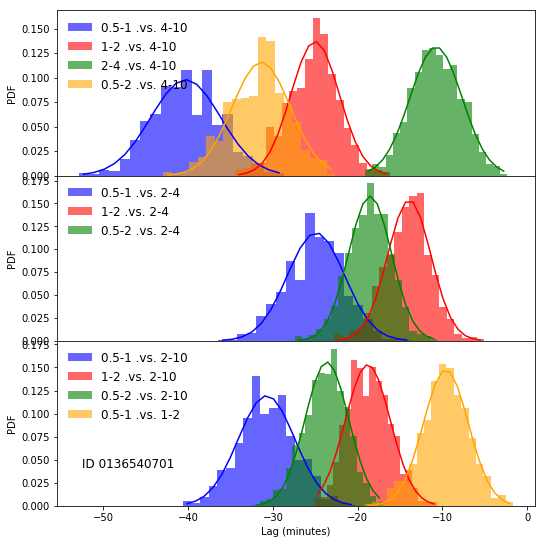} 
\includegraphics[width=0.3\textwidth] {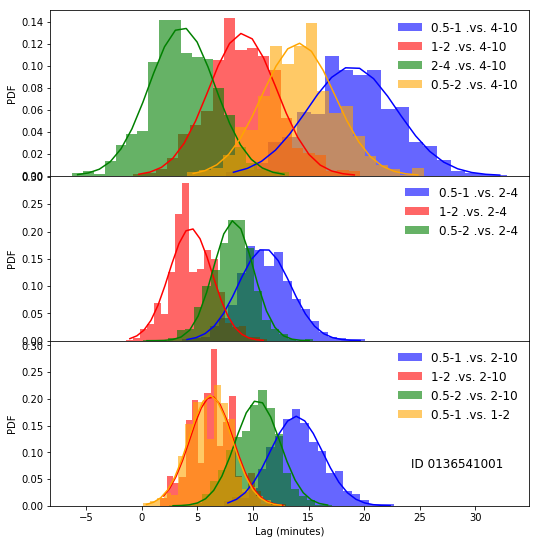} 
\includegraphics[width=0.3\textwidth] {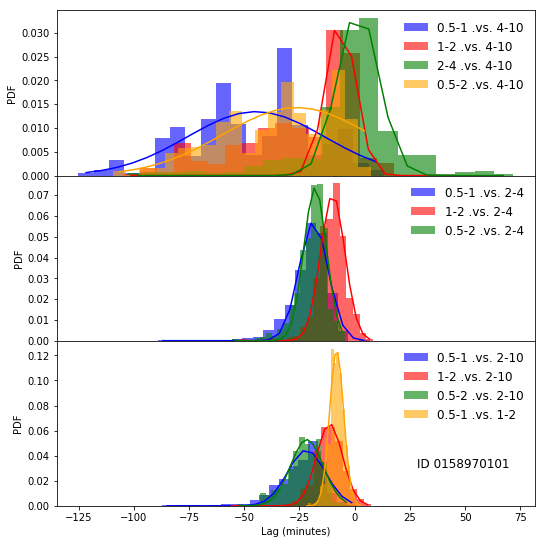} 
\includegraphics[width=0.3\textwidth] {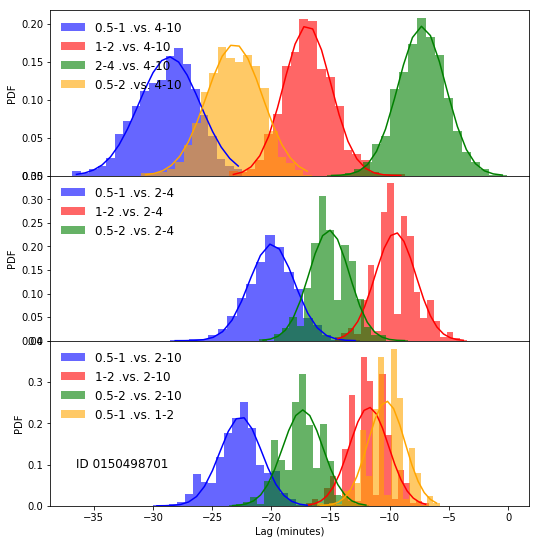} 
\includegraphics[width=0.3\textwidth] {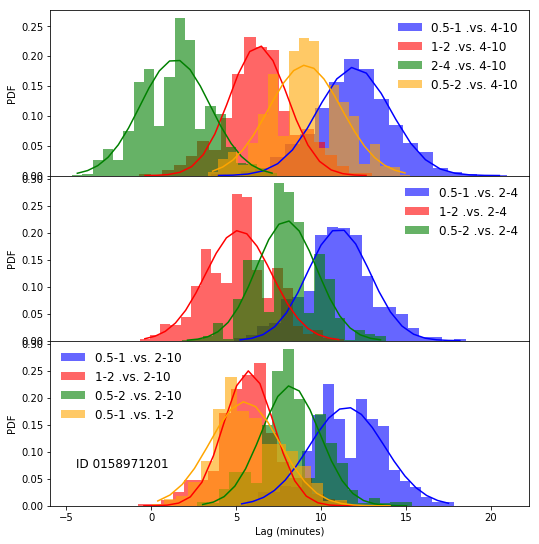} 
\includegraphics[width=0.3\textwidth] {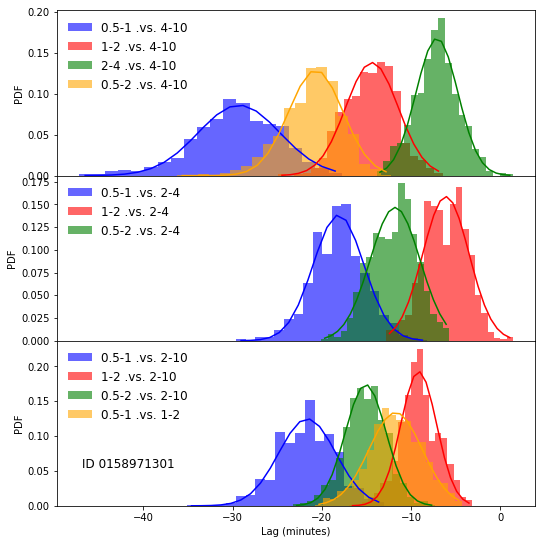} 
\includegraphics[width=0.3\textwidth] {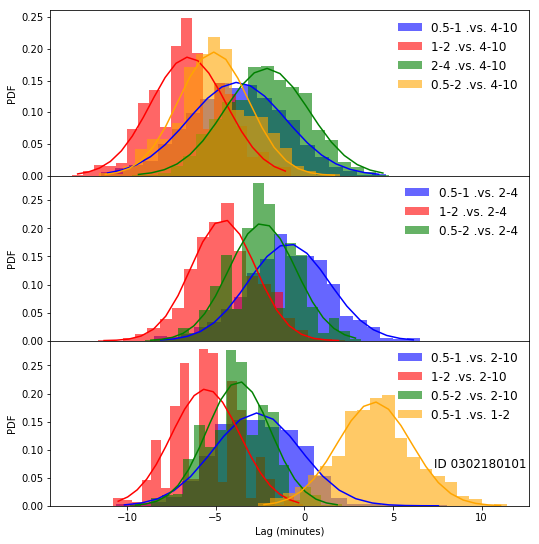} 
\includegraphics[width=0.3\textwidth] {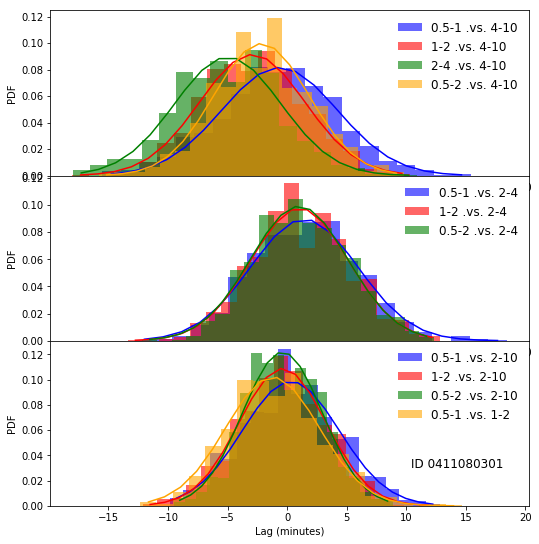} 
\includegraphics[width=0.3\textwidth] {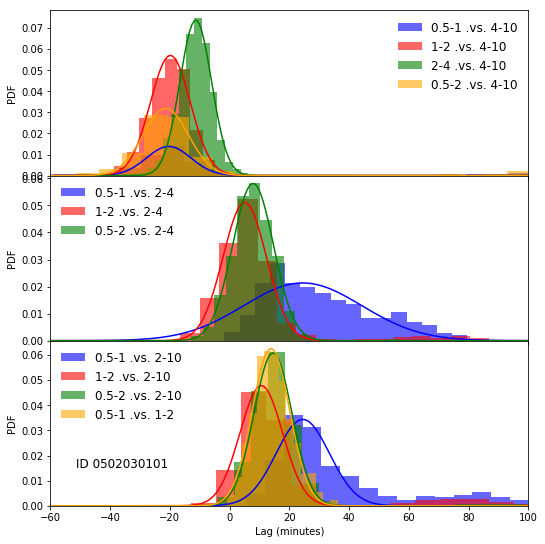}			
\caption{
Cross-correlation centroid distributions from FR/RSS simulations. A Gaussian fit was used to derive the lag and its error.\label{figs:cccd}
}
\vspace{2.2mm}
\end{figure*}

Fig. \ref{figs:cccd} show the CCCDs from FR/RSS simulations for the selected 10 observations with durations of more than 40 ks.

\section{Fourier Transform and Time Lag}\label{append:dft}

Below we provide a brief description of the DFT and the calculation of time lag in the Fourier frequency space.

Consider a light curve denoted as $x(t)$, consisting of $N$ equidistant observations: $\left[t_{\rm j}=j\Delta t, x(t_{\rm j})\right]$ for $j = 0, 1, ..., N-1$,
with a sampling interval $\Delta t$. The DFT of the data set is defined as follows:
\begin{equation}
\tilde{\mathcal{X}}(\omega_{\rm k})=\sum^{N-1}_{j=0}x(t_{\rm j})e^{-i\omega_{\rm k} t_{\rm j}}=\sum^{N-1}_{j=0}x(t_{\rm j})e^{-i2\pi k j/N},
\end{equation}
at each Fourier frequency $\omega_{\rm k}=2\pi\nu_{\rm k}=2\pi k/(N\Delta t)$ with $k=0, 1, 2, ..., N-1$.
Here, $\omega_{\rm k}$ represents the angular frequency in radians per second, and $\nu_{\rm k}$ is in cycles per second (or Hertz).
The minimum frequency is the inverse of the observation duration, i.e., $\nu_{\rm min}=1/N\Delta t$.
The maximum frequency is the Nyquist frequency $\nu_{\rm Ny}=1/2\Delta t$, based on the sampling theorem.
For even $N$, we have $k_{\rm max}=N/2$, while for odd $N$, $k_{\rm max}=(N-1)/2$.

The Fourier time lag associated with variations in the soft-energy and hard-energy channels, 
$\epsilon_{\rm s}$ and $\epsilon_{\rm h}$, is evaluated using the following relationship: 
\begin{equation}
\tau(\omega_{\rm k}) = \frac{\delta\phi(\omega_{\rm k})}{\omega_{\rm k}} = \frac{ Arg\left[ \tilde{\mathcal{S}}(\omega_{\rm k})\tilde{\mathcal{H}}^*(\omega_{\rm k}) \right] }{\omega_{\rm k}},
\label{lageq}
\end{equation}
where $\delta \phi(\omega_{\rm k})$ denotes the phase lag for a given Fourier frequency $\omega_{\rm k}$, 
$\tilde{\mathcal{S}}(\omega_{\rm k})$ and $\tilde{\mathcal{H}}(\omega_{\rm k})$ represent the Fourier transforms of the two light curves $x_{\rm s}(t_{\rm j})$ and $x_{\rm h}(t_{\rm j})$,  
and the asterisk denotes the complex conjugate.
In the given context, a positive value signifies a hard lag, indicating that the hard channel signal lags behind the soft signal. 
Conversely, a negative value indicates a soft lag, with the soft channel signal lagging behind the hard signal.

\section{Numerical and Analytical solution of theoretical Time Lag}\label{append:solution} 

\subsection{Numerical approach} 



Generally, the non-thermal, highly variable X-ray emission from blazars is attributed to the synchrotron emission of high-energy electrons in the jet.
To directly compare with the observed LCs accumulated over certain energy intervals, 
the theoretical LCs integrated over corresponding range $[\epsilon_l, \epsilon_u]$ are calculated through 
 \begin{equation}
 x(t) = \int_{\epsilon_l}^{\epsilon_u} F_{\epsilon}^{\rm syn}(t) d\epsilon,
 \end{equation}
where $F_{\epsilon}^{\rm syn}(t)$ is the synchrotron spectrum at the time $t$ for a given energy $\epsilon m_ec^2$.

Since any variability on a time scale shorter than the light-crossing time of the system with a size $R'$ should not be detectable in principle.
To avoid causality violations, the light crossing effect should be considered in the calculations of the theoretical LCs.
The observed synchrotron spectrum at the time $t$ is calculated by \citep{Zacharias2013ApJ,Finke2014ApJ}
\begin{eqnarray}\label{eqs:lcs}
F_{\epsilon}^{\rm syn}(t) &=& \epsilon' F_0 \int_0^{t'_{\rm lc}}dt' G(t'_{\rm lc},t')\\\nonumber
&\times&\int_1^\infty d\gamma' N'_{\rm e}(\gamma', \frac{t\delta_{\rm D}}{1+z}-t') 
R_{\rm cs}\left(\frac{\epsilon'}{\epsilon'_{\rm c}}\right),\nonumber
\end{eqnarray}
where $\epsilon$ and $\epsilon'$ are related through $\epsilon'=\epsilon(1+z)/\delta_{\rm D}$, $F_0\equiv \frac{\sqrt{3}B' e^3 \delta_D^4}{4\pi hd_L^2}$ with $d_{\rm L}$ 
denoting the luminosity distance of the source at the redshift $z$, $N'_{\rm e}(\gamma', t')$ is the time-dependent solution of Eq. \ref{acc-transport1},
and $G(t'_{\rm lc},t')$ is the geometrical weight function introduced by considering light travel time effects, given by
\begin{equation}
G(t'_{\rm lc},t')=\frac{6}{t'_{\rm lc}}\left[\frac{t'}{t'_{\rm lc}} - \left(\frac{t'}{t'_{\rm lc}}\right)^2 \right],
\end{equation}
where $t'_{\rm lc}\equiv 2R'/c$ denotes the light-travel time across the entire blob.

Here, the function $R_{\rm cs}(\epsilon'/\epsilon'_{\rm c})$ represents the monochromatic emission 
power averaged over a population of electrons with randomly distributed pitch angles \citep{Crusius1986A&A},
and $\epsilon_{\rm c}'=\frac{3}{2}\frac{B'}{B_{\rm cr}}\gamma'^2$ is the dimensionless characteristic synchrotron energy of electrons with the Lorentz factor $\gamma'$.
To minimize computational costs, an efficient approximation provided by \cite{Finke2008} is adopted to calculate the function.

Then, the X-ray time lag associated with the simulated LCs in the soft- and hard-energy channels is evaluated using the standard CCF and cross-spectral methods.

Since the additional Fourier time lag introduced by the light-crossing time effect is independent of the observing energy (see Appendix \ref{appdex:lag}).
The frequency-dependent lag can be calculated by using the standard DFT method to the simulated LCs given by
\begin{eqnarray}\label{eqs:lcs1}
F_{\epsilon}^{\rm syn}(t) &=&\epsilon' F_0 \int_1^\infty d\gamma' N'_{\rm e}(\gamma', \frac{t\delta_{\rm D}}{1+z}) 
R_{\rm cs}\left(\frac{\epsilon'}{\epsilon'_{\rm c}}\right),
\end{eqnarray}
where the light crossing effect is not taken into account for simplicity.

\subsection{Analytical approach}\label{appdex:lag}

Considering that the electron cooling is dominated by synchrotron radiation, and the acceleration, cooling and escape of the electrons are independent of $t'$, 
 taking the Fourier transform of both sides of the transport equation (\ref{acc-transport1}) yields
\begin{eqnarray}
\frac{i\omega'}{D_0}\tilde{\mathscr{N}}'_{\rm e}(\gamma',\omega')&=& \frac{\partial}{\partial \gamma'} \left\{\gamma'^4  \frac{\partial }{\partial \gamma'} \left[\frac{\tilde{\mathscr{N}}'_{\rm e}(\gamma',\omega')}{\gamma'^2}\right] \right\} \\
&-& \frac{\partial}{\partial \gamma'}\left[\left(a\gamma'-b\gamma'^2\right) \tilde{\mathscr{N}}'_{\rm e}(\gamma',\omega') \right] \nonumber\\
&-& \frac{\tilde{\mathscr{N}}'_{\rm e}(\gamma',\omega')}{t'_{\rm esc}D_0} + \frac{\tilde{f}'_{\rm e}(\omega')q_{\rm e}'(\gamma')}{D_0}\nonumber
\label{transport_ft}
\end{eqnarray}
where $a\equiv A_0^{\rm sh}/D_0$, $b\equiv B_0/D_0$ and $B_0\equiv\sigma_{\rm T}B'^2/(6\pi m_{\rm e}c)$ with the strength of the magnetic field $B'$, electron rest mass $m_{\rm e}$, light speed $c$ and Thomson cross-section $\sigma_{\rm T}$, and $\tilde{f}'_{\rm e}(\omega')$ is the Fourier transform of the time-dependent injection profile $f'_{\rm e}(t')$.

Following the standard method in \cite{Schlickeiser1984}, one can obtain the exact solution for the Fourier transform of the electron distribution
\begin{eqnarray}\label{eqs:ntime}
\tilde{\mathscr{N}}'_{\rm G}(\omega',\gamma')=\frac{q_{\rm 0}'f'_{\rm e}(\omega')}{bD_0\gamma'^{2}_{\rm 0}}\left(\frac{\gamma'}{\gamma'_{\rm 0}}\right)^{a/2}e^{-b(\gamma'-\gamma'_{\rm 0})/2}\nonumber\\
\frac{\Gamma(\mu-\kappa+1/2)}{\Gamma(1+2\mu)}M_{\kappa,\mu}(x_{\rm min})W_{\kappa,\mu}(x_{\rm max}),
\end{eqnarray}
for a mono-energetic electrons with Lorentz factor $\gamma_0'$, i.e., $Q'_{\rm e}(\gamma',t')=f'_{\rm e}(t')q_0'\delta(\gamma'-\gamma_0')$.
Here,  the subscript `G'  is used to remind readers that it actually represents the Green function of Eq. \ref{transport_ft}, $\Gamma(\cdot)$ denotes the Gamma function, 
$M_{\kappa,\mu}(\cdot)$ and $W_{\kappa,\mu}(\cdot)$ are the Whittaker's functions with the arguments $x_{\rm min}$ and $x_{\rm max}$ respectively given by
\begin{eqnarray}
x_{\rm min}=\min(b\gamma', b\gamma'_{\rm 0}),~~
x_{\rm max}=\max(b\gamma',b\gamma'_{\rm 0}),
\end{eqnarray}
and the subscript indices $\kappa$ and $\mu$ respectively defined as
\begin{equation}
\kappa =2+\frac{a}{2}, ~~\mu =\sqrt{\left(\frac{a+3}{2}\right)^2 + \frac{1}{D_0t'_{\rm esc}} + i\frac{\omega'}{D_0}}.
\end{equation}

On the other hand, the exact steady-state solution resulting from monogenetic continuous injection can be straightforwardly obtained by simply removing the frequency-dependent factor in Eq. \ref{eqs:ntime}, and is given by
\begin{eqnarray}\label{eqs:steady}
N'_e(\gamma')&=&\frac{q_{\rm 0}'}{bD_0\gamma'^{2}_{\rm 0}}\left(\frac{\gamma'}{\gamma'_{\rm 0}}\right)^{a/2}e^{-b(\gamma'-\gamma'_{\rm 0})/2}\\
&\times&\frac{\Gamma(\mu-\kappa+1/2)}{\Gamma(1+2\mu)}M_{\kappa,\mu}(x_{\rm min})W_{\kappa,\mu}(x_{\rm max}),\nonumber
\end{eqnarray}
and
\begin{equation}
\kappa =2+\frac{a}{2}, ~~\mu =\sqrt{\left(\frac{a+3}{2}\right)^2 + \frac{1}{D_0t'_{\rm esc}} }.
\end{equation}

Given the Fourier transformed electron distribution, the Fourier time lag associated with variations in two energy channels is calculated via Eq.\ref{lageq} and the Fourier transform of synchrotron spectrum given by 
\begin{eqnarray}\label{eqs:freq_spec}
\tilde{\mathscr{F}}_{\epsilon}^{\rm syn}(\omega)&=&\epsilon' F_{\rm 0}\tilde{\mathcal{G}}(t'_{lc},\omega') \frac{1+z}{\delta_{\rm D}} \nonumber\\
&\times&\int_1^\infty d\gamma' \tilde{\mathscr{N}}'_{\rm e}(\gamma', \omega') 
R_{\rm cs}\left(\frac{\epsilon'}{\epsilon'_{\rm c}}\right).
\end{eqnarray}
where $\omega'=\omega(1+z)/\delta_{\rm D}$, and the geometrical weight function in the Fourier domain $\tilde{\mathcal{G}}(t'_{\rm lc},\omega')$ is  
given by
\begin{eqnarray}
\tilde{\mathcal{G}}(t'_{\rm lc},\omega')&=&\int_0^{t'_{lc}} dt'e^{-i\omega't'}G(t'_{\rm lc},t')\\
&=&6\cdot\frac{e^{-i\omega't'_{\rm lc}}(i\omega't'_{\rm lc}+2)+i\omega't'_{\rm lc}-2}{i\omega't'_{\rm lc}\cdot(i\omega't'_{\rm lc})^2}.\nonumber
\end{eqnarray}

The function $\tilde{\mathcal{G}}(t'_{\rm lc},i\omega')$, representing the light-travel time effect, is independent of photon energy. 
Therefore, no observed time lag is related to the light-crossing time.

As indicated by the fact that the PSD of blazars resembles colored noise, it is more reasonable to assume that electrons are injected as a power law in the frequency domain.
However, it can be seen from the time-dependent solution Eq.\ref{eqs:ntime} that the phase lags measured from the argument of the cross-spectrum between two energy channels
are independent of the precise form of $f_{\rm e}'(\omega')$.
Therefore, the form of $f_{\rm e}'(\omega')$ or $f_{\rm e}'(t')$ does not modify the lag-frequency spectra, although it is important to determine the PSD of variability.

For simplicity, we assume an instantaneous electron injection represented by the $\delta$--function: $f_{\rm e}(t')=\delta(t'-t'_0)$, in the present analysis. 
Taking the Fourier transform of $f_{\rm e}(t')$ gives $f'_{\rm e}(\omega')=e^{-i\omega't_0'}$.

\begin{figure*}
\figurenum{10}
\centering
\includegraphics[width=0.8\textwidth]{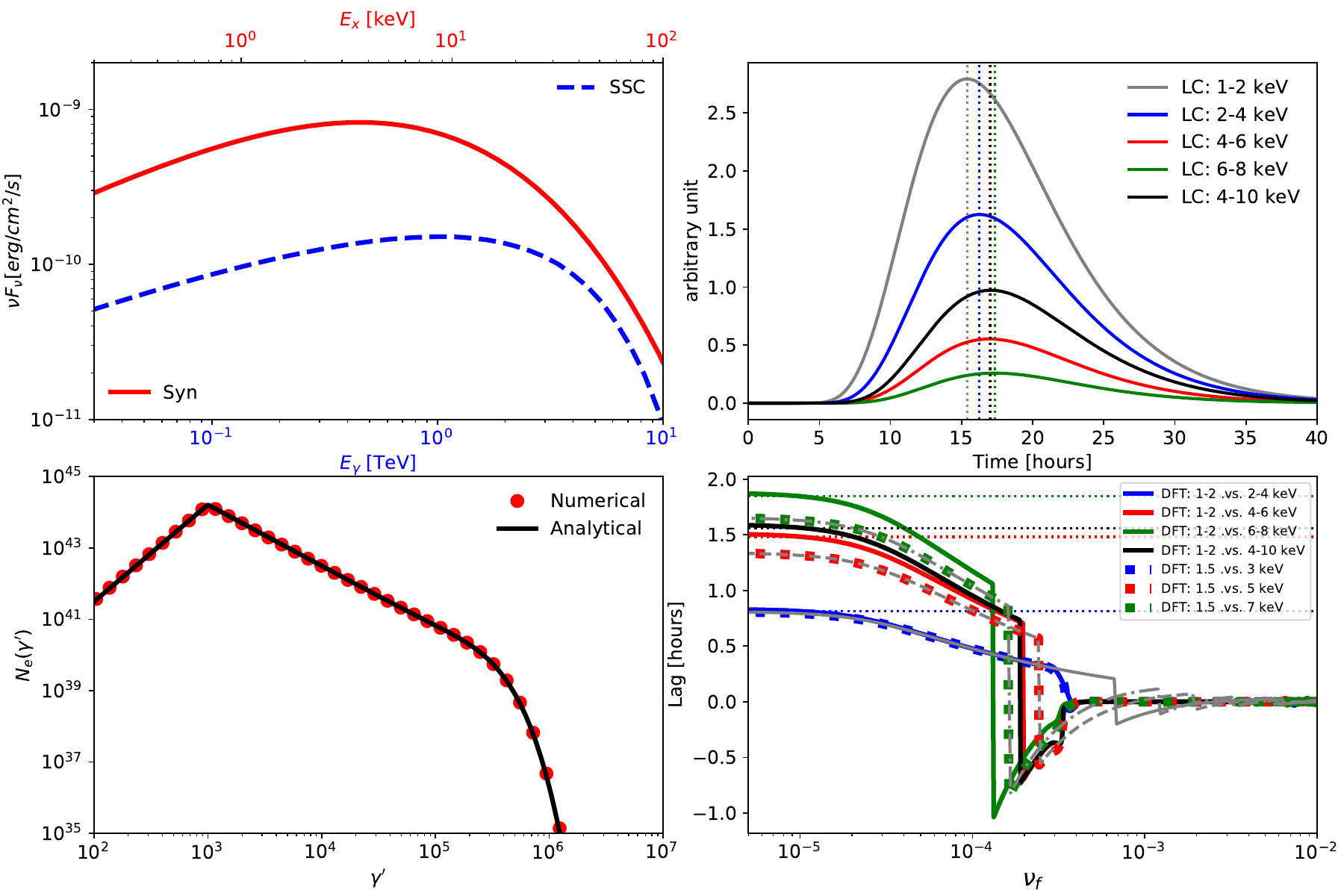}
\caption{Comparison between analytical and numerical solutions.
Upper left: the theoretical synchrotron and SSC spectrum emitted by the equilibrium EED;
Lower left: the analytical and numerical solutions of the emitting EED; 
Upper right: the integrated LCs at five bandpasses; 
Lower right: the Fourier frequency-dependent time lags. 
}\label{fig:steady2}
\end{figure*}

To demonstrate the validity of the numerical approach, comparisons are made with both analytical and numerical results.
In the lower left panel of Fig.\ref{fig:steady2}, we show a comparison of the exact solution for the steady-state EED 
given by Eq.\ref{eqs:steady} with that computed using a bi-directional Runge-Kutta method.
The resulting synchrotron and SSC spectrum are shown in the upper left panel of the figure.
It can be seen that the numerical and analytical solutions are in good agreement. 
The model parameters are selected to produce the typical SED of Mrk 421 during the flaring state 
and are summarized as follows: $B'=7.8\times10^{-2}$ G, $\delta_{\rm D}=45$, $R'=3.3\times10^{15}$ cm, $\nu_{\rm pk}=10^{18}$ Hz, 
$\eta_{\rm esc}=0.67$, $D_0=5.51\times10^{-7} ~s^{-1}$, $q_0'=3.97\times10^{41}~s^{-1}$ and $\gamma'_0=10^3$.
With the given values of $R'$ and $\delta_{\rm D}$, the derived dynamical timescale in the observer's frame $t_{\rm dyn}$ is about $0.7$ hr.

With the parameters set for the steady-state SED, we can calculate the frequency-dependent X-ray time lags 
using the exact solution for the Fourier transformed EED given by Eq.\ref{eqs:ntime}.
For illustrative purposes, we obtain the time delays of the 3, 5, and 7 keV energy channels relative to the 1.5 keV reference energy channel.
The results are shown in the lower right panel of Fig.\ref{fig:steady2}, denoted by the gray solid, dashed, and dash-dotted lines, respectively.

To compare with the exact results, the X-ray LCs at the four mono-energy channels are calculated using the synchrotron 
spectrum given by Eq.(\ref{eqs:lcs}) and the time-dependent EED computed numerically by solving Eq.(\ref{acc-transport1}) 
for an instantaneous injection of mono-energetic electrons.
Then, the frequency-dependent lags are directly calculated by applying the standard DFT to the resulting LCs (see Appendix \ref{append:dft}). 
The frequency-dependent lags between each energy channel and the 1.5 keV reference energy channel are displayed 
in the lower right of Fig.\ref{fig:steady2}, denoted by the thick blue, red, and green dotted lines, respectively. 
Our results indicate that the numerical and analytical Fourier time lag profiles appear in good agreement 
with each other below the frequency $\sim(0.75 t_{\rm lc})^{-1}\simeq 3\times10^{-4}$ Hz. 
It should be noted that Fourier frequencies higher than $t_{\rm lc}^{-1}$ are unphysical in the usual homogeneous model assumed here.

Moreover, we produce the integrated LCs in the five bandpasses: 1--2, 2--4, 4--6, 6--8, and 4--10 keV.
The central energies of the first three bandpasses are 1.5, 3, and 5 keV,  respectively, while both of the central energies of the last two bandpasses are 7 keV.  
The simulated LCs at various energy channels are shown in the upper right panel of Fig.\ref{fig:steady2}.
Similarly, we calculate the time delays of each band with respect to the 1--2 keV reference band, represented by the thick blue, red, green, and black solid, respectively.  
We can see that the difference between the time delays resulting from the integrated LC pairs and 
mono-energy LC pairs increases with the energy difference between the two compared energy channels.  
Although the central energies of both 6--8 and 4--10 keV bandpasses are 7 keV, the time delays of 6--8 and 4--10 keV with respect to 1--2 keV differ 
significantly from the time delay calculated from the mono-energy LC pair: 1.5 and 7 keV.
Thus, the simple estimation for the time lags between two mono-energy channels is not enough to represent the results between two observed LCs.

Using the integrated LCs in various bandpasses, the time lags measured by the CCF centroid are also shown in the lower right panel 
of Fig.\ref{fig:steady2}, denoted by the thin blue, red, green, and black dotted lines, respectively.
The CCF lags between 1--2 and 2--4 keV bands are about 0.8 hr. 
The results indicate that the CCF lags appear in close agreement with the Fourier time lags at the low frequencies, 
where the Fourier lag spectrum weakly depends on the frequency.

\section{Effects of changing in the model parameters}\label{app:est}

\begin{deluxetable*}{cccccccccccccc}
\centerwidetable
\label{tab:para1}
\renewcommand\tabcolsep{6.0pt}
\tablecaption{Input model parameters and derived quantities for the studies of time lags.}
\tablehead{
Model &\colhead{$\delta_{\rm D}$}	&	\colhead{$B'$}      & \colhead{$R'$}  	& \colhead{$\nu_{\rm pk} (E_{\rm pk})$}	  &   \colhead{$q_0'$}  & \colhead{$\gamma_{\rm i,min}'$} & \colhead{$\gamma'_{\rm i,max}$}  & \colhead{$p$} & \colhead{$\eta_{\rm esc}$}  & \colhead{$t'_{\rm acc}$}   & \colhead{$D_0$}  & \colhead{$U_{\rm e}'/U_{\rm B}'$} &$\tau_{\rm c}$\\
	&\colhead{}	 &\colhead{G}	  &\colhead{$10^{15}$cm}	 	&\colhead{$10^{17}$ Hz	(keV)} 	  &\colhead{$\rm s^{-1}$}   &	\colhead{$10^3$}	 &\colhead{$10^5$}	&   & &\colhead{$R'/c$}   &\colhead{$10^{-6}\rm s^{-1}$}    &\colhead{} &\colhead{min}
}  
\decimalcolnumbers
\startdata
		Case A	& 37			 &0.15  		   & $2.5$ 				& $5.5(2.27)$  		  		& $5.10\times10^{45}$& $2.0$			&$5.0$			&$2.23$	& $1.26$ 	   & $2.49$ & $1.20$   &$11.9$	&-16.3 \\
		Case B	& 35			 &  			   & 		 			& $\mathbf{9.0(3.72)}$		& $0.21\times10^{45}$& 				&$\mathbf{1.0}$	&$1.85$	& 		 	   & $1.90$ & $1.58$   &$12.1$	&32.9 \\
		Case C	& 			 &  			   & 		 			& $\mathbf{2.5(1.03)}$	   	& $0.12\times10^{45}$& 				&				&$1.88$	&	 		 & $3.70$ & $0.81$   &$10.3$	&-19.7 \\
		\hline							
		Case A1	& 35			 &			   &					&$5.8(2.40)$	  	  		&$6.16\times10^{45}$ &				&$5.5$			&2.28	&$\mathbf{1.00}$ &$2.36$	&$1.27$	&$12.9$	&1.1   \\
		Case A2	& 38			 &			   &					&$4.0(1.65)$		  		&$2.12\times10^{45}$ &				&$4.5$			&2.11	&$\mathbf{1.69}$ &$2.96$	&$1.01$	&$11.2$	&-12.6 \\
		\hline					
		Case A3	&40			 &			   &					&				  		&$2.85\times10^{45}$&$\mathbf{0.8}$	&				&2.22	&			 &2.59	&1.16	&11.4	&-19.1 \\
		Case A4	&33			 &			   &					&4.5(1.86)			  		&$1.55\times10^{46}$&$\mathbf{10}$	&				&2.26	&			 &2.60	&1.15	&11.8	&-13.5 \\
		\hline					
		Case A5	& 			 &$\mathbf{0.08}$& $\mathbf{4.5}$ 		& 		  		  		& $5.18\times10^{45}$& 				&$6.8$			&		& 		 	   & $3.56$ & $0.47$   &$20.2$	&-43.5 \\
		Case A6	& $\mathbf{52}$&$\mathbf{0.08}$& 		 			& 		  		  		& $1.34\times10^{45}$& 				&$5.8$			&		& 		 	   & $7.59$ & $0.39$   &$34.6$	&-33.1 \\
		\hline\hline				
		Case B1	& 33			 &			   &					&$8.3(3.43)$		  		&$0.38\times10^{45}$ &				&				&1.95	&$\mathbf{1.00}$ &$1.92$	&$1.56$	&$13.0$	&39.4 \\
		Case B2	& 36			 &			   &2.3				&$11.0(4.55)$		  		&$3.73\times10^{43}$ &				&				&1.65	&$\mathbf{1.54}$ &$1.89$	&$1.72$	&$13.2$	&26.4 \\
		\hline					
		Case B3	&36			 &			   &					&9.3(3.85)			  		&$0.16\times10^{45}$&$\mathbf{0.8}$	&				&		&			 &1.89	&1.58	&11.9	&31.8 \\
		Case B4	&33			 &			   &					&8.7(3.60)			  		&$0.35\times10^{45}$&$\mathbf{10}$	&				&		&			 &1.87	&1.60	&11.5	&34.2 \\
		\hline						
		Case B5	& 			 &$\mathbf{0.08}$    & $\mathbf{4.7}$ 	&9.5(3.93) 		  	   	& $0.23\times10^{45}$& 				&				&		&	 		 & $2.52$ & $0.63$   &$18.9$	&89.1 \\
		Case B6	&$\mathbf{50}$ &$\mathbf{0.08}$   &			 	&9.5(3.93) 		  	   	& $5.26\times10^{43}$& 				&				&		&	 		 & $5.77$ & $0.53$   &$33.4$	&71.1 \\
\enddata 
\tablecomments{Column(2-13): same as Tab.\ref{tab:table3}. Column(14): The time lag measured by the CCF centroid.}
\end{deluxetable*}

In this section, we present a detailed discussion about the effects of varying model parameters describing the properties of the turbulence wave and the injected electron spectra, with respect to two baseline models (Case A and B). 
Here, the time lag between two narrow bands: 1--2 and 4--10 keV is used for clarification.

\begin{figure}
\vspace{2.2mm} 
\figurenum{11}
\label{figs:fig11}
\centering
\includegraphics[width=0.45\textwidth] {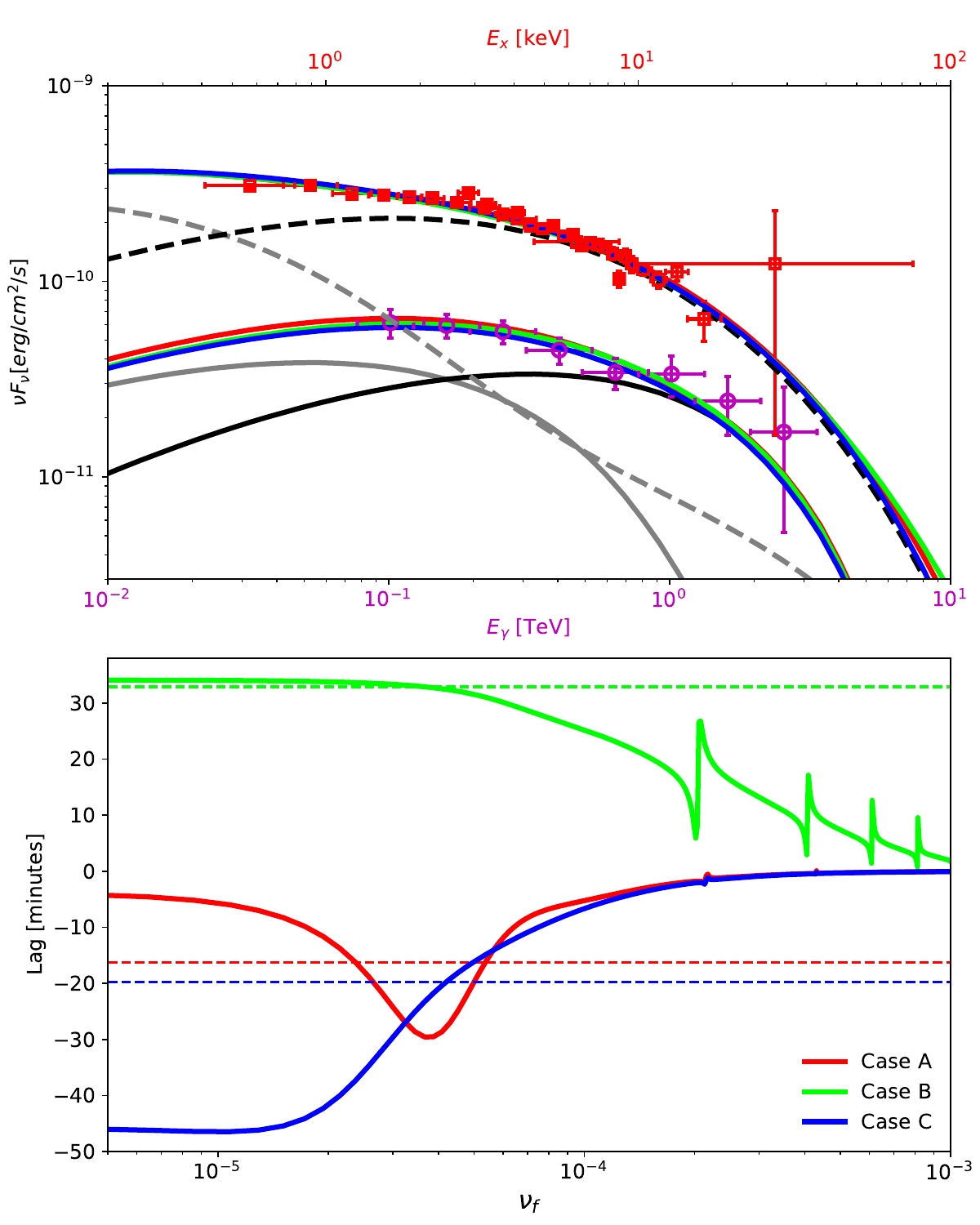} 
\caption{
The theoretical SEDs (upper) and time lags (lower) resulting from variations in $\nu_{\rm pk}$ and $\gamma'_{\rm i,max}$.
In the lower panel, the Fourier time lag profiles for Case A, B and C are denoted by the red, green and blue solid lines, respectively.
The horizontal red, green and blue dashed lines mark the CCF lags for Case A, B and C, respectively.
In the upper panel, the empty red squares and magenta circles represent the simultaneous X-ray and TeV $\gamma$-ray data on 2010 March 22 (MJD 55277), reported in \cite{Aleksic2015}, respectively. 
The gray dashed and solid lines denote the synchrotron and SSC spectrum from the stationary zone, respectively. 
The synchrotron and SSC spectrum from the non-stationary zone for Case A are denoted by the black dashed and solid lines, respectively.
The red, green and blue solid lines are the SEDs from non-stationary zone corresponding to Case A, B and C, 
superimposed with the SED from the (quasi-)stationary zone.
}
\vspace{2.2mm}
\end{figure}

Case A and B are plotted as the red and green solid lines in Fig. \ref{figs:fig11}.
Although it is difficult to distinguish between the two cases based on the produced SED,
both the CCF lags and Fourier time-lag profiles are significantly different in the two cases.
The differences are mainly attributed to the selected values of $\nu_{\rm pk}$ (or $D_0$) and $\gamma'_{\rm i,max}$, 
which are critical in determining whether the lag is detected as either a negative (soft) or a positive (hard) lag.
A hard lag is observed when $\gamma'_{\rm i,max}$ is well below the equilibrium Lorentz factor $\gamma'_{\rm eq}$ 
that is related to $\nu_{\rm pk}$ through the relation $h\nu_{\rm pk}/m_{\rm e}c^2=1.5\epsilon_{\rm B}\gamma'^2_{\rm eq}\delta_{\rm D}/(1+z)$. 
Otherwise, a soft lag is detected.
In the calculations, the values of $\gamma'_{\rm eq}$ are $1.6\times10^{5}$ and $2.2\times10^{5}$ for $h\nu_{\rm pk}\simeq2.3$ and $3.7$ keV, respectively.  
From the lower panel of Fig.\ref{figs:fig11}, it can be observed that the time lags shift from the negative side in Case A (where $\gamma'_{\rm i,max}>\gamma'_{\rm eq}$) 
to the positive side in Case B (where $\gamma'_{\rm i,max}<\gamma'_{\rm eq}$).
In Case A, the time lag measured by the CCF centroid is $\tau_{\rm c, A}=-16.3$ minutes, approximately half of the maximum lags observed in the Fourier domain.
In Case B, the CCF lag is $\tau_{\rm c,B}\simeq32.9$ minutes, approximately equal to the Fourier time lag at low frequencies, 
where the dependence of lag on frequency is very weak.

It should be pointed out that the power-law index of the injection electron spectra also plays an important role in determining the time lags measured between two LCs.
This is due to the fact that the injected electron spectra, which are harder than the equilibrium distribution established by a balance between the various competing processes,
can shelter out the spectral hardening caused by acceleration.
Thus, as $\gamma'_{\rm i,max}$ gradually tends toward $\gamma'_{\rm eq}$ the amount of time lag $|\tau|$ in Case B should approach zero,
and then an increase in $|\tau|$ can be expected as the injected electron spectra gradually become softer.

Moreover, the choice of $\nu_{\rm pk}$ (or $D_0$) is important for determining $|\tau|$ in the cooling-dominated case, 
except for $B'$ and $\delta_{\rm D}$ (see Eq.\ref{lag:cool}).
In the test, we select a smaller $h\nu_{\rm pk}\simeq 1$ keV,
while only $q_0'$ and $p$ are adjusted to produce a similar SED as Case A.
The resulting SED and time lags are displayed in Fig.\ref{figs:fig11},  denoted by blue solid lines.
It is referred to as Case C. 
In the case, the amount of the time lag $|\tau_{\rm f, C}|$ in the Fourier domain increases with decreasing of the frequencies, 
ended by a weakly frequency-dependent spectrum that is similar with that in the acceleration-dominated case.
Meanwhile, a slightly larger $|\tau_{\rm c,C}|=19.7$ minutes is measured using CCF centroid, although it does not significantly differ from $|\tau_{\rm c,A}|=16.3$ in Case A.
This is because the decrease in $\nu_{\rm pk}$ results in the dominance of cooling processes over the electrons 
responsible for the $1-2$ keV X-ray emission, which primarily undergo acceleration in Case A (where $h\nu_{\rm pk}\simeq2.3$ keV).

Given the values of $B'$ and $\delta_{\rm D}$ in Case A, Eq.\ref{lag:cool} yields $|\tau_{\rm A}|\simeq67.7$ minutes, 
which is about $3-4$ times larger than $|\tau_{\rm c,A/C}|$, even about 1.5 times larger than the low-frequency Fourier time lags in Case C.
On the other hand, substituting the values of $t'_{\rm acc}$ and $\delta_{\rm D}$ into Eq.\ref{lag:acc}, one can find $\tau_{\rm B}\simeq120$ minutes in Case B.
The result is about four times larger than the CCF and the low-frequency Fourier time lags.
These imply that the relevant physical quantities inferred using the simple relationships may be significantly overestimated.

In addition, it should be pointed out that the oscillatory behavior at high frequencies is quite complex, and it is difficult to understand 
based on the exact analytical solution given by Eq.\ref{eqs:ntime}.
Our extensive numerical experiments showed that the features may be related to various characteristic timescales associated with various physical processes involved.
However, the features appear at frequencies beyond our interesting frequency range, and we found no significant 
effects on the profiles of the Fourier time lag at low frequencies, as well as on the resulting CCF lag.

\begin{figure*}
\vspace{2.2mm} 
\figurenum{12}
\label{figs:fig12}
\centering
\includegraphics[width=0.45\textwidth] {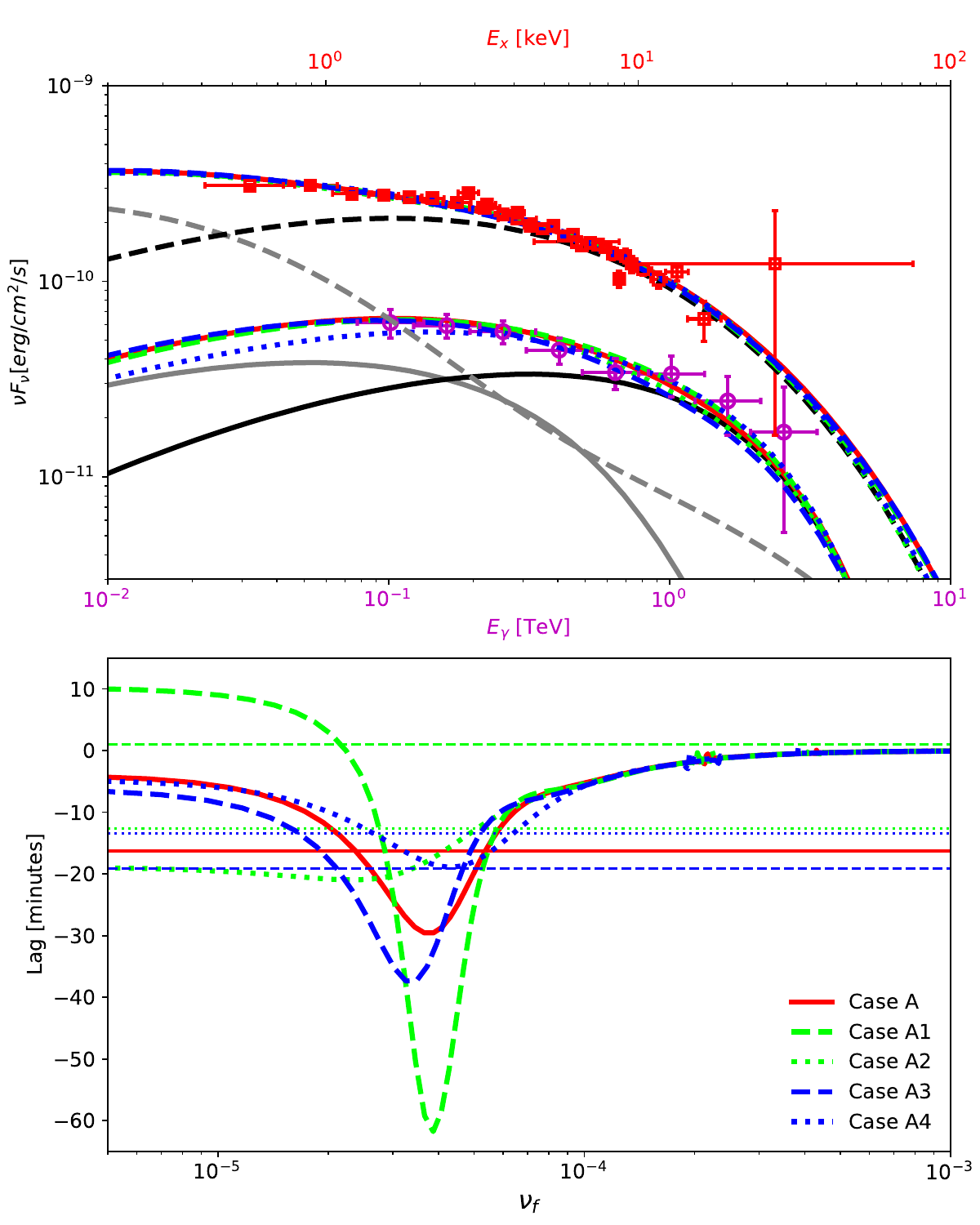} 
\includegraphics[width=0.45\textwidth] {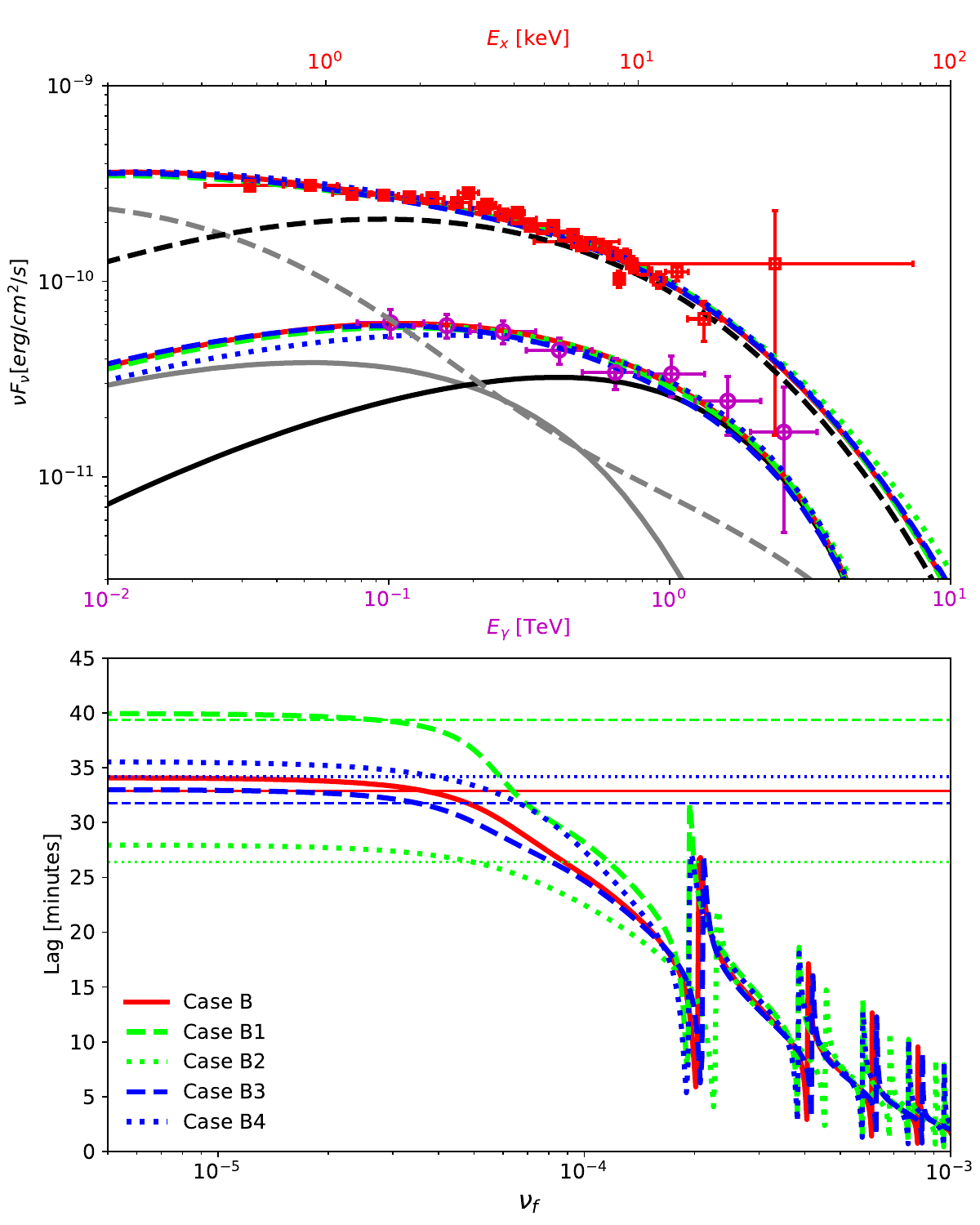} 
\caption{
The theoretical SEDs (upper) and time lags (lower) resulting from variations in $\gamma_{\rm i,min}'$ and $\eta_{\rm esc}$.
In the lower panels, the Fourier time lag profiles for different cases are indicated in the legend, and the corresponding CCF lags are marked with the horizontal line styles.
In the upper panels, the observed data points and theoretical SED from the (quasi-)stationary zone are same as those in Fig.\ref{figs:fig11}.
The synchrotron and SSC spectrum from the non-stationary zone for Case A (left) and B (right) are denoted by the black dashed and solid lines, respectively. 
The SEDs from non-stationary zones are denoted by line styles that are the same as those in the lower panels, 
superimposed with the SED from the (quasi-)stationary zone. }
\vspace{2.2mm}
\end{figure*}

Fig.\ref{figs:fig12} is designed to examine the effects of variations in the parameters $\eta_{\rm esc}$ and $\gamma_{\rm i,min}'$
that are fixed in the previous modeling due to the absence of effective constraints from the observed SED.
For the cooling-dominated case, we have tested two different values of $\eta_{\rm esc}=1.0$ (Case A1) and $1.69$ (Case A2), 
while $\eta_{\rm esc}=1.0$ (Case B1) and $1.54$ (Case B2) are used for the acceleration-dominated case.
For the minimum injected Lorentz factor $\gamma'_{\rm i,min}$, 
we explored the effects of the associated time lags with two different values of $\gamma'_{\rm i,min}=8\times10^2$ (Case A3 and B3) and $10^4$ (Case A4 and B4). 
To minimize the variations in both X-ray and TeV $\gamma$-ray spectra as much as possible, 
several other physical quantities are varied around Case A (or B).

As can be seen from Tab.\ref{tab:para1}, the presence of a small hard CCF lag $\tau_{\rm c,A1}$ in Case A1 cannot be attributed to 
a minor decrease in $\delta_{\rm D}$ due to $|\tau|\propto 1/\sqrt{\delta_{\rm D}}$,
while only a minor increase in $\delta_{\rm D}$ is not enough to produce the smaller soft CCF lag of $\tau_{\rm c, A2}\simeq0.8\tau_{\rm c,A}$ in Case A2.
Similarly, the measured CCF lags of $\tau_{\rm c,A3}\simeq1.2\tau_{\rm c,A}$ in Case A3 is inconsistent with 
the expected smaller soft lags arising from an increase in $\delta_{\rm D}$,
and the measured CCF lags of $\tau_{\rm c,A4}\simeq0.8\tau_{\rm c,A}$ conflict with the expected larger soft lags caused by a decrease in $\delta_{\rm D}$.
In fact, the discrepancies are primarily related to small changes in $D_0$ and $\gamma'_{\rm i,max}$.
More detailed information of the variations in the physical quantities can be observed in the frequency-dependent spectra presented in Fig. \ref{figs:fig11}.

For the acceleration-dominated cases where $\tau\propto 1/\sqrt{ \nu_{\rm pk}\delta_{\rm D}}$, 
our results show that $\tau_{\rm c,B1/B2}/\tau_{\rm c,B}\simeq1.2 (0.8)$ is comparable with 
$\sqrt{[\nu_{\rm pk}\delta_{\rm D}]_{\rm B}/[\nu_{\rm pk}\delta_{\rm D}]_{\rm B1/B2}}\simeq1.1 (0.9)$.
The small discrepancies are due to small adjustments in $p$.
For Case B3 and B4, we find that $\tau_{\rm c,B3/B4}/\tau_{\rm c,B}\simeq\sqrt{[\nu_{\rm pk}\delta_{\rm D}]_{\rm B}/[\nu_{\rm pk}\delta_{\rm D}]_{\rm B3/B4}}\simeq0.97(1.05)$. 
Moreover, the measured CCF lags are very close to the low-frequency Fourier time lags.

It is clear that the time lags resulting from our model exhibit a substantial dependence on the specific choice of $\eta_{\rm esc}$, 
accompanied by small variations in the relevant physical quantities.
These suggest that a modest adjustment to $\eta_{\rm esc}$ will not significantly affect our results inferred from the modeling of the measured time lags.
On the other hand, we find that changes in the theoretical time lags and the relevant physical quantities 
are very small, even though $\gamma'_{\rm i,min}$ varies by an order of magnitude.
This can be easily understood since the X-rays are predominatly emitted by the highest energy electrons.

\bibliography{sample631}{}
\bibliographystyle{aasjournal}


\end{CJK*}
\end{document}